\documentclass[journal, onecolumn]{IEEEtran}
\usepackage{graphicx}
\usepackage{amsmath,amssymb,dsfont,bbm,epstopdf,pgfplots,mathtools,enumitem,mathrsfs, bbm} 
\usepackage{color}
\usepackage{cite}
\usepackage{graphics}
\usepackage{tikz}
\usepackage{pgfplots}
\usepackage{subcaption}
\usetikzlibrary{shapes, arrows, decorations.markings, arrows.meta}
\usetikzlibrary{patterns} 
\allowdisplaybreaks
\graphicspath{{.}{./Figures/}} 
\allowdisplaybreaks[4] 

\newtheorem{corollary}{Corollary}
\newtheorem{theorem}{Theorem}
\newtheorem{proposition}{Proposition}
\newtheorem{remark}{Remark}
\newtheorem{definition}{Definition}

\newcommand{\tk}{\tilde{k}}

\newcommand{\Nk}{{ \mathcal{N}_{\textnormal{k}}}}

\newcommand{\Tf}{{\mathcal K_{\text{URLLC}}} }
\newcommand{\Ts}{{\mathcal K_{\text{eMBB}}} }
\newcommand{\Ta}{{\mathcal K_{\text{active}}} }

\renewcommand{\Pr}{{\mathbb{P}}}

\newcommand{\sRks}{{\{1,\ldots,\lfloor 2^{n R_k^{(\e)}} \rfloor \}}}

\renewcommand{\S}{{\sf{S}}}

\renewcommand{\i}{{\iota}}

\newcommand{\vect}[1]{\boldsymbol{#1}}

\newcommand{\Rx}{\textnormal{Rx}}

\newcommand{\MkF}{M_k^{(\U)}}
\newcommand{\MkS}{M_k^{(\e)}}

\definecolor{ForestGreen}{rgb}{0.0, 0.5, 0.0}

\newcommand{\mw}[1]{{\color{magenta}#1}}

\newcommand{\hn}[1]{{\color{black}#1}}

\renewcommand{\P}{\mathsf{P}}
\newcommand{\D}{\text{D}}
\renewcommand{\P}{\mathsf{P}}

\renewcommand{\Nk}{\mathcal{N}_{k}}

\newcommand{\Ntka}{\mathcal{N}_{k,\textnormal{Tx,active}}}
\newcommand{\Nrka}{\mathcal{N}_{k,\textnormal{Rx,active}}}

\newcommand{\ai}{\mathcal K_i}

\newcommand{\M}{\mathsf{M}}

\newcommand{\U}{\mathsf{U}}
\newcommand{\e}{\mathsf{e}}

\newcommand{\Dt}{\D_{\text {Tx}}}
\newcommand{\Dr}{\D_{\text {Rx}}}

\begin{document}
\title{Interference Networks with Random User Activity and Heterogeneous Delay Constraints}
\author{\hspace{-0.25cm}\IEEEauthorblockN{Homa Nikbakht,~\IEEEmembership{Member, IEEE,}  Mich\`ele Wigger,~\IEEEmembership{Senior Member, IEEE}, Shlomo Shamai (Shitz),~\IEEEmembership{Life Fellow, IEEE}, \\ Jean-Marie Gorce, ~\IEEEmembership{Senior Member, IEEE},  and  H. Vincent Poor, \IEEEmembership{Life Fellow, IEEE }}

\thanks{H.~Nikbakht and H.~V.~Poor are with the Department of Electrical and Computer Engineering, Princeton University, NJ, USA (email: {homa, poor}@princeton.edu). M.   Wigger is with  LTCI, T$\acute{\mbox{e}}$l$\acute{\mbox{e}}$com Paris, IP Paris, 91120 Palaiseau, France (e-mail: michele.wigger@telecom-paris.fr). S. Shamai, is with the Department of Electrical Engineering, Technion---Israel Institute of Technology, Technion City, 32000, Israel, (e-mail: sshlomo@ee.technion.ac.il). J-M.~Gorce is with the INRIA and University of Lyon, CITI, INSA Lyon, 69100 Villeurbanne, France (e-mail: jean-marie.gorce@inria.fr). Part of this work has been presented at IEEE ITW 2020 \cite{HomaITW2020} and IEEE ITW 2021  \cite{HomaITW2021}.
The work of H.~Nikbakht and H.~V.~Poor has been supported by the U.S National Science Foundation under Grants CNS-2128448 and ECCS-2335876. The work of S. Shamai has been supported by the German Research Foundation (DFG) via the German-Israeli Project Cooperation (DIP), under Project SH 1937/1-1. The work of JM Gorce has been supported by the French National Agency for Research (ANR) via the project n°ANR-22-PEFT-0010 of the France 2030 program.
}
}
\maketitle

 \begin{abstract}
To answer the call for a new theoretical framework to  simultaneously accommodate random user activity and heterogeneous delay traffic in Internet of Things (IoT) systems, in 
this paper we propose coding schemes and information-theoretic converse results for the transmission of heterogeneous delay traffic over interference networks with random user 
activity and random data arrivals. The heterogeneous traffic is composed of delay-tolerant traffic and delay-sensitive traffic where only the former can benefit from  transmitter and 
receiver cooperation since the latter is subject to stringent decoding delays. The total number of cooperation rounds at transmitter and receiver sides is limited to $\D$ rounds. Each 
transmitter is active with probability $\rho \in [0,1]$. We  consider  two different models for the arrival of the mixed-delay traffic: in Model~$1$, each active transmitter  sends  a delay-
tolerant  message, and  with probability $\rho_f \in [0,1]$ also transmits an additional delay-sensitive message; in Model~$2$, each active transmitter  sends either a delay-sensitive 
message with  probability $\rho_f$ or a delay-tolerant message with probability $1- \rho_f$.   We derive inner and outer bounds on the fundamental per-user multiplexing gain (MG) region of the 
symmetric Wyner network as well as inner bounds on the fundamental MG region of the hexagonal model.  Our inner and outer bounds are generally very close and coincide in 
special cases. They also show that when both transmitters and receivers can cooperate, then under Model~$1$, transmitting delay-sensitive messages hardly causes any penalty 
on the sum per-user MG, and under Model~$2$,  operating at large delay-sensitive per-user MGs incurs no penalty on the delay-tolerant per-user MG and thus increases the sum per-user MG. However, 
when only receivers can cooperate, the maximum delay-tolerant per-user MG  that our bounds achieve at maximum delay-sensitive per-user MG  is significantly decreased.  
\end{abstract}

\section{Introduction}
A massive Internet of Things (IoT) connectivity system  consists of a large number of connected devices.  At any given time, in order to save energy, a fraction of such devices might be in silence mode \cite{Chettri2020, Ngo2011}, while others send (or receiver)  delay-sensitive or delay-tolerant data to comply with the heterogeneous delay-requirements of modern wireless networks \cite{Khan2022, HomaEntropy2022, Chen2023}. An example of such mixed-delay traffic is the coexistence of enhanced mobile broadband (eMBB) and ultra-reliable low-latency communication (URLLC),  which are two notable services for IoT \cite{Yang2021}. Data-intensive IoT applications such as industrial video surveillance and augmented virtual reality are served by eMBB, whereas, URLLC serves time-sensitive and mission-critical IoT applications such as motion control and autonomous driving \cite{Ren2020, Kassab2019-2, Amjad2019}.   This heterogeneity of purpose calls for a new theoretical framework to  simultaneously accommodate random user activity and heterogeneous delay traffic. 

\subsection{Random User Activity} Networks with random user activities have been previously studied in \cite{ Liu2018, Chen2018, Wang2021, Somekh2008,  Somekh2009, Levy2008}.
 In \cite{Liu2018}, the authors consider an uplink massive device communication scenario in which, in any given  coherence interval only a subset of users are active, and study the performance of such a system in the asymptotic multiple-input multiple-output (MIMO) regime. They show that by using compressed sensing techniques that take into account the sparsity of the user activity pattern \cite{Chen2018},  one can make both the missed-detection and false alarm probabilities go to zero. This result however is based on the assumption that if one device is active for one antenna, it is also active for all the other antennas, which comes at the cost of high computational complexity in practice. To reduce this computational complexity, the work in \cite{Wang2021} proposes to leverage the \emph{ temporal correlation in user activity} which means an active device at the previous time slot is more likely to stay active at the current time slot.

More related to this work are \cite{Somekh2008,  Somekh2009, Levy2008}. In \cite{Somekh2008}, the authors consider a cellular uplink model in which each user is randomly active with probability $\rho \in [0,1]$. The transmission of each active user then is subject to an erasure  due to ergodic shadowing. Erasures are assumed to be i.i.d Bernoulli distributed random variables. As a result of such assumptions, they show that  by using multi-cell processing the throughput of this network is equal to the rate of a two-tap input-erasure channel divided by $\rho$. The work in \cite{Somekh2009} considers a cellular uplink model in which each cell consists of one base station and $K$ users (the case of $K = 1$ has been studied in \cite{Levy2008}). During each transmission slot, users are randomly and independently active with probability $\rho$. The results in \cite{Somekh2009} show that multi-cell processing outperforms single-cell processing in the presence of random user activity.

\subsection{Heterogeneous Delay Traffic}
IoT systems  have to accommodate  heterogeneous traffic composed of delay-sensitive and delay-tolerant data.
 Coding schemes for  such heterogeneous delay traffic are thus of considerable interest to the designers of IoT systems covering a wide range of communication aspects from terahertz (THz) transmissions to vehicular communications  \cite{Zarini2023, Chen2020, Yin2021, Song2019,  Kassab2018, HomaGlobecom2023, Zhao2022, HomaITW2019,  shlomo2012ISIT, Zhang2005, Zhang2008, Zhang2016}.  In particular, the work in \cite{Zarini2023} proposes a reconfigurable intelligent surface (RIS)-aided THz communication system to support the coexistence of URLLC and eMBB services in which the transmission of  eMBB type services are punctured  in favor of the URLLC service. Such a puncturing approach has been also employed in vehicular communications. For example, in \cite{Chen2020},  URLLC transmissions are allowed to puncture the eMBB transmissions upon arrival. The work in \cite{Yin2021} formulates an optimization problem that maximizes the  aggregate utility of the  punctured eMBB users subject to latency constraints for the URLLC users. In  \cite{Song2019}, the time slot allocated to eMBB   transmission is divided into mini-slots and the newly arrived URLLC messages are immediately scheduled in the next mini-slot by puncturing the on-going eMBB transmissions. In \cite{HomaGlobecom2023} we instead use a joint coding approach in which the transmission of URLLC messages is  reinforced  by mitigating the interference of eMBB transmission by means of  dirty paper coding \cite{Costa1983, Scarlett2015, Caire2003}. 

More related to this work are \cite{Kassab2018, HomaITW2019, shlomo2012ISIT}. In particular, \cite{Kassab2018} studies a cloud radio access network (C-RAN) under mixed-delay-constraints traffic where URLLC messages are transmitted only by  users close to  base stations (BS)  and are directly decoded at the BSs, whereas, eMBB messages are transmitted by users that are further away and are decoded at the central processor.
 In \cite{HomaITW2019}, we extended the above C-RAN model to allow each user to send  both  eMBB and URLLC  messages, and to time-varying fading channels. 
The  results in \cite{HomaITW2019} show 
that in most regimes,    the stringent delay constraint on URLLC messages  penalizes the overall performance (sum-rate) of the system. 
The work in  \cite{shlomo2012ISIT} proposes a superposition approach over a fading channel to   communicate  URLLC messages within single coherence blocks and  eMBB messages  over   multiple blocks. 

\subsection{Contributions of this Work}
In this paper, we propose coding schemes and information-theoretic converse results for interference networks with  heterogeneous delay-sensitive and delay-tolerant traffic types under random user activity and random data arrivals. Delay-sensitive data, known as URLLC messages, are subject to stringent delay constraints and their encoding and decoding processes cannot be delayed. Delay-tolerant data, known as eMBB messages, are subject to softer delay constraints and can benefit from transmitter (Tx) and receiver (Rx) cooperation. We specifically consider two setups.  In the first setup, both Rxs and Txs  can cooperate to decode their desired  eMBB messages but not to decode URLLC messages. In the second setup, only Rxs can cooperate to decode their desired  eMBB messages. Cooperation is assumed to take place over dedicated links and during a limited number of  $ \D$ rounds. In both setups, each Tx is active with probability $\rho \in [0,1]$, and the goal is to maximize the average expected eMBB rate of the network, while the rate of  each URLLC  message is fixed to a target value because URLLC messages have to be entirely transmitted in a blocklength and cannot be postponed to future transmissions. For both setups, we study two models for the arrival of the mixed-delay traffic: 
\begin{itemize}
\item  In Model~$1$, each active Tx  transmits  an eMBB message, and  with probability $\rho_f \in [0,1]$ also transmits an additional URLLC message. 
\item In Model~$2$, each active Tx  sends either a URLLC message with  probability $\rho_f$ or an eMBB message with probability $1- \rho_f$.
\end{itemize} The first model is motivated by the random appearance of only control data (i.e., URLLC type data) whereas the second model is motivated by the random appearance of both control data and standard delay-tolerant data (i.e., eMBB type data).

In this work, we propose   coding schemes for general  interference networks for both cooperation setups (Tx- and Rx-cooperation or Rx-cooperation only) and both user activity models (Model 1 or Model 2). In our coding schemes we take advantage of the fact that  the network can be decomposed into smaller non-interfering subnets because non-active Txs remain silent. In the resulting subnets, we time-share a set of schemes so that all Txs can schedule their URLLC transmissions in at least one of the schemes (URLLC transmissions cannot be postponed) and in each scheme we schedule URLLC transmissions  without interfering with each other. We further schedule a maximum number of eMBB transmissions on the other users in a way to allow mitigating their interference using the $\D$ rounds of cooperation. Specifically, the interference of both URLLC and eMBB transmissions on other eMBB transmissions is mitigated at the Rx-side using interference mitigation and Coordinated Multi-Point (CoMP) reception. Since in our model,  CoMP reception can only benefit from a limited number of cooperation rounds, this requires   further silencing some of the Txs to ensure that the network decomposes into sufficiently small subnets \cite{Roy},  each one of a size that allows eMBB transmissions to be decoded by CoMP reception with a limited number of cooperation rounds. When also Txs can cooperate, then interference of eMBB Txs on URLLC Txs is mitigated using dirty-paper coding \cite{Costa1983} and a single round of Tx-cooperation. When only Rxs can cooperate but not Txs, it seems impossible to cancel interference from eMBB transmissions onto URLLC transmissions because the latter have to be decoded without further due. In our scheme  we thus schedule fewer eMBB transmissions under Rx-cooperation  only than when both Txs and Rxs can cooperate, which comes at the expense of a  reduced eMBB multiplexing gain (MG).

The focus of this paper is on  Wyner's symmetric network and the hexagonal network. In Wyner's symmetric network,  Txs are aligned on a one-dimensional grid and cooperation is only between subsequent Txs and subsequent Rxs.  In the hexagonal network, cells are distributed in two dimensions, each cell has a hexagonal shape and six neighbors where its signals interfere and with which it can cooperate. For these two regular network models,  we  present improved coding schemes that better adapt to the random activity and arrival patterns. The per-user MG regions achieved by these adaptive schemes significantly improve over our first general non-adaptive schemes and moreover have performances close to the fundamental limits. Indeed, we also present information-theoretic converse results on the set of all achievable per-user MG pairs in Wyner's symmetric network and we show that for the setup with both Tx-and Rx-cooperation our improved schemes (which we also call \emph{adaptive schemes})  perform close to (and sometimes match)  our converse bounds. 

Our results thus allow us to derive tight approximations of the per-user MG regions for Wyner's symmetric network under mixed-delay traffic and with random user activity and data arrivals as functions of the maximum number of cooperation rounds $\D$. Numerical evaluation of our  results show that for moderate activity parameters $\rho$ a few cooperation rounds suffice to attain performance close to the limit as $\D\to \infty$. We further observe that while under Model 1 the per-user MG region is mostly limited by a single constraint on the URLLC per-user MG and a sum constraint on the URLLC and eMBB per-user MGs, under Model 2 we essentially have two single constraints on the URLLC and the eMBB per-user MG respectively. This is mostly related to the fact that under Model 1 URLLC rate can be transfered to eMBB rate, while this is not possible in Model 2 as any Tx has only one of the two messages to send. We further observe that under Model 1 there is essentially no loss in overall performance of the system (sum per-user MG) whether one sends at low or high URLLC per-user MG.  For Model~2, the overall performance even increases by transmitting at high URLLC per-user MGs, and in particular the eMBB per-user MG does not decrease.  These observations should be put in contrast to the performance of scheduling systems that transmit URLLC and eMBB messages in orthogonal (e.g., time- or frequency) slices and for which the eMBB per-user MG and the sum per-user MG decrease linearly for increasing URLLC per-user MG.

The situation is different when only Rxs can cooperate, in which case the eMBB and/or  sum per-user MGs degrade when one  transmits at high URLLC per-user MG. 
In particular, our converse result  proves that for large URLLC per-user MGs the performance loss in eMBB per-user MG that our schemes (both the adaptive and the non-adaptive ones) incurr compared to the setup with both Tx- and Rx-cooperation is of a fundamental nature in the sense that any  schemes suffers from this degradation.  In fact, if Txs cannot cooperate it does not seem to be possible to mitigate interference of eMBB transmissions on URLLC transmissions because Rx-cooperation cannot serve this purpose as URLLC messages have to be decoded immediately without further delay.

To summarize, the main contributions of this work are:
\begin{itemize}
\item We propose a  coding scheme for general cooperative interference networks with random user activity and random data arrivals under Models 1 and 2 and the two cooperation setups (both  Tx- and Rx-cooperation  or  only Rx-cooperation). 
\item For Wyner's symmetric network and the hexagonal network we propose improved coding schemes that better adapt to the random activity and arrival patterns. For the symmetric Wyner network this improved scheme allows us to show that for moderate activity parameters the optimal performance can already be achieved with a small number of cooperation rounds $\D$. In fact, when only a moderate number of users are active the network can be decomposed into relatively small subnets, and a small number of cooperation rounds suffices to achieve optimal performance for delay-tolerant messages in each subnet.


\item  For Wyner's symmetric network we also present information-theoretic converse results. They are close to the performance of our improved schemes and match them in special cases.

\item Our inner and outer bounds for the symmetric Wyner model allow us to conclude that when both Txs and Rxs can cooperate, then  under Model 1,  transmitting at a high  URLLC per-user MG  does not impose any penalty on the overall performance of the system, i.e., the  sum of URLLC and eMBB per-user MG. For  Model~$2$,  the sum per-user MG increases with the URLLC per-user MG, and more specifically  operating at large URLLC per-user MG does not decrease the achievable eMBB MG. 
\item  Our converse result for the setups with only Rx-cooperation proves that the degradation in eMBB per-user MG at large URLLC MGs incurred by our schemes is inherent to any scheme and not an artifact of our proposition.
\end{itemize}

\subsection{Organization}
The rest of this paper is organized as follows. We end this section with some remarks on notation. Section~\ref{sec:DescriptionOfTheProblem}  describes the general problem setup. Sections~\ref{sec:both} and \ref{sec:onlyRx} present our coding schemes with both Tx and Rx- cooperation and with only Rx-cooperation, respectively. Section~\ref{sec:symmetric} and \ref{sec:hexagonal} discuss our main results for the symmetric Wyner  and  the hexagonal networks, respectively.   Section~\ref{sec:conclusion} concludes the paper. Technical proofs are deferred to  appendices. 

\subsection{Notation}
 The set of all integers is denoted by  $\mathbb Z$, the set of positive integers by $\mathbb Z ^{+}$ and the set of real numbers by $\mathbb R$. For other sets we use calligraphic letters, e.g., $\mathcal{X}$.  Random variables are denoted by uppercase letters, e.g., $X$, and their realizations by lowercase letters, e.g., $x$. For vectors we use boldface notation, i.e., upper case boldface letters such as $\mathbf{X}$  for random vectors and lower case boldface letters such as $\mathbf{x}$ for deterministic vectors. Matrices are depicted with sans serif font, e.g., $\mathsf{H}$. We also write $X^n$ for the tuple of random variables $(X_1, \ldots, X_n)$ and  ${\mathbf X}^n$ for the tuple of random vectors   $( \mathbf{X}_1, \ldots,\mathbf{X}_n)$.

\section{Message Arrival and Communication Model}\label{sec:DescriptionOfTheProblem}



Consider a cellular network with $K$ transmitters (Tx) and ${K}$ corresponding receivers (Rx). Define $\mathcal K \triangleq \{1, \ldots, K\}$. Various network interference structures will be considered and are described later. 

Each Tx~$k \in \mathcal K$ is \emph{active} with probability $\rho \in [0, 1]$. Non-active Txs remain silent and do not participate in the communication. We consider two different models: 
\begin{itemize}
\item  \textit{Model 1:} Each active  Tx sends a so called eMBB message $\MkS$ to its corresponding Rx~$k$. With probability $\rho_{f}\in [0,1]$,  it  also sends an additional URLLC message $\MkF$ to Rx~$k$. These URLLC messages are subject to  stringent delay constraints, as we describe shortly.
\item  \textit{Model 2:} Each active  Tx sends with probability $(1-\rho_f)$ an  eMBB message $\MkS$ to its corresponding Rx~$k$ and otherwise (i.e., with probability $\rho_f$)  it  sends a URLLC message $\MkF$ to Rx~$k$. 
\end{itemize}
 The only difference between the two models is that  under Model 1 any active Tx sends an eMBB message whereas under Model 2 an active Tx  only sends an eMBB message if it does not send a URLLC message. In both models, the various transmitted eMBB messages in the network can be of different rates while the URLLC messages are all of same rate $R^{(\U)}$.

Let  $A_k=1$ if Tx~$k$ is active and $A_k=0$ if Tx~$k$ is not active. Moreover, if Tx~$k$ is active and has a URLLC message to send,  set $B_k=1$, and otherwise 
set $B_k=0$. The random tuple $\mathbf{A}:=(A_1, \ldots, A_K)$  is thus independent and identically distributed (i.i.d.) Bernoulli-$\rho$, and if they exist the random variables $B_1, \ldots, B_K$ are  i.i.d. Bernoulli-$\rho_f$.  Denote by $\mathbf{B}$ the tuple of $B_k$'s that are defined. Further, define  the \emph{active set} and the \emph{URLLC set} as follows:
\begin{IEEEeqnarray}{rCl}\label{eq:Ta}
\Ta&  \triangleq& \{k \in  \mathcal K : A_k = 1\},\\
\Tf &\triangleq& \{ k \in \mathcal K\colon   B_k = 1\}.\label{eq:Tf}
\end{IEEEeqnarray}
Under Model 1, the set  $\Ts$ of Txs transmitting an eMBB message is 
\begin{equation}
\Ts \triangleq \Ta
\end{equation}
and under Model 2, it is
\begin{equation}
\Ts \triangleq \Ta\backslash \Tf.
\end{equation}

We shall assume that each eMBB message  $\MkS$ is uniformly distributed over a set $\mathcal{M}_{k}^{(\e)} \triangleq\sRks$, with $n$ denoting the blocklength and $R_k^{(\e)}$ the rate of message $\MkS$, which we allow to vary across users and can even depend on the random activity parameters $\mathbf{A}$ and $\mathbf{B}$. In contrast, each URLLC message  is  uniformly distributed over the set $\mathcal{M}^{(\U)}\triangleq \{1, \ldots, \lfloor 2^{nR^{(\U)}}  \rfloor \}$, where the constant rate  $R^{(\U)}$ can depend on the network structure but not on the realization of the random activity parameters. The motivation to consider these asymmetric rate constraints is that URLLC messages have stringent delay constraint and have to be transmitted immediately in the next-following block. The rate $R^{(\U)}$ needs to be guaranteed under any circumstances. In contrast, eMBB messages can be postponed to following blocks if required and in each transmission block the goal is to transmit as much eMBB rate as possible. In this sense, one can understand the setup as maximizing the expected eMBB message rate while ensuring a minimum rate for  URLLC messages.

We  describe the encoding at the active Txs. The encoding  starts with a first \emph{Tx-cooperation phase} which consists of $\Dt>0$ rounds and depends only on the eMBB messages in the system. The  URLLC messages, which are subject to stringent delay constraints, are only  generated afterwards, at the beginning of the subsequent   \emph{channel transmission phase}. So, during the first Tx-cooperation phase, neighboring active Txs  communicate with each other over dedicated noise-free links of unlimited capacity over $\Dt>0$ rounds. 

 In each 
 Tx-cooperation round $j\in\{1,\ldots, \Dt\}$,  any active Tx~$k\in \Ta$ sends  a cooperation message to its active neighbors in the network  $\ell \in \Ntka \triangleq \Nk \cap \Ta$, where $\Nk$ indicates the set of  neighboring Txs of Tx~$k$.  (The  sets $\Nk$ depend on the specific network structure  and will be specified latter for the various considered network models.)

 Each cooperation message can depend on the Tx's eMBB message and the cooperation-information received during previous rounds. So, in round $j$, Tx~$k\in \Ta$ sends a message
	\begin{equation}
	T_{k\to \ell}^{(j)}  =\begin{cases} \psi_{k\to \ell}^{(j)} \Big(M_{k}^{(\e)}, \big\{T_{\ell' \to k}^{(1)}, \ldots, T_{\ell'\to k}^{(j-1)} \big\}_{\ell' \in \Ntka},\mathbf{A},\mathbf{B} \Big), &  \quad k\in\Ts \\
 \psi_{k\to \ell}^{(j)} \Big(\big\{T_{\ell' \to k}^{(1)}, \ldots, T_{\ell'\to k}^{(j-1)} \big\}_{\ell' \in \Ntka},\mathbf{A},\mathbf{B} \Big), & \quad k\notin \Ts \end{cases}.
	\end{equation}
	to each Tx~$\ell \in \Ntka$, for some functions $\psi_{k\to \ell}^{(j)}$ on appropriate domains. 
	
At the beginning of the subsequent channel transmission phase, URLLC messages are generated and each active Tx $k \in \Ta$ computes its channel inputs $ X_k^n\triangleq ( X_{k,1},\ldots, X_{k,n}) \in \mathbb R^{n}$ as follows:
\begin{IEEEeqnarray}{rCl}
X_{k}^n = \begin{cases}f_k^{(B)} \big( \MkF, \MkS,  \{T_{\ell' \to k}^{(j)}\},\mathbf{A},\mathbf{B}\big), & k \in \Tf \\
f_k^{(\e)} \big( \MkS,  \{T_{\ell' \to k}^{(j)}\}, \mathbf{A},\mathbf{B}\big), & \hspace{-.7cm} k \in \Ts.
\end{cases},
\end{IEEEeqnarray}
where the sets $ \{T_{\ell' \to k}^{(j)}\}$ run  over indices  $j \in \{1, \ldots, \Dt \}$ and  $\ell' \in \Ntka$, and  $f_k^{(B)}$ and  $f_k^{(\e)}$ are  encoding functions  on appropriate domains satisfying 
the average block-power constraint
\begin{equation}\label{eq:power}
\frac{1}{n} \sum_{t=1}^n X_{k,t}^2
\leq \P, 
\quad \forall\ k \in \Ta, \qquad \textnormal{almost surely.}
\end{equation}
Inactive Txs simply send the all-zero sequence.

We shall assume a Gaussian communication network described as
\begin{equation}\label{Eqn:Channel}
{Y}_{k,t} = \sum_{\tilde{k} \in \{ \mathcal{I}_k\cap \Ta \}} h_{\tilde{k},k} 
{X}_{ \tilde k,t}+ {Z}_{k,t},
\end{equation}
where $\{Z_{k,t}\}$ are i.i.d. standard Gaussian random variables for all $k$ and $t$ and are independent of all messages; $h_{\tilde k , k}\geq 0$ describes  the channel coefficient between Tx~$\tilde k$ and Rx~$k$ and is  a fixed real number; and $X_{0,t} = 0$ for all $t$. In our model we assume that the received signal $Y_{k,t}$ is only interfered with by the signals from  Txs in $\mathcal{I}_k$,  if these neighboring Txs are active. We will  assume short range interference,  i.e., that only signals from Txs that are sufficiently close interfere because the disturbance from further away Txs are below the noise level. This implies that any Tx~$k$ can cooperate with the Txs whose signals interfere at Rx $k$, i.e.,:
\begin{equation}
\mathcal I_k  \subseteq \Nk.
\end{equation}

Decoding also takes place in two phases. In the first \emph{URLLC-decoding phase}, any active  Rx~$k \in \Tf$ decodes the URLLC message $M_k^{(\U)}$ based on its own channel outputs $\vect Y_k^n$ by computing
\begin{align}
	&\hat{{{M}}} _k^{(\U)} ={g_k^{(n)}}\big( \vect Y_k^{n},\mathbf{A}, \mathbf{B}\big),
	\end{align} 
for some decoding function $g_k^{(n)}$ on appropriate domains.	
In the subsequent \emph{eMBB-decoding phase},  active Rxs first communicate with their active neighbors in the network  $\ell \in \Nrka \triangleq \Nk \cap \Ta$,  during $\Dr \ge 0$ rounds over dedicated noise-free links with unlimited capacity, and then they decode their intended eMBB messages based on their outputs and based on this exchanged information.   
Specifically, in each cooperation round $j\in \{1, \ldots, \Dr \}$,  each active Rx~$k\in \mathcal T_{\text{active}}$ sends  a cooperation message  
\begin{equation}
Q^{(j)}_{k\rightarrow \ell}=\phi_{k\to \ell}^{(j)}\Big( \mathbf Y_k^n,\big\{ Q^{(1)}_{\ell' \rightarrow k},\ldots , Q^{(j-1)}_{\ell' \rightarrow k}\}_{\ell' \in \Nrka},\mathbf{A},\mathbf{B}\Big)
\end{equation}
to Rx~$\ell$ if  $\ell \in  \Nrka$ for some appropriate function $\phi_{k\to \ell}^{(j)}$.

After the last cooperation round, each active Rx~$k\in \Ts$ decodes its desired eMBB message as
\begin{equation}\label{mhats}
\hat{{M}}_{k}^{(\e)}={b_{k}^{(n)}}\Big( \mathbf Y_{k}^n, \Big\{ Q^{(1)}_{\ell'\rightarrow k}, \ldots, Q^{(\Dr)}_{\ell'\rightarrow k}\Big\}_{\ell' \in \Nrka},\mathbf{A},\mathbf{B}\Big),
\end{equation}
 where  $b_{k}^{(n)}$ is a decoding function on appropriate domains.

%

\begin{definition}		
Given $\P>0$ and $K>0$, a rate pair $(R^{(\U)}(\P), \bar{R}_K^{(\e)}(\P))$ is said to be $\D$-\emph{achievable} if there exist rates $\{R_{k}^{(\e)}\}_{k=1}^{K}$ satisfying
\begin{IEEEeqnarray}{rCl}
\bar R_K^{(\e)} &\le &  \mathbb E_{\mathbf{A},\mathbf{B}} \left [\sum_{k \in \Ts } \frac{1}{K} R_{k}^{(\e)} \right ] ,\label{eq:RS}
\end{IEEEeqnarray} 
a pair of Tx- and Rx-cooperation rounds $\Dt,\Dr$ summing to $\Dt+\Dr=\D$ and
encoding, cooperation, and decoding functions   satisfying the power constraint \eqref{eq:power} and so that the probability of error
\begin{equation}
\Pr \bigg[\bigcup_{k \in \Tf} \!\!\big( \hat{M}_k^{(\U)} \neq M_k^{(\U)}\big)  \;\text{or} \! \!\bigcup_{k \in \Ts} \big(\hat{M}_k^{(\e)} \neq M_k^{(\e)}\big)  \bigg] 
\end{equation}
tends to $0$ as $n\to \infty$.

An expected average per-user MG pair  $(\S^{(\U)},\S^{(\e)})$ is called $\D$-\emph{achievable}, if for all powers $\P>0$ there exist $\D$-achievable  rates   $\big\{{R}^{(\U)}(\P),\bar R_K^{(\e)}(\P) \big\}_{\P>0}$
satisfying
\begin{IEEEeqnarray} {rCl}
\S^{(\U)}& \triangleq&\varlimsup_{K\rightarrow \infty}\varlimsup_{\P\rightarrow\infty} \;\frac{ R^{(\U)}(\P)}{\frac{1}{2}\log (\P)} \cdot \rho \rho_f,\label{eq:sf}
\end{IEEEeqnarray}
and
\begin{IEEEeqnarray}{rCl}
\S^{(\e)}& \triangleq&\varlimsup_{K\rightarrow \infty}  \varlimsup_{\P\rightarrow\infty} \; \frac{  \bar R_K^{(\e)}(\P)}{\frac{1}{2}\log (\P)} .\label{eq:ss}
\end{IEEEeqnarray}

The  closure of the set of all $\D$-achievable   MG pairs $(\S^{(\U)}, \S^{(\e)})$ is called the ($\D$-\emph{cooperative}) \emph{fundamental MG region} and is denoted by $\mathcal{S}_1^\star(\D, \rho, \rho_f)$ and  $\mathcal{S}_2^\star(\D, \rho, \rho_f)$ for Models 1 and 2, respectively.
\end{definition}
\medskip 

\begin{remark}
The MG in \eqref{eq:ss} measures the average expected eMBB MG on the network. Since the URLLC rate is fixed to $R^{(\U)}$ at all Txs  in $ \Tf$, we multiply the MG in \eqref{eq:sf} by $\rho  \rho_f$ to obtain the average expected URLLC MG of the network.  By the definition in \eqref{eq:RS}, the pair $(\S^{(\U)}, \S^{(\e)})$ thus corresponds to the expected per-user MG  pair of the system.
\end{remark}
\medskip 

 \begin{remark}
 In our model, we assume that  neighboring Txs and neighboring Rxs can only cooperate if they lie in the active set $\Ta$. Txs and Rxs in the \emph{inactive set} $\mathcal{K}\backslash \Ta$ do not participate in the cooperation phases. Notice that all our converse (infeasibility) results remain valid in a setup where inactive Txs and Rxs \emph{do} participate in the cooperation phases. Since our inner and outer bounds are rather close in general (see the subsequent numerical discussion), this indicates that without essential loss in optimality Txs and Rxs in $\mathcal{K}\backslash \Ta$  can entirely be set to sleep mode to conserve their batteries.
 \end{remark}
 
 For simplicity, throughout this manuscript we assume that $\D$ is even. In the following Section~\ref{sec:both}, we propose a coding scheme with both Tx and Rx cooperation under the assumption that $\Dt=1$ and $\Dr=\D-1$. In Section~\ref{sec:onlyRx}, we propose a coding scheme with only Rx cooperation, i.e., when $\Dt = 0$ and $\Dr = \D$. 

 \section{Coding Scheme with both Tx- and Rx-Cooperation} \label{sec:both}
 We assume throughout this section that
 \begin{equation}
 \Dt=1 \quad \textnormal{and} \quad \Dr= \D-1.
 \end{equation}
 
 In Subsection~\ref{sec:basic-both}, we  freely choose  a message assignment and present an achievable total MG for this assignment. In the  subsequent Subsection~\ref{sec:random-both} we then build on these results to  describe and analyze a scheme respecting the random message arrivals.
 \subsection{A Basic Scheme with Chosen Message Assignments} \label{sec:basic-both}

 \subsubsection{Creation of subnets and message assignment}
\begin{figure*}[t]
  \centering
\begin{subfigure}{1\textwidth}
\centering
\begin{tikzpicture}[scale=1.6, >=stealth]
\centering
\tikzstyle{every node}=[draw,shape=circle, node distance=0.5cm];
 \foreach \j in {1,...,20} {
 \draw (-3.5 + 0.5*\j, 1.5) circle (0.1);
\node[draw =none] (s2) at (-3.5+ 0.5*\j,1 ) {\footnotesize$+$};
\draw (-3.5 +0.5*\j, 1) circle (0.1);
 \draw (-3.5 + 0.5*\j, 0.5) circle (0.1);
 \draw   [->] (-3.5+ 0.5*\j,1.9-0.5)-- (-3.5+ 0.5*\j,1.1);
 \draw   [->] (-3.5+ 0.5*\j,0.9)-- (-3.5+ 0.5*\j,0.6);
 }
  \foreach \j in {1,...,6,10,14,15,16,17,18} {
  \draw   [->, dashed] (-3.5+ 0.5*\j,1.9-0.5)-- (-3.5+ 0.5*\j + 0.5,1.1);
  }
    \foreach \j in {2,...,7,11,15,16,17,18,19} {
  \draw   [->, dashed] (-3.5+ 0.5*\j,1.9-0.5)-- (-3.5+0.5*\j - 0.5,1.1);
 }
  \foreach \j in {1,3,7,11,15,19} {
 \draw [fill = yellow](-3.5 + 0.5*\j, 1.5) circle (0.1);
\node[draw =none] (s2) at (-3.5+ 0.5*\j,1 ) {\footnotesize$+$};
\draw (-3.5 +0.5*\j, 1) circle (0.1);
 \draw [fill = yellow] (-3.5 + 0.5*\j, 0.5) circle (0.1);
 \draw   [->] (-3.5+ 0.5*\j,1.9-0.5)-- (-3.5+ 0.5*\j,1.1);
 \draw   [->] (-3.5+ 0.5*\j,0.9)-- (-3.5+ 0.5*\j,0.6);
 }
   \foreach \j in {2,4,5,6,10,14,16,17, 18} {
 \draw [fill = blue](-3.5 + 0.5*\j, 1.5) circle (0.1);
\node[draw =none] (s2) at (-3.5+ 0.5*\j,1 ) {\footnotesize$+$};
\draw (-3.5 +0.5*\j, 1) circle (0.1);
 \draw [fill = blue] (-3.5 + 0.5*\j, 0.5) circle (0.1);
 \draw   [->] (-3.5+ 0.5*\j,1.9-0.5)-- (-3.5+ 0.5*\j,1.1);
 \draw   [->] (-3.5+ 0.5*\j,0.9)-- (-3.5+ 0.5*\j,0.6);
 }
 \node[draw =none] (s2) at (-3.5+0.5,0.2) {$1$};
\node[draw =none] (s2) at (-3.5+1,0.2) {$2$};
\node[draw =none] (s2) at (-3.5+1.5,0.2) {$3$};
\node[draw =none] (s2) at (-3.5+2,0.2) {$4$};
\node[draw =none] (s2) at (-3.5+2.5,0.2) {$5$};
\node[draw =none] (s2) at (-3.5+3,0.2) {$6$};
\node[draw =none] (s2) at (-3.5+3.5,0.2) {$7$};
\node[draw =none] (s2) at (-3.5+4,0.2) {$8$};
\node[draw =none] (s2) at (-3.5+4.5,0.2) {$9$};
\node[draw =none] (s2) at (-3.5+5,0.2) {$10$};
\node[draw =none] (s2) at (-3.5+5.5,0.2) {$11$};
\node[draw =none] (s2) at (-3.5+6,0.2) {$12$};
\node[draw =none] (s2) at (-3.5+6.5,0.2) {$13$};
\node[draw =none] (s2) at (-3.5+7,0.2) {$14$};
\node[draw =none] (s2) at (-3.5+7.5,0.2) {$15$};
\node[draw =none] (s2) at (-3.5+8,0.2) {$16$};
\node[draw =none] (s2) at (-3.5+8.5,0.2) {$17$};
\node[draw =none] (s2) at (-3.5+9,0.2) {$18$};
\node[draw =none] (s2) at (-3.5+9.5,0.2) {$19$};
\node[draw =none] (s2) at (-3.5+10,0.2) {$20$};

\node[draw =none] (s2) at (-3.5+0.5,1.75) {$F$};
\node[draw =none] (s2) at (-3.5+1,0.2+1.55) {$S$};
\node[draw =none] (s2) at (-3.5+1.5,0.2+1.55) {$F$};
\node[draw =none] (s2) at (-3.5+2,0.2+1.55) {$S$};
\node[draw =none] (s2) at (-3.5+2.5,0.2+1.55) {$S$};
\node[draw =none] (s2) at (-3.5+3,0.2+1.55) {$S$};
\node[draw =none] (s2) at (-3.5+3.5,0.2+1.55) {$F$};
\node[draw =none] (s2) at (-3.5+5,0.2+1.55) {$S$};
\node[draw =none] (s2) at (-3.5+5.5,0.2+1.55) {$F$};
\node[draw =none] (s2) at (-3.5+7,0.2+1.55) {$S$};
\node[draw =none] (s2) at (-3.5+7.5,0.2+1.55) {$F$};
\node[draw =none] (s2) at (-3.5+8,0.2+1.55) {$S$};
\node[draw =none] (s2) at (-3.5+8.5,0.2+1.55) {$S$};
\node[draw =none] (s2) at (-3.5+9,0.2+1.55) {$S$};
\node[draw =none] (s2) at (-3.5+9.5,0.2+1.55) {$S$};

 \node[draw =none, rotate =45] (s2) at ( 0.5,1.5) {\huge$+$};
  \node[draw =none, rotate =45] (s2) at ( 0.5,0.5) {\huge$+$};
  \node[draw =none, rotate =45] (s2) at ( 6.5,1.5) {\huge$+$};
  \node[draw =none, rotate =45] (s2) at (6.5,0.5) {\huge$+$};
  \foreach \i in {1,3,15} {
  \draw [<-, very thick, blue ] (-3.5 + 0.5*\i+0.1, 1.5) -- (-3.5 + 0.5*\i + 0.5- 0.1, 1.5);
  }
   \foreach \i in {2,6,10,14,18} {
  \draw [->, very thick, blue ] (-3.5 + 0.5*\i+0.1, 1.5) -- (-3.5 + 0.5*\i + 0.5- 0.1, 1.5);
  }
  
   \foreach \i in {2,6,10,14} {
  \draw [<-, very thick, yellow ] (-3.5 + 0.5*\i+0.1, 0.5) -- (-3.5 + 0.5*\i + 0.5- 0.1, 0.5);
  }
  
    \foreach \i in {1,15} {
  \draw [->, very thick, yellow ] (-3.5 + 0.5*\i+0.1, 0.5) -- (-3.5 + 0.5*\i + 0.5- 0.1, 0.5);
  }
     \foreach \i in {18} {
  \draw [<-, very thick, yellow ] (-3.5 + 0.5*\i+0.1, 0.5) -- (-3.5 + 0.5*\i + 0.5- 0.1, 0.5);
  }
   \foreach \j in {4,16} {
 \draw [very thick, green](-3.5 + 0.5*\j, 0.5) circle (0.12);
  }
\end{tikzpicture}
\caption{The first Tx-cooperation round and the first Rx-cooperation round}
\label{fig:symb}
\end{subfigure}

\begin{subfigure}{1\textwidth}
\centering
\begin{tikzpicture}[scale=1.6, >=stealth]
\centering
\tikzstyle{every node}=[draw,shape=circle, node distance=0.5cm];
 \foreach \j in {2,4,5,6,14,16,17,18} {
 \draw (-3.5 + 0.5*\j, 1.5) circle (0.1);
\node[draw =none] (s2) at (-3.5+ 0.5*\j,1 ) {\footnotesize$+$};
\draw (-3.5 +0.5*\j, 1) circle (0.1);
 \draw (-3.5 + 0.5*\j, 0.5) circle (0.1);
 \draw   [->] (-3.5+ 0.5*\j,1.9-0.5)-- (-3.5+ 0.5*\j,1.1);
 \draw   [->] (-3.5+ 0.5*\j,0.9)-- (-3.5+ 0.5*\j,0.6);
 }
  \foreach \j in {4,5,16,17} {
  \draw   [->, dashed] (-3.5+ 0.5*\j,1.9-0.5)-- (-3.5+ 0.5*\j + 0.5,1.1);
  }
   \foreach \j in {2,14} {
  \draw   [->, dashed] (-3.5+ 0.5*\j,1.9-0.5)-- (-3.5+ 0.5*\j + 0.5+0.5,1.1);
  }
    \foreach \j in {5,6,17,18} {
  \draw   [->, dashed] (-3.5+ 0.5*\j,1.9-0.5)-- (-3.5+0.5*\j - 0.5,1.1);
 }
   \foreach \j in {4,16} {
  \draw   [->, dashed] (-3.5+ 0.5*\j,1.9-0.5)-- (-3.5+0.5*\j - 0.5-0.5,1.1);
 }
  \foreach \j in {1,3,7,11,15,19} {
 \draw [fill = white, draw = none](-3.5 + 0.5*\j, 1.5) circle (0.1);
 }
   \foreach \j in {2,4,5,6,14,16,17, 18} {
 \draw [fill = blue](-3.5 + 0.5*\j, 1.5) circle (0.1);
\node[draw =none] (s2) at (-3.5+ 0.5*\j,1 ) {\footnotesize$+$};
\draw (-3.5 +0.5*\j, 1) circle (0.1);
 \draw [fill = blue] (-3.5 + 0.5*\j, 0.5) circle (0.1);
 \draw   [->] (-3.5+ 0.5*\j,1.9-0.5)-- (-3.5+ 0.5*\j,1.1);
 \draw   [->] (-3.5+ 0.5*\j,0.9)-- (-3.5+ 0.5*\j,0.6);
 }
\node[draw =none] (s2) at (-3.5+1,0.2) {$2$};
\node[draw =none] (s2) at (-3.5+2,0.2) {$4$};
\node[draw =none] (s2) at (-3.5+2.5,0.2) {$5$};
\node[draw =none] (s2) at (-3.5+3,0.2) {$6$};
\node[draw =none] (s2) at (-3.5+7,0.2) {$14$};
\node[draw =none] (s2) at (-3.5+8,0.2) {$16$};
\node[draw =none] (s2) at (-3.5+8.5,0.2) {$17$};
\node[draw =none] (s2) at (-3.5+9,0.2) {$18$};
\node[draw =none] (s2) at (-3.5+1,0.2+1.55) {$S$};
\node[draw =none] (s2) at (-3.5+2,0.2+1.55) {$S$};
\node[draw =none] (s2) at (-3.5+2.5,0.2+1.55) {$S$};
\node[draw =none] (s2) at (-3.5+3,0.2+1.55) {$S$};
\node[draw =none] (s2) at (-3.5+7,0.2+1.55) {$S$};
\node[draw =none] (s2) at (-3.5+8,0.2+1.55) {$S$};
\node[draw =none] (s2) at (-3.5+8.5,0.2+1.55) {$S$};
\node[draw =none] (s2) at (-3.5+9,0.2+1.55) {$S$};

 \node[draw =none, rotate =45] (s2) at ( 0.5,1.5) {\huge${\color{white}+}$};
  \node[draw =none, rotate =45] (s2) at ( 0.5,0.5) {\huge${\color{white}+}$};
  \node[draw =none, rotate =45] (s2) at ( 6.5,1.5) {\huge${\color{white}+}$};
  \node[draw =none, rotate =45] (s2) at (6.5,0.5) {\huge${\color{white}+}$};
%
%
     \foreach \i in {4,5,16,17} {
  \draw [<-, very thick, blue ] (-3.5 + 0.5*\i+0.1, 0.5-0.05) -- (-3.5 + 0.5*\i + 0.5- 0.1, 0.5-0.05);
   \draw [->, very thick, blue ] (-3.5 + 0.5*\i+0.1, 0.5+0.05) -- (-3.5 + 0.5*\i + 0.5- 0.1, 0.5+0.05);
  }
   \foreach \i in {2,14} {
  \draw [<-, very thick, blue ] (-3.5 + 0.5*\i+0.1, 0.5-0.05) -- (-3.5 + 0.5*\i + 0.5, 0.5-0.05);
   \draw [<-, very thick, blue ] (-3.5 + 0.5*\i+0.5, 0.5-0.05) -- (-3.5 + 0.5*\i + 0.5+0.5, 0.5-0.05);
   \draw [->, very thick, blue ] (-3.5 + 0.5*\i, 0.5+0.05) -- (-3.5 + 0.5*\i + 0.5, 0.5+0.05);
    \draw [->, very thick, blue ] (-3.5 + 0.5*\i+0.5, 0.5+0.05) -- (-3.5 + 0.5*\i + 0.5+0.5-0.1, 0.5+0.05);
  }
   \foreach \j in {4,16} {
 \draw [very thick, green](-3.5 + 0.5*\j, 0.5) circle (0.12);
  }
\end{tikzpicture}
\vspace{-0.3cm}
\caption{The last $\D-2$ Rx-cooperation rounds.} 
\end{subfigure}

  \caption{Example for $\D = 6$ : Tx/Rx pairs in yellow  have ``fast'' messages to transmit, Tx/Rx pairs in blue have ``slow''messages to transmit, Tx/Rx pairs in white are deactivated. We deactivated Tx/Rx pairs $8$ and $20$ to satisfy the delay constraint $\D$. Rx~$4$ and Rx~$16$ are master Rxs. Tx/Rx pair $19$ employs the same coding scheme as the ``fast'' transmissions.}
   \label{fig1}
   \vspace{-3mm}
  \end{figure*}
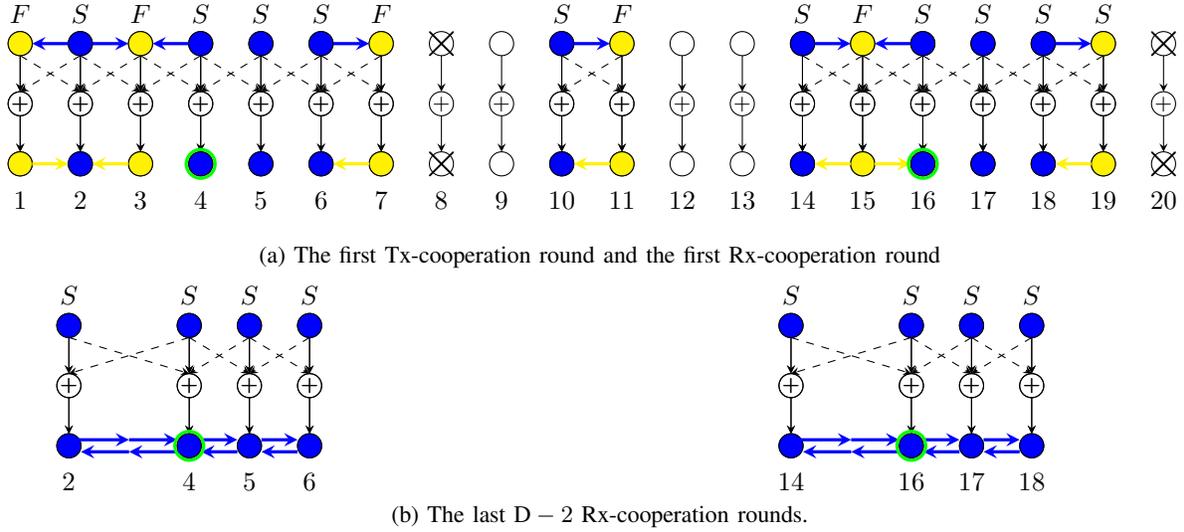

 Each network is decomposed into three subsets of Tx/Rx pairs, $\mathcal T_{\text{silent}}$, $\mathcal T_{\text{URLLC}}$ and $\mathcal T_{\text{eMBB}}$, where 
 \begin{itemize}
 	\item Txs in $\mathcal T_{\text{silent}}$ are silenced and Rxs  in $\mathcal T_{\text{silent}}$ do not take any action.
 	
 	\item Txs in $\mathcal T_{\text{URLLC}}$ send only URLLC messages. The corresponding Txs/Rxs   are  designated as  URLLC Txs/Rxs. 
 	\item Txs in $\mathcal T_{\text{eMBB}}$ send only eMBB messages. The corresponding Txs/Rxs   are  designated as eMBB Txs/Rxs.
 	
 \end{itemize}
 We choose the sets $\mathcal T_{\text{silent}}$, $\mathcal T_{\text{URLLC}}$, and $\mathcal T_{\text{eMBB}}$ in a way that:
 \begin{itemize}
 	\item[C1:] the signals sent by the URLLC Txs  do not interfere with one another; and
 	\item[C2:] silencing the Txs in $\mathcal T_{\text{silent}}$ decomposes the network into non-interfering subnets such that in each subnet there is a dedicated Rx,  called the \emph{master Rx}, that  can send a  cooperation message to any other eMBB Rx  in the same subnet in at most  $\frac{\D}{2}-1$ cooperation rounds.
 \end{itemize}
 For example, consider Wyner's symmetric network (described in detail in Section~\ref{sec:symmetric}) where Txs and Rxs are aligned on a grid and cooperation is possible only between subsequent Txs or Rxs. Interference at a given Rx is only from adjacent Txs. The network is illustrated in Fig.~\ref{fig1}. The figure also shows a possible decomposition of the Tx/Rx pairs into the  sets $\mathcal T_{\text{silent}}$ (in white), $\mathcal T_{\text{URLLC}}$ (in yellow) and $\mathcal T_{\text{eMBB}}$ (in blue) when $\D = 6$. The proposed decomposition creates subnets with $7$ active Tx/Rx pairs where the Rx in the center of any subnet (e.g. Rx~$4$ in the first subnet) can serve as a master Rx as it   reaches any eMBB Rx in the same subnet in at most $\D/2-1= 2$ cooperation rounds.  As required, transmissions from URLLC (yellow) Txs are only interfered with by  transmissions from eMBB (blue) Txs. Notice that in our example, Tx/Rx~$19$ has an eMBB message to send but still is depicted in yellow (URLLC) because in our scheme it acts as if it had a URLLC message to send. In general, any eMBB Tx/Rx can act as if it had a URLLC message to send, because requirements for eMBB messages are less stringent than for  URLLC messages. This observation will play an important role in the following.
 \medskip

 \subsubsection{Precanceling  of eMBB interference at URLLC  Txs}  Any eMBB Tx ${k'}$ quantizes its pre-computed input signal $\vect X_{k'}^n$  (how this signal is generated will be described under item 5)) and describes  the quantized signal $\hat{\vect X}_{k'}^n$ during the single Tx-cooperation round  to all its neighboring URLLC Txs, which then precancel this interference in their transmit signals. Fig.~\ref{fig1} illustrates by blue arrows the sharing of the described quantization information  with neighboring URLLC Txs for Wyner's symmetric network.
 
 To describe this formally, for each $k\in \mathcal K$, we define  the \emph{ eMBB interfering set} 
 \begin{equation}\label{eq:Ik-eMBB}
 \mathcal{I}_{k}^{(\e)}\triangleq \mathcal{I}_{k} \cap \mathcal T_{\text{eMBB}}.
 \end{equation}
 Also, we  denote by $\vect U_{k}^n(M_{k}^{(\U)})$  the  non-precoded input signal precomputed at a given URLLC Tx~$k$, which we will see is of power $\P$. (The following item 3) explains how to obtain  $\vect U_{k}^n(M_{k}^{(\U)})$.) Tx~$k$   sends the inputs  
 \begin{equation} \label{eq:13}
 \vect X_k^n= \gamma_k \left(\vect U_{k}^n(M_{k}^{(\U)}) - \sum_{ k '\in \mathcal I_{k}^{(\e)}}h_{k,k}^{-1}h_{ k',k} \hat{\vect X}_{ k'}^n\right)
 \end{equation} 
 over the channel, where $\gamma_k$ is a factor that ensures that the transmit signal satisfies the power constraint  in \eqref{eq:power}.\footnote{Since all channel coefficients  $h_{k',k}$ are constant and   both $\vect U_{k}^n(M_{k}^{(\U)})$  and  $\hat{\vect X}_{ k'}^n$  are of power $\P$, the prefactor $\gamma_k$ does not  grow with $\P$.}
 Since each  URLLC Rx~$k$ is  not interfered with by the signal sent at any other URLLC Tx,
 the  precoding in \eqref{eq:13} ensures that a URLLC Rx~$k$ observes the almost interference-free signal
 \begin{equation} \label{eq:14}
 \vect Y_{k}^n = \hn{\gamma_k} \mathsf h_{k,k} \vect U_k^n + \underbrace{\sum_{ k' \in \mathcal I_{k}^{(\e)}} h_{ k' ,k} (\vect X_{ k'}^n - \hn{\gamma_k}\hat{\vect X}_{ k'}^n) + \vect Z_{k}^n}_{\textnormal{disturbance}},
 \end{equation}
 where  the variance of the above disturbance is around the noise level and does not grow with $\P$.
 \medskip
 
 \subsubsection{Transmission of   URLLC messages} Each  URLLC Tx~$k$ encodes its desired message $M_k^{(\U)}$  using a codeword $\mathbf{U}_k^{(n)}(M_k^{(\U)})$ from a  Gaussian point-to-point code of power $\P$. The corresponding Rx~$k$ applies a standard point-to-point decoding rule to directly decode this URLLC codeword  without Rx-cooperation   from its ``almost" interference-free outputs $\mathbf{Y}_k$, see \eqref{eq:14}. 
 \medskip
 
 \subsubsection{Canceling  URLLC interference at eMBB Rxs}
 According to the previous item 3),  all URLLC messages are decoded directly from the outputs without any Rx-cooperation. During the first Rx-cooperation round, all URLLC Rxs share their decoded messages  with all their neighboring eMBB Rxs, which can cancel the  corresponding interference  from their receive signals. More formally, we define the \emph{URLLC interference set}
 \begin{equation}\label{eq:IkF}
 \mathcal{I}_{\tk}^{(\U)}\triangleq \mathcal{I}_{\tk} \cap  \mathcal{T}_{\text{URLLC}}
 \end{equation} as   the set of URLLC Txs   whose signals interfere at Rx $\tk$. Each eMBB Rx $\tk$  forms the new signal
 \begin{equation} \label{eq:15}
 \hat{\vect Y}_{\tk}^n: = \vect Y_{\tk}^n -\sum_{\hat k \in \mathcal{I}_{\tk}^{(\U)}} h_{\hat k,\tk}  \hn{\vect U_{\hat k}^n (\hat M_{\hat k}^{(\U)})},
 \end{equation}
 and decodes its desired eMBB message based on this new signal following the steps  described in the following item 5).
 Fig.~\ref{fig:symb} illustrates by yellow arrows the sharing of  decoded URLLC messages with neighboring eMBB Rxs in Wyner's symmetric network.  Recall again that for convenience in this example we treat Message $M_{19}^{(\e)}$ as if it was URLLC.
 \medskip

 \subsubsection{Transmission and reception of eMBB messages using \emph{CoMP} reception}
 Each eMBB Tx~$k$ encodes its message $M_k^{(\e)}$ using a codeword $\vect X_k^n(M_{k}^{(\e)})$ from  a Gaussian point-to-point code of power~$\P$. eMBB messages are decoded based on the new outputs $\hat{\vect Y}_{\tk}^n$ in \eqref{eq:15}. CoMP reception  is employed to decode all eMBB messages in a given subnet. That means that each eMBB Rx~$\tk$ applies a  rate-$\frac{1}{2} \log (1 + \P)$  quantizer to the new output 
 signal  $\hat{\vect Y}_{\tk}^n$, and sends the quantization information over the cooperation links to the {master} Rx in its subnet. Each master Rx reconstructs all the quantized signals and  jointly decodes the eMBB messages, before sending them back to their intended Rxs.  By  item 4)   the influence of URLLC transmissions has been canceled (up to the noise level) from the eMBB receive signals. 
 \medskip
 
 \subsubsection{MG analysis}
 In the described scheme, all transmitted URLLC and  eMBB messages can be sent reliably  at MG $1$ because all interference is canceled (up to the noise level) either at the Tx or the Rx side and because the precoding factors $\gamma_k$ do not vanish as $\P\to \infty$.
 
 The presented coding scheme can  sustain  URLLC rates
 \begin{subequations}\label{eq:rates_Model2}
 \begin{equation}
 R_k^{(\U)} = \frac{1}{2} \log (1+\P) +o(1) , \quad k\in\mathcal{T}_{\text{URLLC}}, 
 \end{equation} 
 and  eMBB rates 
 \begin{equation}
 R^{(\e)}_k =  \frac{1}{2} \log (1+\P) +o(1) , \quad k\in\mathcal{T}_{\text{eMBB}}.
 \end{equation}
 \end{subequations}
 
  \subsubsection{Rate transfer  for Model 1}\label{sec:rate_transfer}

In Model 1,  any URLLC Tx also has an eMBB message to send. Since URLLC messages have more stringent  requirements than eMBB messages, each Tx/Rx pair in $\mathcal{T}_{\text{URLLC}}$ can use part of its rate to send an eMBB message instead of the URLLC message. By this rate-transfer argument, our scheme can achieve any rate-tuple  $(R^{(\U)}_k, R_k^{(\e)}\colon k\in \mathcal{K})$ satisfying
 \begin{subequations}\label{eq:rates_Model1}
 \begin{IEEEeqnarray}{rCl}
 R_k^{(\U)} + R_{k}^{(\e)} & \leq  & \frac{1}{2} \log (1+\P) +o(1) , \quad k\in \mathcal{T}_{\text{URLLC}} ,\\
R_{k}^{(\e)} & \leq  & \frac{1}{2} \log (1+\P) +o(1) , \quad k\in \mathcal{T}_{\text{eMBB}}, \\
R_k^{(\U)} & =&0 , \quad k\in \mathcal{T}_{\text{eMBB}} ,\\
R_{k}^{(\e)}, R_k^{(\U)} & =&0 , \quad k\in \mathcal{T}_{\text{silent}}.
 \end{IEEEeqnarray}
 \end{subequations}
 \subsection{Adaptations to Random Arrivals} \label{sec:random-both}

In our model, we can choose the sets  $\mathcal T_{\text{silent}}$, $\mathcal T_{\text{URLLC}}$, and $\mathcal T_{\text{eMBB}}$ depending on the realizations of the random sets $ \Tf$ and $\Ts$. An ideal choice would obviously be to set  $\mathcal T_{\text{silent}}=\mathcal{K}\backslash \Ta$, $\mathcal T_{\text{URLLC}}=\Tf$, and $\mathcal T_{\text{eMBB}}=\Ts$. However, such a choice is not always possible because the sets $\Ta, \Tf, \Ts$ are random while the sets $\mathcal T_{\text{silent}}$, $\mathcal T_{\text{URLLC}}$, and $\mathcal T_{\text{eMBB}}$ have to satisfy specific requirements, e.g., users in $\mathcal T_{\text{URLLC}}$ cannot be adjacent and users in $\mathcal T_{\text{silent}}$ cannot be too far away from each other so as to ensure that eMBB Rxs in a subnet can reach their master Rx within the required number of cooperation rounds.

As already mentioned, one remedy is to time-share different versions $i=1,2,\ldots$ of the scheme with different choices of the sets $\mathcal T_{\text{silent},i}$, $\mathcal T_{\text{URLLC},i}$, and $\mathcal T_{\text{eMBB},i}$ in version $i$ so that all URLLC users  $\Tf$ can transmit their URLLC messages in at least one of the versions, i.e., so that each $k\in \Tf$ is present in at least one of the sets $\mathcal T_{\text{URLLC},i}$.  

Depending on the network structure, identifying the best choice of the time-shared schemes can be cumbersome. We therefore first propose a non-adaptive choice that works for all realizations of the activity parameters in the sense that  the union of the sets $\bigcup_i \mathcal T_{\text{URLLC},i}$ over all time-shared schemes covers each of the users in $\mathcal{K}$.  

\medskip
 
 \subsubsection{Non-adaptive scheme  under random arrivals}\label{eq:non_adaptive1}

 The networks we consider in this paper have regular connectivity. It is thus possible to find a regular arrangement of the sets $\mathcal T_{\text{silent},1}$, $\mathcal T_{\text{URLLC},1}$, and $\mathcal T_{\text{eMBB},1}$ satisfying the requirements in the scheme described in the previous section, and so that shifted arrangements $\mathcal T_{\text{silent},i}$, $\mathcal T_{\text{URLLC},i}$, and $\mathcal T_{\text{eMBB},i}$ exist, for $i=2,3,\ldots, \kappa$ and \hn{$\kappa=\frac{K}{| \mathcal{T}_{\text{URLLC},1}|}$},  
 with each Tx/Rx pair $k$ lying in exactly one set $\mathcal{T}_{\text{URLLC},i}$. In such a regular arrangement, each URLLC message is transmitted only during a fraction \hn{$\frac{1}{\kappa}$} of the time, and its rate is thus
 \begin{equation}\label{eq:RF}
 R^{(\U)}=\frac{|\mathcal{T}_{\text{URLLC},1}|}{K} \cdot  \frac{1}{2} \log (1+\P) +o(1).
 \end{equation}
  This strategy can thus achieve a URLLC per-user MG of 
 \begin{equation}\label{eq:SFMG}
 \S^{(\U)}   = \rho\rho_f \varlimsup_{K\to \infty}  \frac{|\mathcal T_{\text{URLLC},1}| }{K},
 \end{equation}
 where we recall that the factor $\rho\rho_f$ stems from the definition of $\S^{(\U)}$.

In the $i$-th version of the scheme, the Tx/Rx pairs  $k\in\mathcal{T}_{\text{eMBB},i}$ can transmit eMBB messages at a rate  $\frac{1}{2} \log (1+\P) +o(1)$, if they have such  eMBB messages. Moreover, as already explained, Tx/Rx pairs $k\in \mathcal{T}_{\text{URLLC},i}$ that have an eMBB message but no URLLC message can transmit these eMBB messages also at the same rate $\frac{1}{2} \log (1+\P) +o(1)$. Since by the regularity of the arrangements, each user lies  a fraction $\frac{|\mathcal{T}_{\text{URLLC},1}|}{K}$ of the time in the corresponding set $\mathcal{T}_{\text{URLLC},i}$ and a fraction  $\frac{|\mathcal{T}_{\text{eMBB},1}|}{K}$ of the time in the corresponding set $\mathcal{T}_{\text{eMBB},i}$, we have: 
\begin{itemize}
\item If $k\in (\mathcal{K}_{\textnormal{URLLC}} \cap \mathcal{K}_{\textnormal{eMBB}})$, then $R_k^{(\e)}=  \frac{|\mathcal{T}_{\text{eMBB},1}|}{K} \frac{1}{2} \log (1+\P) +o(1)$;
\item If $k\in (\mathcal{K}_{\textnormal{URLLC}} \backslash \mathcal{K}_{\textnormal{eMBB}})$, then $R_k^{(\e)}= 0$; 
\item If $k\in (\mathcal{K}_{\textnormal{eMBB}} \backslash \mathcal{K}_{\textnormal{URLLC}})$, then $R_k^{(\e)}=  \frac{|\mathcal{T}_{\text{eMBB},1}| + |\mathcal{T}_{\text{URLLC},1}|}{K} \frac{1}{2} \log (1+\P) +o(1)$.
\end{itemize}

Under Model~1, any active user has an eMBB message to transmit and thus the set $\mathcal{K}_{\textnormal{URLLC}} \backslash \mathcal{K}_{\textnormal{eMBB}}$ is empty. The expected per-user eMBB MG is therefore equal to
\begin{IEEEeqnarray}{rCl}
\S^{(\e)} & = & \rho \rho_f   \varlimsup_{K\to \infty}\frac{|\mathcal{T}_{\text{eMBB},1}|}{K} + \rho(1-\rho_f)  \varlimsup_{K\to \infty} \frac{|\mathcal{T}_{\text{eMBB},1}| + |\mathcal{T}_{\text{URLLC},1}|}{K} \\
&=& \rho  \varlimsup_{K\to \infty}\frac{|\mathcal{T}_{\text{eMBB},1}|}{K} + \rho(1-\rho_f)  \varlimsup_{K\to \infty }\frac{ |\mathcal{T}_{\text{URLLC},1}|}{K} . \label{eq:SS_Model1}
\end{IEEEeqnarray}
By the rate-transfer arguments in Subsection~\ref{sec:rate_transfer}, it follows that all  pairs $(\S^{(\U)}, \S^{(\e)})$ satisfying  the following two conditions are achievable: 
\begin{IEEEeqnarray}{rCl}
\label{eq:URLLC1}
 \S^{(\U)} &  \leq & \rho\rho_f \varlimsup_{K\to \infty}  \frac{|\mathcal T_{\text{URLLC},1}| }{K}  \label{eq:SF_Model1}\\
 \S^{(\U)}  +\S^{(\e)}   & \leq & \rho \varlimsup_{K\to \infty}  \frac{|\mathcal T_{\text{URLLC},1}| +|\mathcal T_{\text{eMBB},1}| }{K}.\label{eq:SS_Model1}
\end{IEEEeqnarray}

Under Model~2 any active user that has a URLLC message to send  has no eMBB message and thus the set $\mathcal{K}_{\textnormal{URLLC}} \cap \mathcal{K}_{\textnormal{eMBB}}$ is empty.  The expected per-user eMBB MG is therefore equal to
\begin{IEEEeqnarray}{rCl}
\S^{(\e)} & = &\rho(1-\rho_f)   \varlimsup_{K\to \infty}\frac{|\mathcal{T}_{\text{eMBB},1}| + |\mathcal{T}_{\text{URLLC},1}|}{K}.\label{eq:SS_Model2}
\end{IEEEeqnarray}
The URLLC MG is as in \eqref{eq:SFMG}.

\subsubsection{Non-adaptive scheme  under random arrivals transmitting at  $\S^{(\U)}=0$}
If  we are  interested in transmitting at the largest possible total eMBB MG, we can set $\mathcal{T}_{\textnormal{URLLC}}=\emptyset$ and simply choose a set $\mathcal{T}_{\textnormal{eMBB}}$ satisfying Condition C2 in the previous section. Under Model~1, the expected per-user eMBB MG is then equal to
\begin{IEEEeqnarray}{rCl}
\S^{(\e)} & = &\rho \varlimsup_{K\to \infty}\frac{|\mathcal{T}_{\text{eMBB}}|}{K}\label{eq:SS_high1}
\end{IEEEeqnarray}
and under Model 2 it is 
\begin{IEEEeqnarray}{rCl}
\S^{(\e)} & = &\rho( 1-\rho_f) \varlimsup_{K\to \infty}\frac{|\mathcal{T}_{\text{eMBB}}|}{K}. \label{eq:SS_high2}
\end{IEEEeqnarray}

 \medskip
   
    \subsubsection{Adaptive schemes under random arrivals} 
As the realizations of the activity sets are random,  proposing a general formulation for this scheme is only possible when the interference graph of the network is known. In Appendix~\ref{App:A} we explain this scheme for the symmetric Wyner network. 

\section{Coding Scheme with only  Rx-Cooperation} \label{sec:onlyRx}
We assume throughout this section that
 \begin{equation}
 \Dt=0 \quad \textnormal{and} \quad \Dr= \D.
 \end{equation}

\subsection{The Basic Scheme}\label{sec: IV-A}

 The only difference between this scheme and the one proposed in Section~\ref{sec:basic-both} is that without Tx-cooperation, item 2) cannot be executed. More specifically, in this scheme, precanceling of eMBB interference at URLLC Txs is not allowed. Therefore, Txs in $\mathcal T_{\text{eMBB}}$ whose transmissions interfere with URLLC transmissions are deactivated. 

With Rx-cooperation only, we therefore choose sets $\mathcal T_{\text{silent}}$, $\mathcal T_{\text{URLLC}}$, and $\mathcal T_{\text{eMBB}}$  that satisfy the two conditions C1 and C2 in Subsection~\ref{sec:basic-both} and the additional condition
 \begin{itemize}
		\item[C3:] eMBB transmit signals do not interfere at URLLC Rxs:
	\begin{equation}
	k \notin \mathcal{I}_{k'}, \quad \forall k\in \mathcal{T}_{\text{eMBB}}, \; k' \in \mathcal{T}_{\text{URLLC}},
	\end{equation}
	where we recall that $\mathcal{I}_{k'}$ denotes the set of Txs whose signals interfere at Rx $k'$.
 \end{itemize}

The coding scheme in Section~\ref{sec:both} (without item 2) then again achieves the rates in \eqref{eq:rates_Model2}. Moreover, with an appropriate rate-transfer argument from URLLC to eMBB messages, under Model 1 it  even achieves the rates in \eqref{eq:rates_Model1}. Notice that in our schemes, the penalty from having only Rx-cooperation but no Tx-cooperation thus only stems from the additional  requirement C3, compared to the case with Tx- and Rx-cooperation.

 \subsection{Adaptations to Random Arrivals} \label{sec:random-onlyRx}
 
  \subsubsection{Non-adaptive scheme under random arrivals} \label{sec:non-adaptive-onlyRx}

 Unlike with Tx- and Rx-cooperation, when only Rxs can cooperate  there does not seem to be any interesting non-adaptive scheme that is not degenerate, i.e., that includes a non-trivial combination of URLLC and eMBB Tx/Rx pairs. The reason is that Requirement C3 is very stringent and for many networks  does not seem to allow for meaningful choices where both $\mathcal{T}_{\textnormal{URLLC}}$ and $\mathcal{T}_{\textnormal{eMBB}}$ are  non-empty. In these networks the best option seems to time-share schemes with  either  only URLLC or only eMBB Txs. The scheme with only URLLC Txs should  time-share different schemes with different choices of  $\mathcal{T}_{\textnormal{URLLC},i}$ so that each Tx lies in $\mathcal{T}_{\textnormal{URLLC},i}$ for at least one phase $i$. Each of the sets $\mathcal{T}_{\textnormal{URLLC},i}$ has to satisfy Condition C1, and we set $\mathcal{T}_{\text{silent},i}= \mathcal K \backslash \mathcal{T}_{\text{URLLC},i}$ so that $\mathcal{T}_{\textnormal{eMBB},i}=\emptyset$. The scheme with only eMBB Txs sets  $\mathcal{T}_{\textnormal{URLLC}}=\emptyset$ and then chooses $\mathcal{T}_{\textnormal{silent}}$ so as to decompose the network in an appropriate way so that  the remaining set $\mathcal{T}_{\textnormal{eMBB}}= \mathcal K \backslash \mathcal{T}_{\text{silent}}$    satisfies Condition~C2. 
 
 After time-sharing, the overall scheme then achieves  the expected per-user  MG pair 
 \begin{IEEEeqnarray}{rCl}
 \S^{(\U)} &= & \alpha \cdot \rho \rho_f \varlimsup_{K\to \infty} \frac{ | \mathcal{T}_{\textnormal{URLLC}}'|}{K} \label{eq:Mod2-onlyRx1}\\
  \S^{(\e)} &= &\alpha\rho(1-\rho_f) \varlimsup_{K\to \infty} \frac{ | \mathcal{T}_{\textnormal{URLLC}}'|}{K}+  (1- \alpha) \cdot \rho (1-\rho_f )\varlimsup_{K\to \infty}\frac{ | \mathcal{T}_{\textnormal{eMBB}}'|}{K}, \label{eq:Mod2-onlyRx2}
 \end{IEEEeqnarray}
 where $\alpha\in[0,1]$ is the time-sharing parameter and the set $\mathcal{T}_{\textnormal{URLLC}}'\subseteq \mathcal K $ has to satisfy Condition~C1 while the set $\mathcal{T}_{\textnormal{eMBB}}' \subseteq \mathcal K$ has to satisfy Condition C2. (Notice that here the sets $\mathcal{T}_{\textnormal{URLLC}}'$ and $\mathcal{T}_{\textnormal{eMBB}}' $ do not need to be disjoint; for this reason we use the prime-notation.)
 
 Under Model~1 each Tx has an eMBB message to send with probability $\rho$, and   moreover  we can  apply a rate-transfer argument from URLLC to eMBB messages. The scheme can then  achieve any expected per-user MG pair satisfying
  \begin{IEEEeqnarray}{rCl}
 \S^{(\U)} &\leq & \alpha \cdot \rho \rho_f  \varlimsup_{K\to \infty}\frac{ | \mathcal{T}_{\textnormal{URLLC}}'|}{K} \label{eq:Mod1-onlyRx_1}  \\
 \S^{(\U)} + \S^{(\e)} &\leq  & \alpha \cdot \rho \varlimsup_{K\to \infty} \frac{ | \mathcal{T}_{\textnormal{URLLC}}'|}{K} + (1- \alpha) \cdot \rho  \varlimsup_{K\to \infty} \frac{ | \mathcal{T}_{\textnormal{eMBB}}'|}{K}, \label{eq:Mod1-onlyRx}
 \end{IEEEeqnarray}
 where $\alpha\in[0,1]$, and the sets $\mathcal{T}_{\textnormal{URLLC}}'$   and $\mathcal{T}_{\textnormal{eMBB}}'$ satisfy Conditions C1 and C2, respectively.

 Under Model~2 any active user that has a URLLC message to send  has no eMBB message and thus the set $\mathcal{K}_{\textnormal{URLLC}} \cap \mathcal{K}_{\textnormal{eMBB}}$ is empty.  The expected per-user eMBB MG is therefore equal to
\begin{IEEEeqnarray}{rCl}
\S^{(\e)} & = &\rho(1-\rho_f)   \varlimsup_{K\to \infty}\frac{(1-\alpha)|\mathcal{T}_{\text{eMBB}}'| + \alpha|\mathcal{T}_{\text{URLLC}}'|}{K}.\label{eq:SS_Model2-OnlyRx}
\end{IEEEeqnarray}
\medskip

 \subsubsection{Adaptive schemes under random arrivals} 
In Appendices~\ref{App:B}  and \ref{App:C} we describe  adaptive schemes for the symmetric Wyner and hexagonal networks.

\begin{figure}[t]
  \centering
  \small
\begin{tikzpicture}[scale=1.5, >=stealth]
\centering
\tikzstyle{every node}=[draw,shape=circle, node distance=0.5cm];
\foreach \j in {0,1,...,15} {
 \draw [fill= pink](-3.5 + 0.5*\j, 2) circle (0.1);
  \draw [fill= gray](-3.5 + 0.5*\j, 0.5) circle (0.1);
\node[draw =none] (s2) at (-3.5+ 0.5*\j,1 ) {\footnotesize$+$};
\draw (-3.5 +0.5*\j, 1) circle (0.1);
\draw   [->] (-3.5+ 0.5*\j,1.9)-- (-3.5+ 0.5*\j,1.1);
 \draw   [->] (-3.5+ 0.5*\j,0.9)-- (-3.5+ 0.5*\j,0.1+0.5);
  \draw   [->, dashed] (-3.5+ 0.5*\j,1.9)-- (-3.5+ 0.5*\j + 0.5,1.1);
   \draw   [->, dashed] (-3.5+ 0.5*\j,1.9)-- (-3.5+ 0.5*\j - 0.5,1.1);
}
\foreach \j in {0,1,...,14} {
\draw[->, purple] (-3.5 +0.5*\j+0.1, -0.05+0.5) --(-3.5 +0.5*\j + 0.5-0.1, -0.05+0.5);
\draw[<-, purple] (-3.5 +0.5*\j+0.1, +0.05+0.5) --(-3.5 +0.5*\j + 0.5-0.1, +0.05+0.5);

\draw[->, purple] (-3.5 +0.5*\j+0.1, -0.05+2) --(-3.5 +0.5*\j + 0.5-0.1, -0.05+2);
\draw[<-, purple] (-3.5 +0.5*\j+0.1, +0.05+2) --(-3.5 +0.5*\j + 0.5-0.1, +0.05+2);
 }
 \node[draw =none] (s2) at (-3.5+ 0.5*0,-0.25+0.5 ) {\footnotesize$1$};
 \node[draw =none] (s2) at (-3.5+ 0.5*1,-0.25+0.5 ) {\footnotesize$2$};
 \node[draw =none] (s2) at (-3.5+ 0.5*3,-0.25+0.5 ) {\footnotesize$\ldots$};
 \node[draw =none] (s2) at (-3.5+ 0.5*15,-0.25+0.5 ) {\footnotesize$K$};

\end{tikzpicture}
\vspace*{-2ex}
 \caption{Wyner's symmetric  network.  Black  dashed arrows show interference links and  purple arrows show cooperation links. }
  \label{fig2}
 \vspace*{-4ex}
\end{figure}
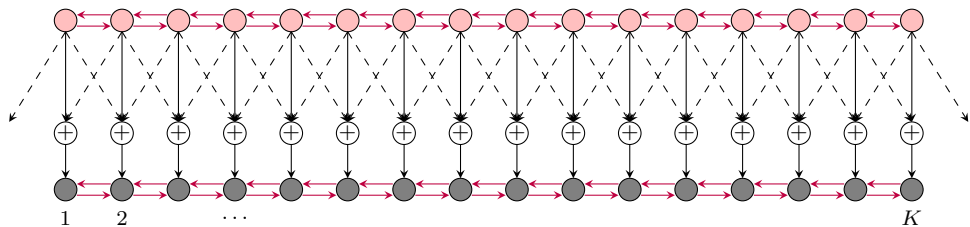
\section{The Symmetric Wyner Network}\label{sec:symmetric}

\subsection{Network and Cooperation Model}\label{sec:setupSym}

Consider Wyner's symmetric linear cellular model where  cells are aligned in a single dimension and  signals of users that lie in a given cell interfere only with signals sent in the two adjacent cells. See Fig.~\ref{fig2} where the interference pattern is illustrated by black dashed lines. We assume that the various mobile users in a cell are scheduled on different frequency bands, and focus on a single mobile user per cell (i.e., on a single frequency band).  

  The input-output relation of the network is   
\begin{IEEEeqnarray}{rCl}\label{Eqn:Channel}
\vect{Y}_{k,t} = h_{k,k} \vect{X}_{k,t} +\sum_{\tilde k \in \{k-1, k+1\}}h_{\tilde k,k} \vect{X}_{\tilde k,t} +\vect{Z}_{k,t},\IEEEeqnarraynumspace
\end{IEEEeqnarray}
where $\vect{X}_{0,t} = \vect{0}$ for all $t$, 
and the interference set at a given user  $k$ is
\begin{equation}
\mathcal{I}_{k}=\{k-1,k+1\},
\end{equation}where indices out of the range $\mathcal K$ should be ignored. 
In this model, Rxs and Txs can cooperate with the two Rxs and Txs  in the adjacent cells, so 
\begin{equation}
\Nk =\{k-1,k+1\} 
\end{equation}
Fig.~\ref{fig2} illustrates the interference pattern of the network and the available cooperation links. 

\subsection{Results with Both Tx- and Rx-Cooperation}

We first present our results for Model 1.

 \begin{proposition}[Non-Adaptive Scheme, Model 1] \label{theorem1}
 For $\rho\in (0,1]$, the  fundamental per-user MG region $\mathcal{S}_1^\star(\D, \rho, \rho_f)$  includes all nonnegative pairs $(\S^{(\U)}, \; \S^{(\e)})$ satisfying
 \begin{IEEEeqnarray}{rCl}
 \S^{(\U)} &\le& \frac{\rho \rho_f}{2}, \label{eq:a} \\
 \S^{(\e)} +   \S^{(\U)} & \le & \rho\frac{\D+1}{\D+2}. \label{eq:b}
 \end{IEEEeqnarray}
 \end{proposition}
 \begin{IEEEproof}
Achievability of the rate-pair $\S^{(\U)}=  \frac{\rho \rho_f}{2}$ and $ \S^{(\e)} =  \rho\frac{\D+1}{\D+2}- \frac{\rho \rho_f}{2}$ holds by \eqref{eq:SF_Model1} and \eqref{eq:SS_Model1} and the choice of the sets 
\begin{subequations}\label{eq:set}
\begin{IEEEeqnarray}{rCl}
\mathcal{T}_{\text{silent},1}& = & \left \{ \ell \left (\D+2 \right ) \colon \ell=1,\ldots, \left \lfloor \frac{K}{\D+2} \right \rfloor \right \} , \label{eq:Tsilent1}\\
\mathcal{T}_{\text{URLLC},1}& = &\{1,3,\ldots, K-1\},\\
\mathcal{T}_{\text{eMBB},1}& = &\mathcal K \backslash (\mathcal{T}_{\text{silent},1} \cup \mathcal{T}_{\text{URLLC},1}).
\end{IEEEeqnarray}
\end{subequations}
 \end{IEEEproof}

  \begin{theorem}[Adaptive Scheme, Model 1] \label{theorem1}
 For $\rho\in (0,1)$, the  fundamental per-user MG region $\mathcal{S}_1^\star(\D, \rho, \rho_f)$ includes all nonnegative pairs $(\S^{(\U)}, \; \S^{(\e)})$ satisfying
 \begin{IEEEeqnarray}{rCl}
 \S^{(\U)} &\le& \frac{\rho \rho_f}{2},  \\
 \S^{(\e)} + M_{\text{both},1}^{(\text{A})} \cdot  \S^{(\U)} & \le & \rho - \frac{(1-\rho) \rho^{\D+2}}{1-\rho^{\D+2}},
 \end{IEEEeqnarray}
 where 
 \begin{IEEEeqnarray}{rCl} \label{eq:M}
 M_{\text{both},1}^{(\text{A})}  \triangleq 1+ \frac{(1-\rho)^2 \rho^{\D}}{\rho_f(1-\rho^{\D+2})} -\frac{(1-\rho)^2  \rho^{\D} (1-\rho_f)^{2 }  }{\rho_f (1- \rho^{\D+2}(1-\rho_f)^{2})}. \IEEEeqnarraynumspace
 \end{IEEEeqnarray}
 \end{theorem}
 \begin{IEEEproof}
 The Results follows by rate- and time-sharing arguments and the achievability of the pairs $(\S^{(\U)}, \S^{(\e)})$, with $\S^{(\e)}$ given in \eqref{eq:ssmax0} and the pair $(\S^{(\U)}, \S^{(\e)})$ as given in \eqref{eq:SFF1} and \eqref{eq:sslast}; see Appendix~\ref{App:A}.
 \end{IEEEproof} 
 \medskip 
 We also have the following outer bound.
 \begin{theorem}[Outer Bound, Model 1] \label{theorem2}
 For $\rho \in (0,1)$, all achievable MG pairs  $(\S^{(\U)}, \; \S^{(\e)})$  of Model 1 satisfy \eqref{eq:a} and 
 \begin{IEEEeqnarray}{rCl}
 \S^{(\e)} + \S^{(\U)} & \le & \rho - \frac{(1-\rho) \rho^{\D+2}}{1-\rho^{\D+2}}. \label{eq:conv2}
 \end{IEEEeqnarray}
 For $\rho =1$ they satisfy \eqref{eq:a} and \eqref{eq:b}.
 \end{theorem}
 \begin{IEEEproof}
 See Appendix~\ref{App: A-D}. \end{IEEEproof}
 \medskip 
 
Inner and outer bounds are generally very close. In particular, they determine the largest achievable URLLC per-user MG, which is $\S_{\max}^{(\U)}=\frac{\rho\rho_f}{2}$, and the largest eMBB per-user MG,  which is
\begin{equation}\label{eq:SS_max1}
S^{(\e)}_{\max} =  \rho - \frac{(1-\rho) \rho^{\D+2}}{1-\rho^{\D+2}}.
\end{equation}

Our inner and outer bounds completely coincide in the extreme cases $\rho=1$ and $\D \to \infty$.
 \begin{corollary}
For $\rho = 1$ or when $\D\to \infty$, Theorem~\ref{theorem2} is exact. For $\rho=1$, 
the fundamental per-user MG region  $\mathcal{S}_1^\star(\D,\rho, \rho_f)$ \emph{is} the set of  all nonnegative MG pairs $(\S^{(\U)}, \; \S^{(\e)})$ satisfying \eqref{eq:a} and \eqref{eq:b}, 
and for $\rho\in (0,1)$ and $\D\to \infty$ it \emph{is} the set of all MG pairs  $(\S^{(\U)}, \; \S^{(\e)})$ satisfying \eqref{eq:a} and 
 \begin{IEEEeqnarray}{rCl}\label{eq:rhoequal1}
 \S^{(\e)} +\S^{(\U)} & \le & \rho.
 \end{IEEEeqnarray}
\end{corollary}
 \medskip
 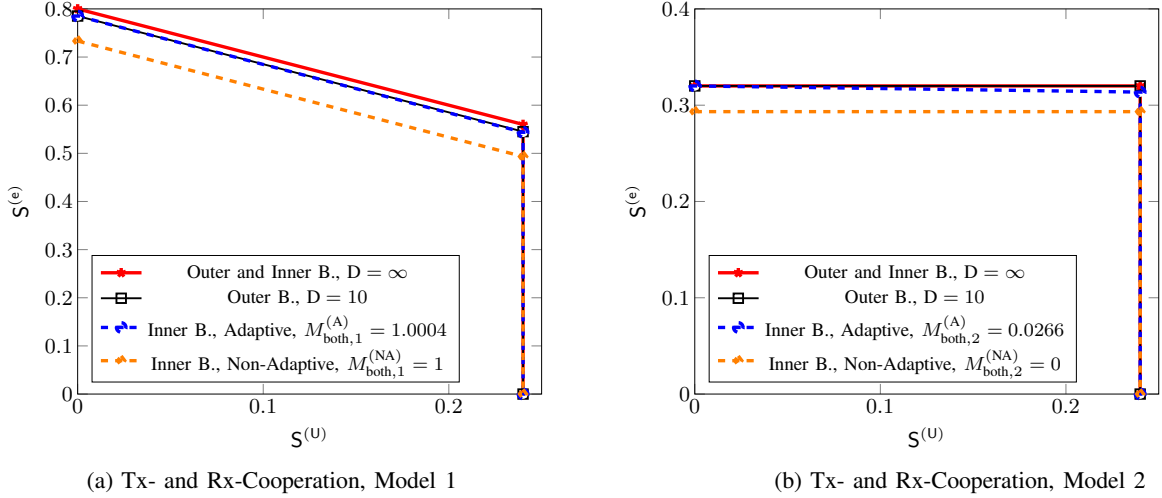
\begin{figure}[h!]

\begin{subfigure}{0.5\textwidth}
\centering
\begin{tikzpicture}[scale=.9]
\begin{axis}[
    xlabel={\small {$\S^{(\U)}$ }},
    ylabel={\small {$\S^{(\e)}$ }},
     xlabel style={yshift=.5em},
     ylabel style={yshift=-1.25em},
    xmin=0, xmax=0.25,
    ymin=0, ymax=0.8,
    xtick={0,0.1,0.2,0.3,0.4},
    ytick={0,0.1,0.2,0.3,0.4,0.5,0.6,0.7,0.8,0.9,1},
    yticklabel style = {font=\small,xshift=0.25ex},
    xticklabel style = {font=\small,yshift=0.25ex},
    legend pos=south west,
]

 \addplot[ color=red,   mark=star, line width = 0.5mm] coordinates {  (0,0.8) (0.24, 0.56)(0.24,0) };
 \addplot[ color=black,   mark=square, thick] coordinates {  (0,0.7852) (0.24, 0.5451)(0.24,0) };
\addplot[ color=blue,   mark=halfcircle, line width = 0.5mm, dashed] coordinates {  (0,0.7852) (0.24, 0.543748535079553)(0.24,0) };
  \addplot[ color=orange,   mark=diamond,  line width = 0.5mm, dashed] coordinates {  (0,0.733333333333333) (0.24, 0.493333333333333)(0.24,0) };

\footnotesize
      \legend{{Outer and Inner B., $\D = \infty$}, {Outer B., $ \D = 10$}, {Inner B., Adaptive, $M_{\text{both},1}^{(\text A)} = 1.0004$}, {Inner B., Non-Adaptive, $M_{\text{both},1}^{(\text {NA})} = 1$}}  
\end{axis}


\vspace{-0.4cm}
\end{tikzpicture}
\caption{Tx- and Rx-Cooperation, Model 1}
\end{subfigure}
\begin{subfigure}{0.5\textwidth}
\begin{tikzpicture}[scale=.9]
\begin{axis}[
    xlabel={\small {$\S^{(\U)}$ }},
    ylabel={\small {$\S^{(\e)}$ }},
     xlabel style={yshift=.5em},
     ylabel style={yshift=-1.25em},
    xmin=0, xmax=0.25,
    ymin=0, ymax=0.4,
    xtick={0,0.1,0.2,0.3,0.4},
    ytick={0,0.1,0.2,0.3,0.4,0.5,0.6,0.7,0.8,0.9,1},
    yticklabel style = {font=\small,xshift=0.25ex},
    xticklabel style = {font=\small,yshift=0.25ex},
    legend pos=south west,
]

\addplot[ color=red,   mark=star, line width = 0.5mm] coordinates {  (0,0.3200) (0.24, 0.3200)(0.24,0) };
 \addplot[ color=black,   mark=square, thick] coordinates {  (0,0.319999216012473) (0.24, 0.319999216012473)(0.24,0) };
 \addplot[ color=blue,   mark=halfcircle, line width = 0.5mm, dashed] coordinates {  (0,0.319999216012473) (0.24, 0.3136)(0.24,0) };
 \addplot[ color=orange,   mark=diamond,  line width = 0.5mm, dashed] coordinates {  (0,0.2933) (0.24, 0.2933)(0.24,0) };

\footnotesize
      \legend{{Outer and Inner B., $\D = \infty$}, {Outer B., $ \D = 10$},{Inner B., Adaptive, $M_{\text{both},2}^{(\text A)} = 0.0266$}, {Inner B., Non-Adaptive, $M_{\text{both},2}^{(\text {NA})} = 0$}}  
\end{axis}


\vspace{-0.4cm}
\end{tikzpicture}
\caption{Tx- and Rx-Cooperation, Model 2}
\end{subfigure}
\caption{Inner and outer bounds on the fundamental per-user  MG regions $\mathcal{S}_1^\star(\D,\rho, \rho_f)$ and $\mathcal{S}_2^\star(\D,\rho, \rho_f)$  for $\rho= 0.8$, $\rho_f = 0.6$ and $\D = 10$ for Tx- and Rx-Cooperation. }
\label{fig3}
\end{figure}
 
\begin{figure}[h!]
\begin{subfigure}{0.5\textwidth}
\centering
\begin{tikzpicture}[scale=.9]
\begin{axis}[
    xlabel={\small {$\S^{(\U)}$ }},
    ylabel={\small {$\S^{(\e)}$ }},
     xlabel style={yshift=.5em},
     ylabel style={yshift=-1.25em},
    xmin=0, xmax=0.065,
    ymin=0, ymax=0.4,
    xtick={0,0.01,0.02,0.03,0.04,0.05,0.06},
    ytick={0,0.1,0.2,0.3,0.4,0.5,0.6,0.7,0.8,0.9,1},
    yticklabel style = {font=\small,xshift=0.25ex},
    xticklabel style = {font=\small,yshift=0.25ex},
    legend pos=south west,
]

 \addplot[ color=red,   mark=star, line width = 0.5mm] coordinates {  (0,0.4) (0.06, 0.34)(0.06,0) };
 \addplot[ color=black,   mark=square, thick] coordinates {  (0,0.3975) (0.06, 0.3375)(0.06,0) };
\addplot[ color=blue,   mark=halfcircle, line width = 0.5mm, dashed] coordinates {  (0,0.3975) (0.06, 0.3375)(0.06,0) };
  \addplot[ color=orange,   mark=diamond,  line width = 0.5mm, dashed] coordinates {  (0,0.273333333333333) (0.06, 0.213333333333333)(0.06,0) };

\footnotesize
      \legend{{Outer and Inner B., $\D = \infty$}, {Outer B., $ \D = 4$}, {Inner B., Adaptive, $M_{\text{both},1}^{(\text A)} = 1$}, {Inner B., Non-Adaptive, $M_{\text{both},1}^{(\text {NA})} = 1$}}  
\end{axis}


\vspace{-0.4cm}
\end{tikzpicture}
\caption{Tx- and Rx-Cooperation, Model 1}
\end{subfigure}
\begin{subfigure}{0.5\textwidth}
\begin{tikzpicture}[scale=.9]
\begin{axis}[
    xlabel={\small {$\S^{(\U)}$ }},
    ylabel={\small {$\S^{(\e)}$ }},
     xlabel style={yshift=.5em},
     ylabel style={yshift=-1.25em},
    xmin=0, xmax=0.065,
    ymin=0, ymax=0.3,
    xtick={0,0.01,0.02,0.03,0.04,0.05,0.06},
    ytick={0,0.1,0.2,0.3,0.4,0.5,0.6,0.7,0.8,0.9,1},
    yticklabel style = {font=\small,xshift=0.25ex},
    xticklabel style = {font=\small,yshift=0.25ex},
    legend pos=south west,
]

\addplot[ color=red,   mark=star, line width = 0.5mm] coordinates {  (0,0.28) (0.06, 0.28)(0.06,0) };
 \addplot[ color=black,   mark=square, thick] coordinates {  (0,0.279652871703360) (0.06, 0.279652871703360)(0.06,0) };
 \addplot[ color=blue,   mark=halfcircle, line width = 0.5mm, dashed] coordinates {  (0,0.279652871703360) (0.06, 0.2714)(0.06,0) };
 \addplot[ color=orange,   mark=diamond,  line width = 0.5mm, dashed] coordinates {  (0,0.233333333333333) (0.06, 0.233333333333333)(0.06,0) };

\footnotesize
      \legend{{Outer and Inner B., $\D = \infty$}, {Outer B., $ \D = 4$},{Inner B., Adaptive, $M_{\text{both},2}^{(\text A)} = 0.13$}, {Inner B., Non-Adaptive, $M_{\text{both},2}^{(\text {NA})} = 0$}}  
\end{axis}


\vspace{-0.4cm}
\end{tikzpicture}
\caption{Tx- and Rx-Cooperation, Model 2}
\end{subfigure}
\caption{Inner and outer bounds on the fundamental per-user  MG regions $\mathcal{S}_1^\star(\D,\rho, \rho_f)$ and $\mathcal{S}_2^\star(\D,\rho, \rho_f)$   for $\rho=0.4 $, $\rho_f =0.3 $ and $\D = 4$ for Tx- and Rx-Cooperation.}
\label{figitw4}
\end{figure}
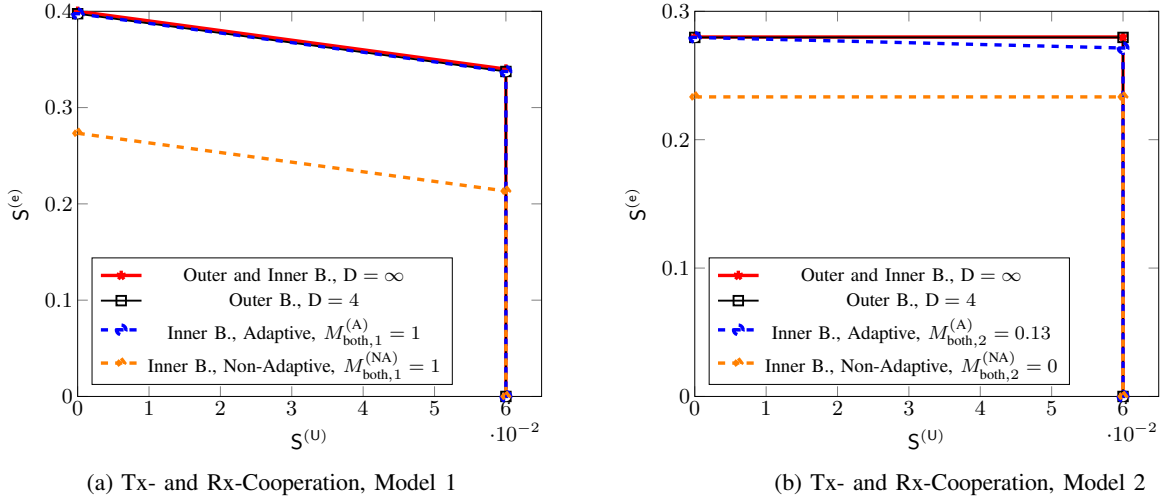

For Model 2 we have the following results. 

 \begin{proposition}[Non-Adaptive Scheme, Model 2] \label{proposition2}
 For $\rho\in (0,1]$, the  fundamental per-user MG region $\mathcal{S}_2^\star(\D, \rho, \rho_f)$  includes all nonnegative pairs $(\S^{(\U)}, \; \S^{(\e)})$ satisfying
 \begin{IEEEeqnarray}{rCl}
 \S^{(\U)} &\le& \frac{\rho \rho_f}{2}, \label{eq:a2} \\
 \S^{(\e)} & \le & \rho(1-\rho_f)\frac{\D+1}{\D+2}.\label{eq:b2}
 \end{IEEEeqnarray}
 \end{proposition}
 \begin{IEEEproof}
The pair $\S^{(\U)}=  \frac{\rho \rho_f}{2}$ and $ \S^{(\e)} =  \rho(1- \rho_f)\frac{\D+1}{\D+2}$ is achievable by \eqref{eq:RF} and \eqref{eq:SS_Model2} and the choice of the sets in \eqref{eq:set}. On the other hand, the pair $\S^{(\U)}=0$ and $\S^{(\e)} = \rho(1-\rho_f) \frac{\D+1}{\D+2}$ is achievable by \eqref{eq:SS_high2} and the choice  $\mathcal{T}_{\text{eMBB}}=\mathcal K\backslash \mathcal{T}_{\text{silent},1}$ for $\mathcal{T}_{\text{silent},1}$ as in \eqref{eq:Tsilent1}. The proposition then follows by  time-sharing arguments.
 \end{IEEEproof}
 
 \medskip
 
  \begin{theorem}[Adaptive Scheme, Model 2] \label{theorem3}
 For $\rho\in (0,1)$ and $\rho_f\in(0,1]$, the  fundamental per-user MG region $\mathcal{S}_2^\star(\D, \rho, \rho_f)$  includes all nonnegative pairs $(\S^{(\U)}, \; \S^{(\e)})$ satisfying
 \begin{IEEEeqnarray}{rCl}
 \S^{(\U)} &\le& \frac{\rho \rho_f}{2}, \label{eq:a2} \\
 \S^{(\e)} +  M_{\text{both},2}^{(\text{A})} \cdot  \S^{(\U)} & \le &\rho(1-\rho_f) - \frac{(1-\rho(1-\rho_f))\rho^{\D+2}(1-\rho_f)^{\D+2}}{ 1- \rho^{\D+2} (1- \rho_f)^{\D+2} }, \label{eq:achiev-Model2-both}
 \end{IEEEeqnarray}
 where 
\begin{subequations}
 \begin{IEEEeqnarray}{rCl} \label{eq:Mboth2}
  M_{\text{both},2}^{(\text{A})} & \triangleq&  1 + \frac{2(1-\rho_f)}{\rho_f} - \frac{2(1-\rho(1-\rho_f))\rho^{\D+1}(1-\rho_f)^{\D+2}}{ \rho_f( 1- \rho^{\D+2} (1- \rho_f)^{\D+2} )}\notag \\
&& -  \frac{2(1-\rho_f)}{\rho_f(1-\rho^2(1-\rho_f))^2} \left ( (1+ \rho)(1- \rho(1-\rho_f))^2 - \rho^3 \rho_f^2 \right) \notag \\
&& - \frac{ \left ((1- \rho(1-\rho_f))^2 + (1- \rho)^2 (1- \rho_f) (2 \rho + 2\rho^2 (1- \rho_f) + 1) \right)}{\rho_f( 1- \rho^2(1- \rho_f))} - \frac{(1-\rho)^2 \rho^{\D+1} (1-\rho_f)^{\frac{\D+6}{2}}}{2(1- \rho^{\D+2} (1- \rho_f)^{\frac{\D+6}{2}})} \notag \\
&& + \frac{\rho^{\D+2} (1- \rho_f)^{\frac{\D+2}{2}}}{2 (1- \rho^2 (1-\rho_f)) (1- \rho^{\D+2} (1- \rho_f)^{\frac{\D+2}{2}})} \left ( (1+\rho) \left ( (1- \rho(1- \rho_f))^2 + \frac{(1-\rho)^2}{\rho}\right)\right).
 \end{IEEEeqnarray}
\end{subequations}
For $\rho_f=0$, the left-hand side of \eqref{eq:achiev-Model2-both} has to be replaced by $\S^{(\e)}$ only.
 \end{theorem}
 \begin{IEEEproof}
 The results hold by time-sharing arguments and the achievability of the pairs $(\S^{(\U)}, \S^{(\e)})$, with $\S^{(\e)}$ given in \eqref{eq:lastsss} and the pair $(\S^{(\U)}, \S^{(\e)})$ as given in \eqref{eq:SFF} and \eqref{eq:198}; see Appendix~\ref{App:B1}.
 \end{IEEEproof}
 \medskip 
 
 We also have the following outer bound.
 \begin{theorem}[Outer Bound, Model 2] \label{theorem4}
 For  $\rho \in (0,1]$, all achievable MG pairs  $(\S^{(\U)}, \; \S^{(\e)})$  of Model 2 satisfy \eqref{eq:a2} and 
\begin{IEEEeqnarray}{rCl}\label{eq:b22}
 \S^{(\e)} & \le & \rho(1-\rho_f) - \frac{(1-\rho(1-\rho_f))\rho^{\D+2}(1-\rho_f)^{\D+2}}{ 1- \rho^{\D+2} (1- \rho_f)^{\D+2} }.
 \end{IEEEeqnarray}
 \end{theorem}
 \begin{IEEEproof}
See Appendix~\ref{App: A-G}. \end{IEEEproof}

Inner and outer bounds are generally very close also for this Model 2. As under Model 1, they also determine the largest achievable URLLC per-user MG, which is again $\S_{\max}^{(\U)}=\frac{\rho\rho_f}{2}$, and the largest eMBB per-user MG,  which now under Model 2 is
\begin{equation}\label{eq:SS_max2}
S^{(\e)}_{\max} =\rho(1-\rho_f) - \frac{(1-\rho(1-\rho_f))\rho^{\D+2}(1-\rho_f)^{\D+2}}{ 1- \rho^{\D+2} (1- \rho_f)^{\D+2} }.
\end{equation}
Notice that Expression \eqref{eq:SS_max2} can be obtained from \eqref{eq:SS_max1}, which determines the largest eMBB per-user under Model 1, by replacing $\rho$ by $\rho(1-\rho_f)$. The reason is that when only  eMBB messages are to be transmitted, then only the probability of each user having an eMBB message matters, and this probability equals $\rho$ under Model 1 and it equals $\rho(1-\rho_f)$ under Model 2.

\medskip

\begin{corollary}\label{cor:2}
For $\D\to \infty$, Theorem~\ref{theorem4} is exact. In this case,
the fundamental MG region  $\mathcal{S}_2^\star(\D=\infty,\rho, \rho_f)$ \emph{is} the set of  all non-negative MG pairs $(\S^{(\U)}, \; \S^{(\e)})$  satisfying \eqref{eq:a2} and 
 \begin{IEEEeqnarray}{rCl}\label{eq:rhoequal2}
 \S^{(\e)} & \le & \rho(1-\rho_f).
 \end{IEEEeqnarray}
\end{corollary}
\begin{IEEEproof} The results follow from Proposition~\ref{proposition2} and Theorem \ref{theorem4}.
\end{IEEEproof}

Fig.~\ref{fig3} illustrates the outer and inner bounds on the MG region for Models 1 and 2 under both Tx- and Rx-Cooperation for  $\rho = 0.8, \rho_f = 0.6$ and $\D = 10$.  We generally notice that the best inner bound is quite close to the best outer bound, thus providing a good approximation of the fundamental per-user MG regions. Comparing the different inner bounds, we further remark that the adaptive scheme significantly improves over the non-adaptive scheme, and for the presented set of parameters and for $\D=10$ the fundamental per-user MG is already close to the one for an unlimited number of cooperation rounds $\D\to \infty$. Fig.~\ref{figitw4}, which shows similar plots but for $\rho =0.4, \rho_f = 0.3$ and $\D=4$, allows for the same conclusions, and moreover shows that for a small user activity parameter $\rho=0.4$ even $\D=4$ cooperation rounds suffice to well approximate the asymptotic fundamental per-user MG region for $\D\to \infty$.  The reason is that a large number of cooperation rounds is only useful in subnets with a large number of consecutive Txs that are  active, and such subnets are extraordinarily rare when $\rho$ is small.  In our achievability and converse results this phenomenon can be observed by noting that the influence of $\D$ in the exponent decreases with the value of  $\rho$ (for Model 1) and with $\rho (1-\rho_f)$   for Model 2.

 In Model~$1$,  the bounds all have the shapes of right-angled trapezoids with parallel sides at $\S^{(\U)}=0$ and at the maximum URLLC MG $\S^{(\U)}=\frac{\rho \rho_f}{2}$.  The  most interesting part of the bounds is the upper side of the trapezoids, which lies opposite the two right angles. In particular, the slope of this side is $-1$ for the outer bounds which indicates that sum per-user MG along this  line stays constant for all values  of the URLLC per-user  MG $\S^{(\U)}\leq \frac{\rho\rho_f}{2}$. For the inner bounds  the slope is  $-M_{\text{both},1}^{(i)}$ with $i \in \{\text{A}, \text{NA}\}$ indicating that  their sum per-user MG is reduced by $(M_{\text{both},1}^{(i)}-1) \beta$ when the URLLC per-user MG $\S^{(\U)}$ is increased by $\beta$.  

In Model~$2$,  the outer bounds  have rectangular shapes and the inner bounds are nearly-rectangular. These shapes indicate that the sum per-user MG increases with the URLLC per-user MG, and thus operating at large URLLC per-user MG $\S^{(\U)}$ does not penalize the achievable eMBB per-user MG $\S^{(\e)}$. In fact, under Model 2  one cannot  trade URLLC messages for eMBB messages because each Tx only has one of the two to send.

The shapes of the per-user MG regions in the two models should also be compared to a triangular shape which corresponds to a scheme that schedules the eMBB and URLLC transmissions into orthogonal slices (e.g., frequency bands or time-slots). Under both models, our schemes significantly outperform such simple scheduling schemes.


\subsection{Results with only Rx-Cooperation}
\begin{proposition}[Non-Adaptive Scheme, Model 1] \label{proposition3}
For $\rho\in (0,1]$ and $\rho_f\in(0,1]$, the  fundamental MG region $\mathcal{S}_1^\star(\D, \rho, \rho_f)$ for Model 1 includes all nonnegative pairs $(\S^{(\U)}, \; \S^{(\e)})$ satisfying
\begin{IEEEeqnarray}{rCl}
 \S^{(\U)} &\le&   \frac{\rho \rho_f}{2}, \label{eq:a-onlyRx} \\
 \S^{(\e)} +   M_{\text{Rx},1}^{(\text{NA})}\S^{(\U)} & \le & \rho\frac{\D+1}{\D+2},\label{eq:b-onlyRx}
 \end{IEEEeqnarray}
where
\begin{IEEEeqnarray}{rCl}
M_{\text{Rx},1}^{(\text{NA})} \triangleq \frac{\D+1}{\D+2} \frac{2}{\rho_f} - \frac{1-\rho_f}{\rho_f}.
\end{IEEEeqnarray}
For $\rho\in(0,1]$ and $\rho_f=0$, above result continues to hold if the left-hand side of \eqref{eq:b-onlyRx} is replaced by the single term $\S^{(\e)}$.
 \end{proposition}
 \begin{IEEEproof}
 Choose 
 \begin{subequations}\label{choice}
 \begin{IEEEeqnarray}{rCl}
  \mathcal{T}_{\text{URLLC}}' &=& \{1,3, \ldots, K-1\} \\
  \mathcal{T}_{\textnormal{silent}}' &= & \{c(\D+2)\}_{c=1}^{\lceil \frac{\ell}{\D+2} \rceil} \\
  \mathcal{T}_{\text{eMBB}}' &= & \mathcal K \backslash \mathcal{T}_{\textnormal{silent}}'.
   \end{IEEEeqnarray}
   \end{subequations}
  Substituting  this choice into \eqref{eq:Mod1-onlyRx_1} and \eqref{eq:Mod1-onlyRx} proves achievability of  the pair
 $\S^{(\U)}=\alpha \frac{\rho_f \rho}{2}$ and  $\S^{(\e)}=\alpha \frac{(1-\rho_f) \rho}{2}+(1-\alpha)\rho\frac{\D+1}{\D+2}$  for any $\alpha\in[0,1]$. The proposition is obtained by time-sharing the schemes for $\alpha=0$ and $\alpha=1$.
 \end{IEEEproof}
 \medskip

 \begin{theorem}[Adaptive Scheme, Model 1] \label{theorem5}
 For $\rho\in (0,1)$ and $\rho_f \in(0,1]$, the  fundamental MG region $\mathcal{S}_1^\star(\D, \rho, \rho_f)$ of Model~1 includes all nonnegative pairs $(\S^{(\U)}, \; \S^{(\e)})$ satisfying
 \begin{IEEEeqnarray}{rCl}
 \S^{(\U)} &\le& \frac{\rho \rho_f}{2}, \label{eq:SF-Model1}  \\
 \S^{(\e)} + M_{\text{Rx},1}^{(\text A)} \cdot  \S^{(\U)} & \le & \rho - \frac{(1-\rho) \rho^{\D+2}}{1-\rho^{\D+2}}, \label{eq:da}
 \end{IEEEeqnarray}
 where 
 \begin{IEEEeqnarray}{rCl} \label{eq:MRx1}
 M_{\text{Rx},1}^{(\text A)} & \triangleq&  \frac{2}{\rho_f} - \frac{2(1-\rho) \rho^{\D+1}}{\rho_f(1-\rho^{\D+2})} +  \frac{\rho^3 \rho_f(1-\rho_f)}{(1-\rho^2(1-\rho_f)))} \left ( \frac{2}{(1-\rho^2(1-\rho_f))} - \frac{\rho^{\D}(1-\rho_f)^{\frac{\D}{2}}}{1-\rho^{\D+2}(1-\rho_f)^{\frac{\D+2}{2}}} \right)\nonumber \\
&& -   \frac{ (1-\rho_f)(1- \rho(1-\rho_f))^2}{\rho_f(1- \rho^2(1-\rho_f))} \left ( 2 \rho + 2 \rho^2(1- \rho_f) + 1 - \frac{\rho^{\D+1}(1-\rho_f)^{\frac{\D}{2}}(1+\rho(1-\rho_f))}{(1- \rho^{\D+2}(1-\rho_f)^{\frac{\D+2}{2}})}\right) \notag \\
&& - \frac{ (1-\rho)^2}{\rho_f(1-\rho^2(1-\rho_f))} \left ( \frac{2\rho(1-\rho_f)}{(1-\rho^2(1-\rho_f))} + 1- \frac{\rho^{\D+1}(1-\rho_f)^{\frac{\D+2}{2}}(1+\rho)}{(1- \rho^{\D+2}(1-\rho_f)^{\frac{\D+2}{2}})}\right).
 \end{IEEEeqnarray}
 For $\rho\in (0,1)$ and $\rho_f =0$, above result holds if the left-hand side of \eqref{eq:da} is replaced  by $\S^{(\e)}$.
 \end{theorem}
\begin{IEEEproof}
 See Appendix~\ref{App:B}.
 \end{IEEEproof}

\begin{theorem}[Outer Bound, Model 1] \label{theorem6-Rx}
 For $\rho \in (0,1)$, all achievable MG pairs  $(\S^{(\U)}, \; \S^{(\e)})$  of Model 1 satisfy \eqref{eq:SF-Model1} 
  \begin{IEEEeqnarray}{rCl}
 \S^{(\e)} + \S^{(\U)} & \le & \rho - \frac{(1-\rho) \rho^{\D+2}}{1-\rho^{\D+2}}. \label{eq:conv2b}\\
 \S^{(\e)} +(1+ \rho)   \S^{(\U)} & \le & \rho. \label{eq:conv2d}
 \end{IEEEeqnarray}
 \end{theorem}
 \begin{IEEEproof}
 The bounds in \eqref{eq:SF-Model1} and \eqref{eq:conv2b} coincide with the bounds for Tx- and Rx-cooperation in Theorem~\ref{theorem2} and can be proved in the same way.  The bound in  \eqref{eq:conv2b} is proved in Appendix~\ref{App: B-conv}. \end{IEEEproof}

Our inner and outer bounds determine again the largest URLLC and the largest eMBB per-user MGs. They are the same as under both Tx- and Rx-cooperation. Having only Rx-cooperation thus does not harm the individual largest per-user MGs.  In contrast, the largest eMBB per-user MG that is achievable for $\S^{(\U)}=\frac{\rho\rho_f}{2}$ is significantly deteriorated when only Rxs can cooperate. For example, in the extreme case $\D\to \infty$, with Tx- and Rx-cooperation $\S^{(\e)}=\rho \left(1- \frac{\rho \rho_f}{2}\right)$ is achievable when $\S^{(\U)}=\frac{\rho \rho_f}{2}$. Our converse  result in Theorem~\ref{theorem6-Rx} indicates that in this case no eMBB per-user MG above $\rho \left(1-(1+\rho) \frac{\rho \rho_f}{2}\right)$ is achievable.
The reason for this degradation is  that  cancelling interference of eMBB transmissions on URLLC transmissions requires Tx-cooperation and seems unfeasible otherwise. As a consequence, when Txs cannot cooperate, eMBB transmissions that would interfere URLLC transmissions cannot be scheduled. 

\medskip

 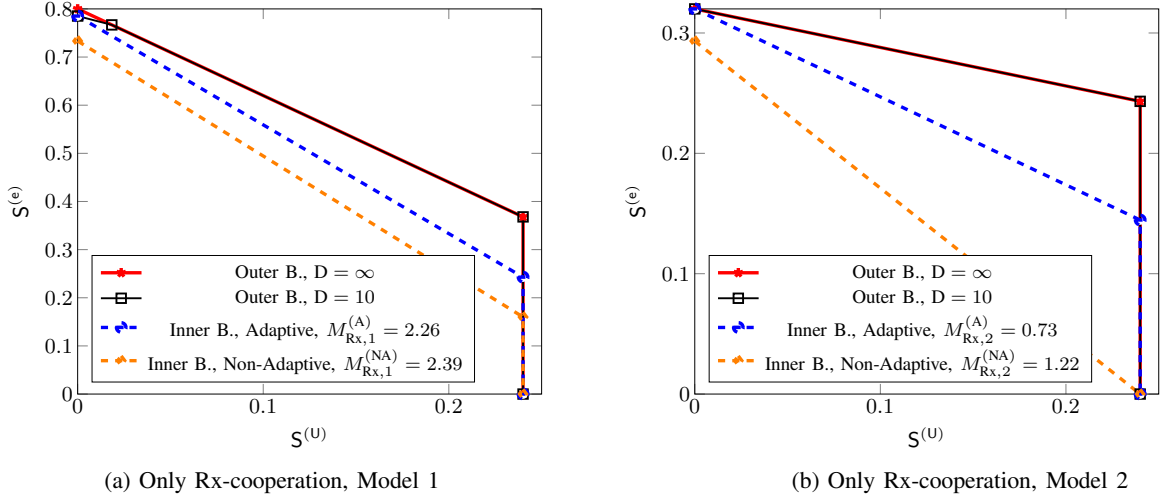
\begin{figure}[t!]
\begin{subfigure}{0.5\textwidth}
\centering
\begin{tikzpicture}[scale=.9]
\begin{axis}[
    xlabel={\small {$\S^{(\U)}$ }},
    ylabel={\small {$\S^{(\e)}$ }},
     xlabel style={yshift=.5em},
     ylabel style={yshift=-1.25em},
    xmin=0, xmax=0.25,
    ymin=0, ymax=0.8,
    xtick={0,0.1,0.2,0.3,0.4},
    ytick={0,0.1,0.2,0.3,0.4,0.5,0.6,0.7,0.8,0.9,1},
    yticklabel style = {font=\small,xshift=0.25ex},
    xticklabel style = {font=\small,yshift=0.25ex},
    legend pos=south west,
]

 \addplot[ color=red,   mark=star, line width = 0.5mm] coordinates {  (0,0.8) (0.24, 0.3680)(0.24,0) };
 \addplot[ color=black,   mark=square, thick] coordinates {  (0,0.7852) (0.0184,0.7668) (0.24,0.3680)(0.24,0) };
 \addplot[ color=blue,   mark=halfcircle, line width = 0.5mm, dashed] coordinates {  (0,0.7852) (0.24,0.2428)(0.24,0) };
  \addplot[ color=orange,   mark=diamond,  line width = 0.5mm, dashed] coordinates {  (0,0.733333333333333) (0.24, 0.1600)(0.24,0) };


\footnotesize
      \legend{{Outer B., $\D = \infty$}, {Outer B., $\D = 10$}, {Inner B., Adaptive, $M_{\text{Rx},1}^{(\text A)} = 2.26$}, {Inner B., Non-Adaptive, $M_{\text{Rx},1}^{(\text{NA})} = 2.39$}}  
\end{axis}


\vspace{-0.4cm}
\end{tikzpicture}
\caption{Only Rx-cooperation, Model 1}
\end{subfigure}
\begin{subfigure}{0.5\textwidth}
\begin{tikzpicture}[scale=.9]
\begin{axis}[
    xlabel={\small {$\S^{(\U)}$ }},
    ylabel={\small {$\S^{(\e)}$ }},
     xlabel style={yshift=.5em},
     ylabel style={yshift=-1.25em},
    xmin=0, xmax=0.25,
    ymin=0, ymax=0.32,
    xtick={0,0.1,0.2,0.3,0.4},
    ytick={0,0.1,0.2,0.3,0.4,0.5,0.6,0.7,0.8,0.9,1},
    yticklabel style = {font=\small,xshift=0.25ex},
    xticklabel style = {font=\small,yshift=0.25ex},
    legend pos=south west,
]

\addplot[ color=red,   mark=star, line width = 0.5mm] coordinates {  (0,0.32) (0.24,0.2432)(0.24,0) };
 \addplot[ color=black,   mark=square, thick] coordinates {  (0,0.319999216012473) (0.24,0.2432)(0.24,0) };
 \addplot[ color=blue,   mark=halfcircle, line width = 0.5mm, dashed] coordinates {  (0,0.319999216012473) (0.24, 0.1446)(0.24,0) };
  \addplot[ color=orange,   mark=diamond,  line width = 0.5mm, dashed] coordinates {  (0,0.2933) (0.24,0) };

\footnotesize
      \legend{{Outer B., $\D = \infty$}, {Outer B., $\D = 10$},  {Inner B., Adaptive, $M_{\text{Rx},2}^{(\text A)} = 0.73$},{Inner B., Non-Adaptive, $M_{\text{Rx},2}^{(\text{NA})} =1.22$}}  
\end{axis}


\vspace{-0.4cm}
\end{tikzpicture}
\caption{Only  Rx-cooperation, Model 2}
\end{subfigure}
\caption{Inner and outer bounds on the fundamental per-user  MG regions $\mathcal{S}_1^\star(\D,\rho, \rho_f)$ and $\mathcal{S}_2^\star(\D,\rho, \rho_f)$ for $\rho= 0.8$ and $\rho_f = 0.6$, $\D = 10$ for Rx-cooperation only.}
\label{fig3b}
\vspace{-0.5cm}
\end{figure}

We next present our results under Model 2.
\begin{proposition}[Non-Adaptive Scheme, Model 2] \label{proposition4}
 For $\rho\in (0,1]$ and $\rho_f\in(0,1]$, the  fundamental MG region $\S_2^*(\D, \rho, \rho_f)$  includes all nonnegative pairs $(\S^{(\U)}, \; \S^{(\e)})$ satisfying
 \begin{IEEEeqnarray}{rCl}
 \S^{(\U)} &\le& \frac{\rho \rho_f}{2}, \label{eq:a-onlyRx} \\
 \S^{(\e)} +   M_{\text{Rx},2}^{(\text{NA})} \cdot \S^{(\U)} & \le &\rho(1-\rho_f)\frac{\D+1}{\D+2}.,\label{eq:Model2-onlyRx}
 \end{IEEEeqnarray}
where
\begin{equation}\label{eq:lastss}
M_{\text{Rx},2}^{(\text{NA})} \triangleq \frac{1-\rho_f}{\rho_f} \cdot \frac{2(\D+1)}{\D+2}.
\end{equation}
For $\rho\in(0,1]$ and $\rho_f =1$ the same result holds if the left-hand side of \eqref{eq:a-onlyRx} is replaced by the single term $\S^{(\e)}$.
 \end{proposition}

 \begin{IEEEproof}
This result follows by substituting  the choice  \eqref{choice} into \eqref{eq:Mod2-onlyRx1} and  \eqref{eq:Mod2-onlyRx2}.  For $\alpha=0$ this proves achievability of the pair $\S^{(\U)}=0$ and $\S^{(\e)}= \rho (1-\rho_f) \frac{ \D+1}{\D+2}$ and for $\alpha=1$ achievability of the pair $\S^{(\U)}=\frac{\rho\rho_f}{2}$ and $\S^{(\e)}= \frac{\rho (1-\rho_f) }{2}$. The proposition then holds by time-sharing arguments.
 \end{IEEEproof}

 \begin{theorem}[Adaptive Scheme, Model 2] \label{theorem6}
For $\rho\in (0,1)$ and $\rho_f \in(0,1]$, the  fundamental MG region $\mathcal{S}_2^\star(\D, \rho, \rho_f)$  includes all nonnegative pairs $(\S^{(\U)}, \; \S^{(\e)})$ satisfying
 \begin{IEEEeqnarray}{rCl}
 \S^{(\U)} &\le& \frac{\rho \rho_f}{2},  \\
 \S^{(\e)} + M_{\text{Rx},2}^{(\text A)} \cdot  \S^{(\U)} & \le &\rho(1-\rho_f) - \frac{(1-\rho(1-\rho_f))\rho^{\D+2}(1-\rho_f)^{\D+2}}{ 1- \rho^{\D+2} (1- \rho_f)^{\D+2} },\label{ee}
 \end{IEEEeqnarray}
 where 
 \begin{IEEEeqnarray}{rCl} \label{eq:MRx1}
 M_{\text{Rx},2}^{(\text A)}  &\triangleq&  \frac{2(1-\rho_f)}{\rho_f} - \frac{2(1-\rho(1-\rho_f))\rho^{\D+1}(1-\rho_f)^{\D+2}}{ \rho_f(1- \rho^{\D+2} (1- \rho_f)^{\D+2}) } \notag \\
&& +\frac{\rho^3\rho_f(1-\rho_f)^2}{(1-\rho^2(1-\rho_f)^2)} \left ( \frac{2}{1-\rho^2(1-\rho_f)^2} - \frac{\rho^{\D}(1-\rho_f)^{\D}}{1-\rho^{\D+2}(1-\rho_f)^{\D+2}}\right)\nonumber \\
&& -  \frac{(1-\rho_f) \left ( (1-\rho(1-\rho_f))^2 + (1-\rho)^2 \right)}{\rho_f(1- \rho(1-\rho_f))}\left (\frac{1}{1-\rho(1-\rho_f)} - \frac{\rho^{\D+1}(1-\rho_f)^{\D+1}}{1-\rho^{\D+2}(1-\rho_f)^{\D+2}} \right). 
\nonumber \\
 \end{IEEEeqnarray}
 For $\rho\in(0,1)$ and $\rho_f =1$ the same result holds if the left-hand side of \eqref{ee} is replaced by the single term $\S^{(\e)}$.
 \end{theorem}
\begin{IEEEproof}
 See Appendix~\ref{App:B}.
 \end{IEEEproof}

\begin{theorem}[Outer Bound, Model 2] \label{theorem8-Rx}
 For $\rho \in (0,1)$, all achievable MG pairs  $(\S^{(\U)}, \; \S^{(\e)})$  of Model 2 satisfy \eqref{eq:SF-Model1} and the two bounds 
 \begin{IEEEeqnarray}{rCl}
 \S^{(\e)} & \le & \rho(1-\rho_f) - \frac{(1-\rho(1-\rho_f))\rho^{\D+2}(1-\rho_f)^{\D+2}}{ 1- \rho^{\D+2} (1- \rho_f)^{\D+2} }.\label{eq:b22dnew}
 \end{IEEEeqnarray}
 and
 \begin{IEEEeqnarray}{rCl}
 \S^{(\e)} +\rho(1-\rho_f)\cdot \S^{(\U)} & \le & \rho(1- \rho_f). \label{eq:conv22}
 \end{IEEEeqnarray}
 \end{theorem}
 \begin{IEEEproof}
 The bounds in \eqref{eq:SF-Model1} and \eqref{eq:b22dnew} coincide with the bounds for  Tx- and Rx-cooperation in Theorem~\ref{theorem4} and can be proved in the same way. The bound in \eqref{eq:conv22} is proved in Appendix~\ref{App: B-conv2}.
  \end{IEEEproof}
  \medskip

Fig.~\ref{fig3b} plots our inner and outer bounds on the fundamental per-user MG regions $\mathcal{S}_1^\star(\D,\rho, \rho_f)$ and $\mathcal{S}_2^\star(\D,\rho, \rho_f)$ for Rx-cooperation under both models and  for $\rho= 0.8$, $\rho_f = 0.6$ and $\D = 10$. As discussed above, the  bounds show a significant degradation of the largest eMBB per-user MG that is simultaneously achievable as the URLLC per-user MG $\S^{(\U)}=\frac{\rho \rho_f}{2}$ compared to the case with both Tx- and Rx- cooperation.

\section{Hexagonal Network} \label{sec:hexagonal}
\begin{figure}[t]
\begin{subfigure}{0.5\textwidth}
\centering
  	\begin{tikzpicture}[scale=0.6, >=stealth]
  	\centering
  	\foreach \j in {1,0,-1}
  	\foreach \i in {-2}{ 	       
	       \draw (2+3 +0.5 + 1.5+3*\j,1.7321+1*1.7321+0.8660+ 0.5*1.7321+1.7321*\i) --(3*1+3*cos{60}+ 1.5+3*\j,2*sin{60}*3+sin{60}-1.7321+ 0.5*1.7321+1.7321*\i)--(3*1+1+ 1.5+3*\j,2*sin{60}*3-1.7321+ 0.5*1.7321+1.7321*\i)--+(-60:1)--(2+3 +0.5+ 1.5+3*\j,1.7321+1*1.7321+0.8660-1.7321+ 0.5*1.7321+1.7321*\i)--(3*1+1+2+ 1.5+3*\j,2*sin{60}*3-1.7321+ 0.5*1.7321+1.7321*\i) --(2+3 +0.5+ 1.5+3*\j,1.7321+2*1.7321+0.8660-1.7321+ 0.5*1.7321+1.7321*\i);
	       }
	       
	       \foreach \j in {1,0,-1}
  	\foreach \i in {-1}{ 	       
	       \draw (2+3 +0.5 + 1.5+3*\j,1.7321+1*1.7321+0.8660+ 0.5*1.7321+1.7321*\i) --(3*1+3*cos{60}+ 1.5+3*\j,2*sin{60}*3+sin{60}-1.7321+ 0.5*1.7321+1.7321*\i)--(3*1+1+ 1.5+3*\j,2*sin{60}*3-1.7321+ 0.5*1.7321+1.7321*\i)--+(-60:1)--(2+3 +0.5+ 1.5+3*\j,1.7321+1*1.7321+0.8660-1.7321+ 0.5*1.7321+1.7321*\i)--(3*1+1+2+ 1.5+3*\j,2*sin{60}*3-1.7321+ 0.5*1.7321+1.7321*\i) --(2+3 +0.5+ 1.5+3*\j,1.7321+2*1.7321+0.8660-1.7321+ 0.5*1.7321+1.7321*\i);
	       }
	       \foreach \j in {1,0,-1}
  	\foreach \i in {0}{ 	       
	       \draw (2+3 +0.5 + 1.5+3*\j,1.7321+1*1.7321+0.8660+ 0.5*1.7321+1.7321*\i) --(3*1+3*cos{60}+ 1.5+3*\j,2*sin{60}*3+sin{60}-1.7321+ 0.5*1.7321+1.7321*\i)--(3*1+1+ 1.5+3*\j,2*sin{60}*3-1.7321+ 0.5*1.7321+1.7321*\i)--+(-60:1)--(2+3 +0.5+ 1.5+3*\j,1.7321+1*1.7321+0.8660-1.7321+ 0.5*1.7321+1.7321*\i)--(3*1+1+2+ 1.5+3*\j,2*sin{60}*3-1.7321+ 0.5*1.7321+1.7321*\i) --(2+3 +0.5+ 1.5+3*\j,1.7321+2*1.7321+0.8660-1.7321+ 0.5*1.7321+1.7321*\i);
	       }
	       \foreach \j in {1,0,-1}
  	\foreach \i in {1}{ 	       
	       \draw  (2+3 +0.5 + 1.5+3*\j,1.7321+1*1.7321+0.8660+ 0.5*1.7321+1.7321*\i) --(3*1+3*cos{60}+ 1.5+3*\j,2*sin{60}*3+sin{60}-1.7321+ 0.5*1.7321+1.7321*\i)--(3*1+1+ 1.5+3*\j,2*sin{60}*3-1.7321+ 0.5*1.7321+1.7321*\i)--+(-60:1)--(2+3 +0.5+ 1.5+3*\j,1.7321+1*1.7321+0.8660-1.7321+ 0.5*1.7321+1.7321*\i)--(3*1+1+2+ 1.5+3*\j,2*sin{60}*3-1.7321+ 0.5*1.7321+1.7321*\i) --(2+3 +0.5+ 1.5+3*\j,1.7321+2*1.7321+0.8660-1.7321+ 0.5*1.7321+1.7321*\i);
	       }
	        \foreach \j in {1,0,-1}
\foreach \i in {1}{
  	  \draw  (2+3 +0.5 + 3*\j,1.7321+2*1.7321+0.8660+1.7321*\i) --(3*1+3*cos{60}+3*\j,2*sin{60}*3+sin{60}+1.7321*\i)--(3*1+1+3*\j,2*sin{60}*3+1.7321*\i)--+(-60:1)--(2+3 +0.5+3*\j,1.7321+1*1.7321+0.8660+1.7321*\i)--(3*1+1+2+3*\j,2*sin{60}*3+1.7321*\i) --(2+3 +0.5+3*\j,1.7321+2*1.7321+0.8660+1.7321*\i);
  	  }
  	  
  	      \foreach \j in {1,0,-1}
\foreach \i in {0}{
  	  \draw  (2+3 +0.5 + 3*\j,1.7321+2*1.7321+0.8660+1.7321*\i) --(3*1+3*cos{60}+3*\j,2*sin{60}*3+sin{60}+1.7321*\i)--(3*1+1+3*\j,2*sin{60}*3+1.7321*\i)--+(-60:1)--(2+3 +0.5+3*\j,1.7321+1*1.7321+0.8660+1.7321*\i)--(3*1+1+2+3*\j,2*sin{60}*3+1.7321*\i) --(2+3 +0.5+3*\j,1.7321+2*1.7321+0.8660+1.7321*\i);
  	  }
  	      \foreach \j in {1,0,-1}
\foreach \i in {-1}{
  	  \draw (2+3 +0.5 + 3*\j,1.7321+2*1.7321+0.8660+1.7321*\i) --(3*1+3*cos{60}+3*\j,2*sin{60}*3+sin{60}+1.7321*\i)--(3*1+1+3*\j,2*sin{60}*3+1.7321*\i)--+(-60:1)--(2+3 +0.5+3*\j,1.7321+1*1.7321+0.8660+1.7321*\i)--(3*1+1+2+3*\j,2*sin{60}*3+1.7321*\i) --(2+3 +0.5+3*\j,1.7321+2*1.7321+0.8660+1.7321*\i);
  	  }
  	      \foreach \j in {1,0,-1}
\foreach \i in {-2}{
  	  \draw (2+3 +0.5 + 3*\j,1.7321+2*1.7321+0.8660+1.7321*\i) --(3*1+3*cos{60}+3*\j,2*sin{60}*3+sin{60}+1.7321*\i)--(3*1+1+3*\j,2*sin{60}*3+1.7321*\i)--+(-60:1)--(2+3 +0.5+3*\j,1.7321+1*1.7321+0.8660+1.7321*\i)--(3*1+1+2+3*\j,2*sin{60}*3+1.7321*\i) --(2+3 +0.5+3*\j,1.7321+2*1.7321+0.8660+1.7321*\i);
  	  }
  	\foreach \j in {-1,0,1}
  	\foreach \i in {-2,-1,0,1}{
  	  \draw  (2+3 +0.5 + 3*\j,1.7321+2*1.7321+0.8660+1.7321*\i) --(3*1+3*cos{60}+3*\j,2*sin{60}*3+sin{60}+1.7321*\i)--(3*1+1+3*\j,2*sin{60}*3+1.7321*\i)--+(-60:1)--(2+3 +0.5+3*\j,1.7321+1*1.7321+0.8660+1.7321*\i)--(3*1+1+2+3*\j,2*sin{60}*3+1.7321*\i) --(2+3 +0.5+3*\j,1.7321+2*1.7321+0.8660+1.7321*\i);

  	       \draw [fill=green] (2+3 + 3*\j,1.7321+2*1.7321+1.7321*\i)  circle (0.15);
  	       
  	        \foreach \a in {-30, 30, 150,-150, -90} {   
  	     \draw[dashed] (2+3 + 3*\j,1.7321+2*1.7321+1.7321*\i)--+(\a:1.7321);
  	     }
  	     
  	      \foreach \a in {90} {   
  	     \draw[dashed] (2+3 + 3*\j,1.7321+2*1.7321+1.7321*\i)--+(\a:1.7321*0.5);
  	     }
}

\foreach \j in {-1,0,1}
  	\foreach \i in {-2,-1,0,1}{ 	       
	       \draw  (2+3 +0.5 + 1.5+3*\j,1.7321+1*1.7321+0.8660+ 0.5*1.7321+1.7321*\i) --(3*1+3*cos{60}+ 1.5+3*\j,2*sin{60}*3+sin{60}-1.7321+ 0.5*1.7321+1.7321*\i)--(3*1+1+ 1.5+3*\j,2*sin{60}*3-1.7321+ 0.5*1.7321+1.7321*\i)--+(-60:1)--(2+3 +0.5+ 1.5+3*\j,1.7321+1*1.7321+0.8660-1.7321+ 0.5*1.7321+1.7321*\i)--(3*1+1+2+ 1.5+3*\j,2*sin{60}*3-1.7321+ 0.5*1.7321+1.7321*\i) --(2+3 +0.5+ 1.5+3*\j,1.7321+2*1.7321+0.8660-1.7321+ 0.5*1.7321+1.7321*\i);
	     
 \foreach \a in {-30, 30, 90, 150,-150} {
	        \draw[dashed] (2+3 + 1.5+3*\j,1.7321*\i+2*1.7321+0.5*1.7321)--+(\a:1.7321);
	        }
	         \foreach \a in {-90} {
	        \draw[dashed] (2+3 + 1.5+3*\j,1.7321*\i+2*1.7321+0.5*1.7321)--+(\a:1.7321*0.5);
	        }
	          \draw [fill= green] (2+3 + 1.5+3*\j,1.7321*\i+2*1.7321+0.5*1.7321)  circle (0.15);
	       }

    \draw[red, very thick, ->] (2+3 + 1.5+3*0,1.7321*0+2*1.7321+0.5*1.7321)--+(90:1.7321);
	             \draw[red, very thick, ->] (2+3 + 1.5+3*0,1.7321*0+2*1.7321+0.5*1.7321)--+(-30:1.7321);
\node[draw =none] at (2+3 + 1.5+3*0 + 0.4,1.7321*0+2*1.7321+0.5*1.7321-0.45-0.1) {\large ${\color{red}\mathbf{e}_x}$};
\node[draw =none] at (2+3 + 1.5+3*0 + 0.35+0.1,1.7321*0+2*1.7321+0.5*1.7321+0.5) {\large ${\color{red}\mathbf{e}_y}$};      
  	\end{tikzpicture} 
\caption{}
\label{fig6.a}
\end{subfigure}
\begin{subfigure}{0.5\textwidth}
\centering
\begin{tikzpicture}[scale=0.6, >=stealth]
  	\centering
  	\foreach \j in {1,0,-1}
  	\foreach \i in {-2}{ 	       
	       \draw[fill=blue]  (2+3 +0.5 + 1.5+3*\j,1.7321+1*1.7321+0.8660+ 0.5*1.7321+1.7321*\i) --(3*1+3*cos{60}+ 1.5+3*\j,2*sin{60}*3+sin{60}-1.7321+ 0.5*1.7321+1.7321*\i)--(3*1+1+ 1.5+3*\j,2*sin{60}*3-1.7321+ 0.5*1.7321+1.7321*\i)--+(-60:1)--(2+3 +0.5+ 1.5+3*\j,1.7321+1*1.7321+0.8660-1.7321+ 0.5*1.7321+1.7321*\i)--(3*1+1+2+ 1.5+3*\j,2*sin{60}*3-1.7321+ 0.5*1.7321+1.7321*\i) --(2+3 +0.5+ 1.5+3*\j,1.7321+2*1.7321+0.8660-1.7321+ 0.5*1.7321+1.7321*\i);
	       }
	       
	       \foreach \j in {1,0,-1}
  	\foreach \i in {-1}{ 	       
	       \draw[fill= pink]  (2+3 +0.5 + 1.5+3*\j,1.7321+1*1.7321+0.8660+ 0.5*1.7321+1.7321*\i) --(3*1+3*cos{60}+ 1.5+3*\j,2*sin{60}*3+sin{60}-1.7321+ 0.5*1.7321+1.7321*\i)--(3*1+1+ 1.5+3*\j,2*sin{60}*3-1.7321+ 0.5*1.7321+1.7321*\i)--+(-60:1)--(2+3 +0.5+ 1.5+3*\j,1.7321+1*1.7321+0.8660-1.7321+ 0.5*1.7321+1.7321*\i)--(3*1+1+2+ 1.5+3*\j,2*sin{60}*3-1.7321+ 0.5*1.7321+1.7321*\i) --(2+3 +0.5+ 1.5+3*\j,1.7321+2*1.7321+0.8660-1.7321+ 0.5*1.7321+1.7321*\i);
	       }
	       \foreach \j in {1,0,-1}
  	\foreach \i in {0}{ 	       
	       \draw[fill=gray]  (2+3 +0.5 + 1.5+3*\j,1.7321+1*1.7321+0.8660+ 0.5*1.7321+1.7321*\i) --(3*1+3*cos{60}+ 1.5+3*\j,2*sin{60}*3+sin{60}-1.7321+ 0.5*1.7321+1.7321*\i)--(3*1+1+ 1.5+3*\j,2*sin{60}*3-1.7321+ 0.5*1.7321+1.7321*\i)--+(-60:1)--(2+3 +0.5+ 1.5+3*\j,1.7321+1*1.7321+0.8660-1.7321+ 0.5*1.7321+1.7321*\i)--(3*1+1+2+ 1.5+3*\j,2*sin{60}*3-1.7321+ 0.5*1.7321+1.7321*\i) --(2+3 +0.5+ 1.5+3*\j,1.7321+2*1.7321+0.8660-1.7321+ 0.5*1.7321+1.7321*\i);
	       }
	       \foreach \j in {1,0,-1}
  	\foreach \i in {1}{ 	       
	       \draw[fill=blue]  (2+3 +0.5 + 1.5+3*\j,1.7321+1*1.7321+0.8660+ 0.5*1.7321+1.7321*\i) --(3*1+3*cos{60}+ 1.5+3*\j,2*sin{60}*3+sin{60}-1.7321+ 0.5*1.7321+1.7321*\i)--(3*1+1+ 1.5+3*\j,2*sin{60}*3-1.7321+ 0.5*1.7321+1.7321*\i)--+(-60:1)--(2+3 +0.5+ 1.5+3*\j,1.7321+1*1.7321+0.8660-1.7321+ 0.5*1.7321+1.7321*\i)--(3*1+1+2+ 1.5+3*\j,2*sin{60}*3-1.7321+ 0.5*1.7321+1.7321*\i) --(2+3 +0.5+ 1.5+3*\j,1.7321+2*1.7321+0.8660-1.7321+ 0.5*1.7321+1.7321*\i);
	       }
	        \foreach \j in {1,0,-1}
\foreach \i in {1}{
  	  \draw[fill=gray]  (2+3 +0.5 + 3*\j,1.7321+2*1.7321+0.8660+1.7321*\i) --(3*1+3*cos{60}+3*\j,2*sin{60}*3+sin{60}+1.7321*\i)--(3*1+1+3*\j,2*sin{60}*3+1.7321*\i)--+(-60:1)--(2+3 +0.5+3*\j,1.7321+1*1.7321+0.8660+1.7321*\i)--(3*1+1+2+3*\j,2*sin{60}*3+1.7321*\i) --(2+3 +0.5+3*\j,1.7321+2*1.7321+0.8660+1.7321*\i);
  	  }
  	  
  	      \foreach \j in {1,0,-1}
\foreach \i in {0}{
  	  \draw[fill=pink]  (2+3 +0.5 + 3*\j,1.7321+2*1.7321+0.8660+1.7321*\i) --(3*1+3*cos{60}+3*\j,2*sin{60}*3+sin{60}+1.7321*\i)--(3*1+1+3*\j,2*sin{60}*3+1.7321*\i)--+(-60:1)--(2+3 +0.5+3*\j,1.7321+1*1.7321+0.8660+1.7321*\i)--(3*1+1+2+3*\j,2*sin{60}*3+1.7321*\i) --(2+3 +0.5+3*\j,1.7321+2*1.7321+0.8660+1.7321*\i);
  	  }
  	      \foreach \j in {1,0,-1}
\foreach \i in {-1}{
  	  \draw[fill=blue]  (2+3 +0.5 + 3*\j,1.7321+2*1.7321+0.8660+1.7321*\i) --(3*1+3*cos{60}+3*\j,2*sin{60}*3+sin{60}+1.7321*\i)--(3*1+1+3*\j,2*sin{60}*3+1.7321*\i)--+(-60:1)--(2+3 +0.5+3*\j,1.7321+1*1.7321+0.8660+1.7321*\i)--(3*1+1+2+3*\j,2*sin{60}*3+1.7321*\i) --(2+3 +0.5+3*\j,1.7321+2*1.7321+0.8660+1.7321*\i);
  	  }
  	      \foreach \j in {1,0,-1}
\foreach \i in {-2}{
  	  \draw[fill=gray]  (2+3 +0.5 + 3*\j,1.7321+2*1.7321+0.8660+1.7321*\i) --(3*1+3*cos{60}+3*\j,2*sin{60}*3+sin{60}+1.7321*\i)--(3*1+1+3*\j,2*sin{60}*3+1.7321*\i)--+(-60:1)--(2+3 +0.5+3*\j,1.7321+1*1.7321+0.8660+1.7321*\i)--(3*1+1+2+3*\j,2*sin{60}*3+1.7321*\i) --(2+3 +0.5+3*\j,1.7321+2*1.7321+0.8660+1.7321*\i);
  	  }
  	\foreach \j in {-1,0,1}
  	\foreach \i in {-2,-1,0,1}{
  	  \draw  (2+3 +0.5 + 3*\j,1.7321+2*1.7321+0.8660+1.7321*\i) --(3*1+3*cos{60}+3*\j,2*sin{60}*3+sin{60}+1.7321*\i)--(3*1+1+3*\j,2*sin{60}*3+1.7321*\i)--+(-60:1)--(2+3 +0.5+3*\j,1.7321+1*1.7321+0.8660+1.7321*\i)--(3*1+1+2+3*\j,2*sin{60}*3+1.7321*\i) --(2+3 +0.5+3*\j,1.7321+2*1.7321+0.8660+1.7321*\i);

  	       \draw [fill=green] (2+3 + 3*\j,1.7321+2*1.7321+1.7321*\i)  circle (0.15);
  	       
  	        \foreach \a in {-30, 30, 150,-150, -90} {   
  	     \draw[dashed] (2+3 + 3*\j,1.7321+2*1.7321+1.7321*\i)--+(\a:1.7321);
  	     }
  	     
  	      \foreach \a in {90} {   
  	     \draw[dashed] (2+3 + 3*\j,1.7321+2*1.7321+1.7321*\i)--+(\a:1.7321*0.5);
  	     }
}

\foreach \j in {-1,0,1}
  	\foreach \i in {-2,-1,0,1}{ 	       
	       \draw  (2+3 +0.5 + 1.5+3*\j,1.7321+1*1.7321+0.8660+ 0.5*1.7321+1.7321*\i) --(3*1+3*cos{60}+ 1.5+3*\j,2*sin{60}*3+sin{60}-1.7321+ 0.5*1.7321+1.7321*\i)--(3*1+1+ 1.5+3*\j,2*sin{60}*3-1.7321+ 0.5*1.7321+1.7321*\i)--+(-60:1)--(2+3 +0.5+ 1.5+3*\j,1.7321+1*1.7321+0.8660-1.7321+ 0.5*1.7321+1.7321*\i)--(3*1+1+2+ 1.5+3*\j,2*sin{60}*3-1.7321+ 0.5*1.7321+1.7321*\i) --(2+3 +0.5+ 1.5+3*\j,1.7321+2*1.7321+0.8660-1.7321+ 0.5*1.7321+1.7321*\i);
	     
 \foreach \a in {-30, 30, 90, 150,-150} {
	        \draw[dashed] (2+3 + 1.5+3*\j,1.7321*\i+2*1.7321+0.5*1.7321)--+(\a:1.7321);
	        }
	         \foreach \a in {-90} {
	        \draw[dashed] (2+3 + 1.5+3*\j,1.7321*\i+2*1.7321+0.5*1.7321)--+(\a:1.7321*0.5);
	        }
	          \draw [fill= green] (2+3 + 1.5+3*\j,1.7321*\i+2*1.7321+0.5*1.7321)  circle (0.15);
	       }

  	\end{tikzpicture} 
\caption{}
\label{fig6.b}
\end{subfigure}
   \caption{An illustration of the hexagonal a) Small circles indicate Txs and Rxs,  black solid lines  the cell borders, and  black dashed lines interference between cells. b) the Tx/Rx pairs in $\mathcal K_1$ are colored in gray, the Tx/Rx pairs in $\mathcal K_2$ in blue and the Tx/Rxs in $\mathcal K_3$ in pink. }
  	
  	\vspace*{-4mm}

  \end{figure}
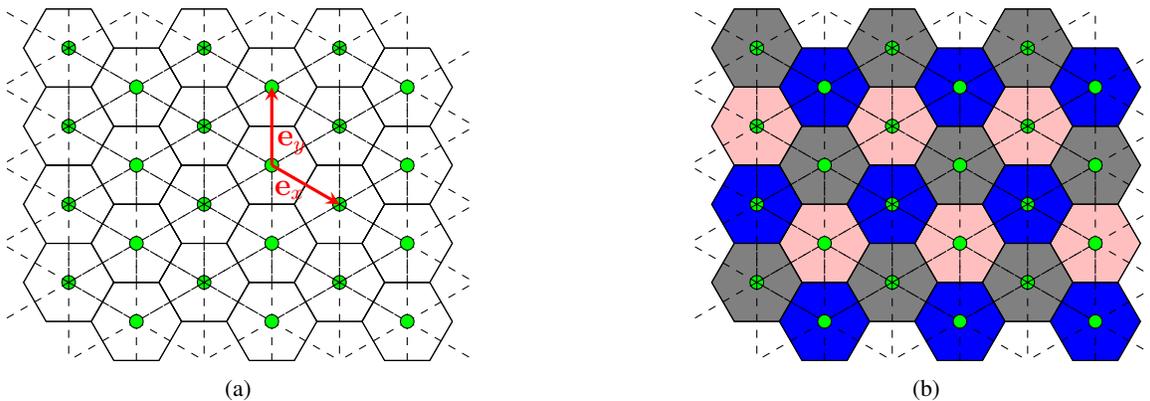

\subsection{Network Model and Cooperation}
Consider a  network with $K$ hexagonal cells, {where each cell consists of one single mobile user (MU) and one BS}. The signals of users that lie in a given cell interfere with the signals sent in the $6$ adjacent cells. 
The interference pattern of our network  is depicted by the black dashed lines in Fig.~\ref{fig6.a}, i.e., 
 the interference set $\mathcal I_k$ contains the indices of the $6$ neighboring cells whose signals interfere with cell $k$. 
 Each Tx~$k$ and each  Rx~$k$ can cooperate with the six Txs or  Rxs in the adjacent cells, i.e., $|\Nk| = 6$. 

To describe the setup and our schemes in  detail, we parametrize the locations of the Tx/Rx pair in the $k$-th cell by a  number $o_k$ in the complex plane $\mathbb{C}$. Introducing the coordinate vectors 
 \begin{equation}
 \mathbf{e}_x = \frac{\sqrt{3}}{2}- \frac{1}{2}i \quad \text{and} \quad  \mathbf{e}_y = i,
 \end{equation}
as in Fig.~\ref{fig6.a}, the position $o_k$ of  Tx/Rx pair $k$ is than associated with a pair of  integers   $(a_k,b_k)$ satisfying
  \begin{equation}
  o_k \triangleq a_k\cdot  \mathbf{e}_x + b_k \cdot  \mathbf{e}_y.
  \end{equation}
The interference set $\mathcal I_k$ and the neighbour set $\mathcal{N}_k$ of Tx/Rx pair $k$ can then be expressed as
\begin{IEEEeqnarray}{rCl}\label{eq:neighbour_Hex}
\Nk=\mathcal I_k= \left \{ k' \colon \quad  |a_k-a_{k'}|=1 \quad \textnormal{and} \quad  |b_k-b_{k'}|=1 \quad  \textnormal{ and} \quad |a_k-a_{k'}-b_k + b_{k'}|= 1\right\}. \IEEEeqnarraynumspace
\end{IEEEeqnarray}

For simplicity, in this section we assume 
\begin{equation}\D =\infty. \label{eq:D_infty}
\end{equation} Similar results can be derived for finite number of cooperation round $\D<\infty$. 


\subsection{Results with both Tx- and Rx-Cooperation}
 Notice that when $\D \to \infty$ 
 there is no difference between the adaptive and  non-adaptive schemes we proposed in the previous section. In this section therefore we only propose a single achievability result under Tx- and Rx-cooperation based on a non-adaptive scheme. 

 \begin{proposition}[Achievability Result, Model 1] \label{propos4}
 For $\rho\in (0,1]$, the  fundamental MG region $\mathcal{S}_1^\star(\D=\infty, \rho, \rho_f)$ for Model 1 includes all non-negative pairs $(\S^{(\U)}, \; \S^{(\e)})$ satisfying
 \begin{IEEEeqnarray}{rCl}
 \S^{(\U)} &\le& \frac{\rho \rho_f}{3}, \label{eq:ahexa} \\
 \S^{(\e)} +   \S^{(\U)} & \le & \rho. \label{eq:bhexa}
 \end{IEEEeqnarray}
 \end{proposition}
 \begin{IEEEproof}
We partition $\mathcal K$  into three subsets $\mathcal K_1, \mathcal K_2, \mathcal K_{3}$  as in Figure~\ref{fig6.b} so that all the signals sent by  Txs in a given subset $\mathcal{K}_i$ do not interfere with one another, i.e., for each  $i \in \{1,2,3 \}$:
\begin{equation}\label{eq:cond_subsets}
k' \notin \mathcal I_{k''} \quad \text{and}\quad  k'' \notin \mathcal I_{k'}, \quad  \forall k', k'' \in \ai.
\end{equation}

Substituting  for  $i\in\{1,2,3\}$ the choice $\mathcal{T}_{\textnormal{silent},i}=\emptyset$ and
\begin{subequations}\label{eq:sethexa}
\begin{IEEEeqnarray}{rCl}
\mathcal{T}_{\textnormal{URLLC},i}& = &\mathcal K_{i}\\
\mathcal{T}_{\textnormal{eMBB},i}& = &\mathcal K \backslash \mathcal{T}_{\textnormal{URLLC},i}
\end{IEEEeqnarray}
\end{subequations}
 into  \eqref{eq:URLLC1} and \eqref{eq:SS_Model1} proves the proposition.
 \end{IEEEproof}

 \begin{proposition}[Achievability Result, Model 2] \label{propos5}
 For $\rho\in (0,1]$, the  fundamental MG region $\mathcal{S}_2^\star( \D=\infty,\rho, \rho_f)$ for Model 2 includes all nonnegative pairs $(\S^{(\U)}, \; \S^{(\e)})$ satisfying
 \begin{IEEEeqnarray}{rCl}
 \S^{(\U)} &\le& \frac{\rho \rho_f}{3}, \label{eq:a2hexa} \\
 \S^{(\e)} & \le & \rho(1-\rho_f).\label{eq:b2hexa}
 \end{IEEEeqnarray}
 \end{proposition}
 \begin{IEEEproof}We partition $\mathcal K$  into three subsets $\mathcal K_1, \mathcal K_2, \mathcal K_{3}$  as in Figure~\ref{fig6.b} so that all the signals sent by  Txs in a given subset $\mathcal{K}_i$ do not interfere with one another, i.e., for each  $i \in \{1,2,3 \}$:
\begin{equation}\label{eq:cond_subsets}
k' \notin \mathcal I_{k''} \quad \text{and}\quad  k'' \notin \mathcal I_{k'}, \quad  \forall k', k'' \in \ai.
\end{equation}

Substituting  the choice \eqref{eq:sethexa} 
 into  \eqref{eq:URLLC1} and \eqref{eq:SS_Model2} proves the proposition.
 \end{IEEEproof}


\subsection{Results with Only Rx-Cooperation}

\begin{proposition}[Non-Adaptive Scheme, Model 1] \label{propos6}
 For $\rho\in (0,1]$, the  fundamental MG region $\mathcal{S}_1^\star(\D=\infty, \rho, \rho_f)$ under Model 1 includes all nonnegative pairs $(\S^{(\U)}, \; \S^{(\e)})$ satisfying
 \begin{IEEEeqnarray}{rCl}
 \S^{(\U)} &\le&   \frac{\rho \rho_f}{3}, \label{eq:a-onlyRxHexa} \\
 \S^{(\e)} +   L_{\text{Rx},1}^{(\text{NA})}\S^{(\U)} & \le & \rho,\label{eq:b-onlyRxHexa}
 \end{IEEEeqnarray}
where
\begin{IEEEeqnarray}{rCl}
L_{\text{Rx},1}^{(\text{NA})} \triangleq \frac{2+\rho_f}{\rho_f}.
\end{IEEEeqnarray}
 \end{proposition}
 \begin{IEEEproof}
Substitute    \eqref{eq:sethexa} into   \eqref{eq:Mod1-onlyRx_1} and \eqref{eq:Mod1-onlyRx}. 
 \end{IEEEproof}
  \begin{theorem}[Adaptive Scheme, Model 1] \label{theorem7}
 For $\rho\in (0,1)$, the  fundamental MG region $\mathcal{S}_1^\star(\D=\infty, \rho, \rho_f)$ of Model 1 includes all nonnegative pairs $(\S^{(\U)}, \; \S^{(\e)})$ satisfying
 \begin{IEEEeqnarray}{rCl}
 \S^{(\U)} &\le& \frac{\rho \rho_f}{3},  \\
 \S^{(\e)} + L_{\Rx,1}^{(\text A)}\S^{(\U)}  & \le \rho.
 \end{IEEEeqnarray}
where 
\begin{equation}
L_{\Rx,1}^{(\text A)} \triangleq \frac{2+ \rho_f - 2(1-\rho\rho_f)^3}{\rho_f}.
\end{equation}
 \end{theorem}
 \begin{IEEEproof}
See Appendices~\ref{App: C-A} and \ref{App: C-B}.
 \end{IEEEproof} 
 
 For Model 2 we have the following results.

\begin{proposition}[Non-Adaptive Scheme, Model 2] \label{propos7}
 For $\rho\in (0,1]$, the  fundamental MG region $\mathcal{S}_2^\star(\D=\infty, \rho, \rho_f)$ for Model 2 includes all nonnegative pairs $(\S^{(\U)}, \; \S^{(\e)})$ satisfying
 \begin{IEEEeqnarray}{rCl}
 \S^{(\U)} &\le&  \frac{\rho \rho_f}{3}, \label{eq:a-onlyRxHexa} \\
 \S^{(\e)} +   L_{\text{Rx},2}^{(\text{NA})}\S^{(\U)} & \le & \rho(1-\rho_f),\label{eq:b-onlyRxHexa}
 \end{IEEEeqnarray}
where
\begin{IEEEeqnarray}{rCl}
L_{\text{Rx},2}^{(\text{NA})} \triangleq \frac{2(1-\rho_f)}{\rho_f}.
\end{IEEEeqnarray}
 \end{proposition}
 \begin{IEEEproof}
 Substitute    \eqref{eq:sethexa} into  \eqref{eq:Mod2-onlyRx1} and \eqref{eq:Mod2-onlyRx2}. 
 \end{IEEEproof}

 \begin{theorem}[Adaptive Scheme, Model 2] \label{theorem8}
 For $\rho\in (0,1)$, the  fundamental MG region $\mathcal{S}_2^\star(\D=\infty, \rho, \rho_f)$ of Model 2 includes all nonnegative pairs $(\S^{(\U)}, \; \S^{(\e)})$ satisfying
 \begin{IEEEeqnarray}{rCl}
 \S^{(\U)} &\le& \frac{\rho \rho_f}{3},  \\
 \S^{(\e)} + L_{\Rx,2}^{(\text A)}\S^{(\U)}  & \le & \rho(1-\rho_f).
 \end{IEEEeqnarray}
where 
\begin{equation}
L_{\Rx,2}^{(\text A)} \triangleq \frac{ 2(1-\rho_f)(1-(1-\rho\rho_f)^3)}{\rho_f}.\end{equation}
 \end{theorem}
 \begin{IEEEproof}
See Appendices~\ref{App: C-D} and \ref{App: C-E}.
 \end{IEEEproof}

\subsection{Numerical Analysis}

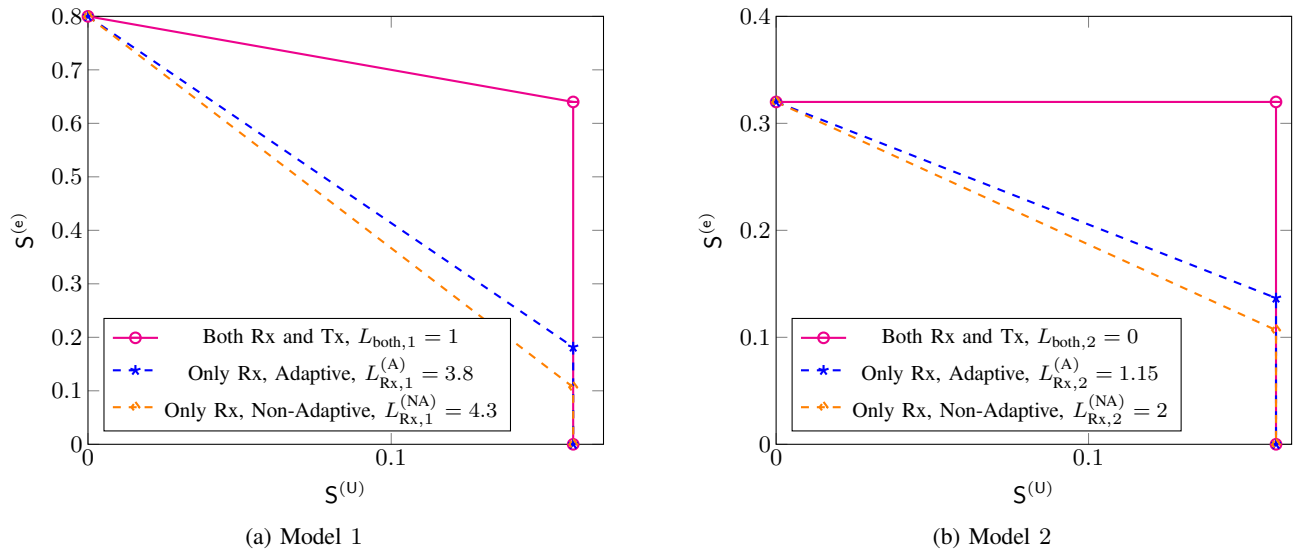
\begin{figure}[t!]
\begin{subfigure}{0.5\textwidth}
\centering
\small
\begin{tikzpicture}[scale=1]
\begin{axis}[
    xlabel={\small {$\S^{(\U)}$ }},
    ylabel={\small {$\S^{(\e)}$ }},
     xlabel style={yshift=.5em},
     ylabel style={yshift=-1.25em},
    xmin=0, xmax=0.17,
    ymin=0, ymax=0.8,
    xtick={0,0.1,0.2,0.3,0.4,0.5},
    ytick={0,0.1,0.2,0.3,0.4,0.5,0.6,0.7,0.8,0.9,1},
    yticklabel style = {font=\small,xshift=0.25ex},
    xticklabel style = {font=\small,yshift=0.25ex},
legend pos=south west,
]


    \addplot[ color=magenta,   mark=halfcircle, thick] coordinates {  (0,0.8) (0.16,0.64) (0.16,0) };
\addplot[ color=blue,   mark= star, thick, dashed] coordinates {  (0,0.8) (0.16,0.1817) (0.16,0) };
 \addplot[ color=orange,   mark= diamond, thick, dashed] coordinates {  (0,0.8) (0.16,0.1067) (0.16,0) };

   \footnotesize 
   \legend{{Both Rx and Tx, $L_{\text{both},1}= 1$}, {Only Rx, Adaptive, $L_{\text{Rx},1}^{(\text{A})} = 3.8$},{Only Rx, Non-Adaptive, $L_{\text{Rx},1}^{(\text{NA})} = 4.3$}}  
\end{axis}


\vspace{-0.4cm}
\end{tikzpicture}
\caption{Model~$1$}
\end{subfigure}
\begin{subfigure}{0.5\textwidth}
\centering 
\small
\begin{tikzpicture}[scale=1]
\begin{axis}[
    xlabel={\small {$\S^{(\U)}$ }},
    ylabel={\small {$\S^{(\e)}$ }},
     xlabel style={yshift=.5em},
     ylabel style={yshift=-1.25em},
    xmin=0, xmax=0.165,
    ymin=0, ymax= 0.4,
    xtick={0,0.1,0.2,0.3,0.4,0.5},
    ytick={0,0.1,0.2,0.3,0.4,0.5,0.6,0.7,0.8,0.9,1},
    yticklabel style = {font=\small,xshift=0.25ex},
    xticklabel style = {font=\small,yshift=0.25ex},
legend pos=south west, 
]

 \addplot[ color=magenta,   mark = halfcircle, thick] coordinates {  (0,0.32) (0.16,0.32) (0.16,0)};
\addplot[ color=blue,   mark=star, thick, dashed] coordinates {  (0,0.32) (0.16,0.1367)(0.16,0)};
     \addplot[ color=orange,   mark=diamond, thick, dashed] coordinates {  (0,0.32)(0.16, 0.1067) (0.16,0)};
\footnotesize

   \legend{{Both Rx and Tx,  $L_{\text{both},2}= 0$},{Only Rx, Adaptive, $L_{\text{Rx},2}^{(\text{A})} = 1.15$},{Only Rx,  Non-Adaptive, $L_{\text{Rx},2}^{(\text{NA})} = 2$}}  
\end{axis}

\end{tikzpicture}
\caption{Model~$2$}
\end{subfigure}
\caption{Inner bounds on the fundamental per-user MG regions $\mathcal{S}_1^\star(\D=\infty, \rho, \rho_f)$ and $\mathcal{S}_2^\star(\D=\infty, \rho, \rho_f)$ under the two models for the hexagonal network with $\rho= 0.8$ and $\rho_f = 0.6$.}
\label{fig7}
\end{figure}

 \begin{figure}[t!]
\begin{subfigure}{0.5\textwidth}
\centering
\small
\begin{tikzpicture}[scale=1]
\begin{axis}[
    xlabel={\small {$\S^{(\U)}$ }},
    ylabel={\small {$\S^{(\e)}$ }},
     xlabel style={yshift=.5em},
     ylabel style={yshift=-1.25em},
    xmin=0, xmax=0.03,
    ymin=0, ymax=0.8,
    xtick={0,0.01,0.02,0.03,0.4,0.5},
    ytick={0,0.1,0.2,0.3,0.4,0.5,0.6,0.7,0.8,0.9,1},
    yticklabel style = {font=\small,xshift=0.25ex},
    xticklabel style = {font=\small,yshift=0.25ex},
legend pos=south west,
]


    \addplot[ color=magenta,   mark=halfcircle, thick] coordinates {  (0,0.8) (0.0267,0.8-0.0267) (0.0267,0) };
\addplot[ color=blue,   mark= star, thick, dashed] coordinates {  (0,0.8) (0.0267,0.6553) (0.0267,0) };
 \addplot[ color=orange,   mark= diamond, thick, dashed] coordinates {  (0,0.8) (0.0267,0.24) (0.0267,0) };

   \footnotesize 
   \legend{{Both Rx and Tx, $L_{\text{both},1}= 1$}, {Only Rx, Adaptive, $L_{\text{Rx},1}^{(\text{A})} = 5.42$},{Only Rx, Non-Adaptive, $L_{\text{Rx},1}^{(\text{NA})} = 21$}}  
\end{axis}


\vspace{-0.4cm}
\end{tikzpicture}
\caption{Model~$1$}
\end{subfigure}
\begin{subfigure}{0.5\textwidth}
\centering 
\small
\begin{tikzpicture}[scale=1]
\begin{axis}[
    xlabel={\small {$\S^{(\U)}$ }},
    ylabel={\small {$\S^{(\e)}$ }},
     xlabel style={yshift=.5em},
     ylabel style={yshift=-1.25em},
    xmin=0, xmax=0.03,
    ymin=0, ymax= 0.8,
    xtick={0,0.01,0.02,0.03,0.4,0.5},
    ytick={0,0.1,0.2,0.3,0.4,0.5,0.6,0.7,0.8,0.9,1},
    yticklabel style = {font=\small,xshift=0.25ex},
    xticklabel style = {font=\small,yshift=0.25ex},
legend pos=south west, 
]

 \addplot[ color=magenta,   mark = halfcircle, thick] coordinates {  (0,0.72) (0.0267,0.72) (0.0267,0)};
\addplot[ color=blue,   mark=star, thick, dashed] coordinates {  (0,0.72) (0.0267,0.6138)(0.0267,0)};
     \addplot[ color=orange,   mark=diamond, thick, dashed] coordinates {  (0,0.72)(0.0267, 0.24) (0.0267,0)};
\footnotesize

   \legend{{Both Rx and Tx,  $L_{\text{both},2}= 0$},{Only Rx, Adaptive, $L_{\text{Rx},2}^{(\text{A})} = 3.98$},{Only Rx,  Non-Adaptive, $L_{\text{Rx},2}^{(\text{NA})} = 18$}}  
\end{axis}

\end{tikzpicture}
\caption{Model~$2$}
\end{subfigure}
\caption{Inner bounds on the fundamental per-user MG regions $\mathcal{S}_1^\star(\D=\infty, \rho, \rho_f)$ and $\mathcal{S}_2^\star(\D=\infty, \rho, \rho_f)$ under the two models for the hexagonal network with $\rho= 0.8$ and $\rho_f = 0.1$.}
\label{fig8}
\end{figure}
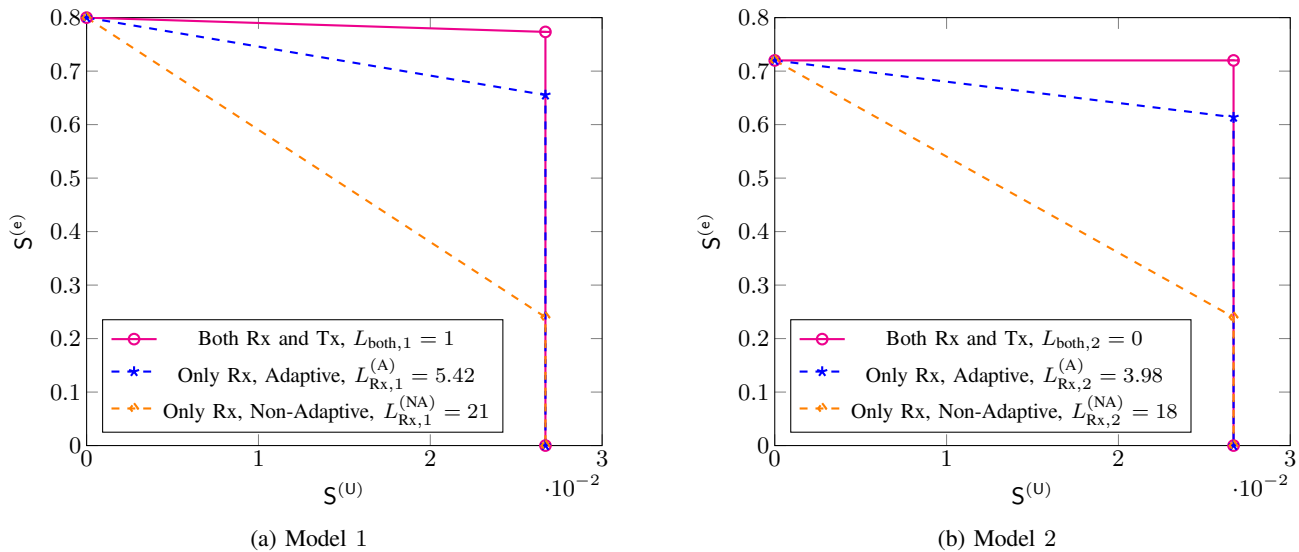

Figures \ref{fig7} and \ref{fig8} illustrate the inner bounds on the MG region $\mathcal{S}_1^\star(\D=\infty, \rho, \rho_f)$ and $\mathcal{S}_2^\star(\D=\infty, \rho, \rho_f)$ for $\rho = 0.8$, $\rho_f = 0.6$ and $\rho_f = 0.1$. As can be seen from this figures, when both Txs and Rxs can cooperate, in Model~$1$, the sum per-user MG stays constant with the URLLC per-user MG, and in Model~$2$, the sum per-user MG increases with the URLLC per-user MG, and thus operating at large URLLC per-user MG $\S^{(\U)}$ does not penalize the achievable eMBB per-user MG $\S^{(\e)}$. However, when only Rxs can cooperate,  large URLLC per-user MG $\S^{(\U)}$  penalizes the achievable eMBB per-user MG $\S^{(\e)}$ under both Model~$1$ and Model~$2$. 

\section{Conclusions and Outlook} \label{sec:conclusion}
We have proposed coding schemes to simultaneously transmit delay-sensitive and delay-tolerant traffic over interference networks with randomly activated users and random data arrivals in a setup with both Tx- and Rx-cooperation or with  only Rx-cooperation. In both setups only delay-tolerant messages can benefit from Tx- or Rx-cooperation  but not delay-sensitive messages. We further considered two user data arrival models. In Model 1 any active Tx has a delay-tolerant message to send and with probability $\rho_f$ it has also a delay-sensitive message to send. In Model 2 a Tx only has a delay-tolerant message to send if it is active but has no delay-sensitive message to send.

 We have derived inner and outer bounds on the per-user multiplexing gain (MG) region of the symmetric Wyner network for both setups  and  both models. Our bounds are tight in  general and coincide in special cases. Moreover, they show that  transmitting delay-sensitive data hardly penalizes the sum per-user MG (for Model 1) or the delay-tolerant per-user MG (for Model 2) as long as both Txs and Rxs can cooperate.  This should in particular be considered in view of scheduling algorithms \cite{Bairagi} where transmission of delay-sensitive data inherently causes a penalty on the sum-MG that is linear in the delay-sensitive  and the sum per-user MG. We further observe that the influence of the number of cooperation rounds $\D$ vanishes with decreasing  probability for a Tx to have a delay-tolerant message. In other words, for small $\rho$ (under Model 1) or small $\rho(1-\rho_f)$ (under Model 2) a small number of cooperation rounds $\D$ allows the system to achieve the same performance as a larger number of rounds.
 
 When only Rxs can cooperate, then the delay-tolerant per-user MG and the sum per-user MG degrade significantly for large delay-sensitive per-user MGs compared to the setup with both Tx- and Rx-cooperation. The reason is that Tx-cooperation seems required to cancel interference of delay-tolerant transmissions on delay-sensitive transmissions, without which the performance degrades. Notice that an  analogous result can also be proved when only Txs but not the Rxs can cooperate, in which case the performance degradation comes from the fact that interference of delay-sensitive transmissions cannot be canceled at  delay-sensitive Rxs. 
 
We further have derived inner bounds on the MG region of the hexagonal model for both cooperation setups (Tx- and Rx-cooperation or only Rx-cooperation) and under both Models~$1$ and  $2$. The results have shown that when both Txs and Rxs can cooperate, in Model~$1$, the sum per-user MG stays constant with increasing delay-sensitive per-user MG and in Model~$2$, the delay-tolerant  per-user MG remains constant while the sum per-user MG increases with increasing delay-sensitive per-user MG.  As for the Wyner symmetric network, we observe that in our inner bounds,  when only Rxs can cooperate, a  large delay-sensitive per-user MG  penalizes the achievable delay-sensitive per-user MG  under both Models~$1$ and~$2$.

Future interesting research directions include the sectorized model studied in \cite{HomaSPAWC2019, Promponas2023}. We conjecture that  also for the sectorized hexagonal model,  a combination of  Tx- and Rx-cooperation allows to mitigate most of the interference  and  essentially eliminate any penalty caused by transmission of  large delay-sensitive data rates. Excellent interference cancellation performance is also expected for  multi-antenna setups.   
\appendices

\section{Useful Summation Formulas}
We start with the well-known formula for geometric sums. For any $d >0$,
\begin{IEEEeqnarray}{rCl} \label{eq:equalities2}
\sum_{\ell = 0}^K d^\ell &=& \frac{1-d^{K+1}}{1-d}, \\
\sum_{\ell = 0}^K \ell d^\ell &=& \frac{d(K d^{K+1} - (K+1)d^K+1)}{(d-1)^2}
\end{IEEEeqnarray}
Moreover, for any  $c \in (0,1)$,
\begin{IEEEeqnarray}{rCl}
\sum_{\ell=1}^{\infty} \ell c ^{\ell} &=&  \frac{c}{(1-c)^2 },  \label{eq:sum2} \\
\sum_{\ell=1}^{\infty} \ell^2 c ^{\ell} &=&  \frac{c(c+1) }{(1-c)^3}. \label{eq:sumx}
\end{IEEEeqnarray} 
Since both sums converge we have,
\begin{IEEEeqnarray}{rCl}
\lim_{K\to \infty}  \frac{1}{K} \sum_{\ell=1}^{K} \ell c^{\ell} =  0 \label{eq:sum_divide}\\
\end{IEEEeqnarray}
and
\begin{IEEEeqnarray}{rCl}
\lim_{K\to \infty}\frac{1}{K}   \sum_{\ell=1}^{K} \ell^2 c^{\ell} =  0  \label{eq:sum_sq_Kdivide}.
\end{IEEEeqnarray}

We next prove that for any
 positive integers $A,B$ so that $B$ divides $A$ and any $c\in[0,1)$ and $d\in[0,1]$,
  \begin{equation} \label{eq:eq2}
  \sum_{\ell = 1}^{\infty}  c^{\ell}  \left \lfloor \frac{\ell}{A} \right \rfloor d^{ \left \lfloor \frac{\ell}{B} \right \rfloor } =\frac{c^A d^{A/B}}{ (1 -c^{A} d^{A/B})(1-c^B d)}  \cdot  \frac{ 1- c^{B}}{1-c}  \end{equation}
  and 
    \begin{equation} \label{eq:eq3}
  \sum_{\ell = 1}^{\infty}  c^{\ell}  \left \lfloor \frac{\ell+1}{A} \right \rfloor d^{ \left \lfloor \frac{\ell}{B} \right \rfloor } -\sum_{\ell = 1}^{\infty}  c^{\ell}  \left \lfloor \frac{\ell}{A} \right \rfloor d^{ \left \lfloor \frac{\ell}{B} \right \rfloor } = c^{A -1}d^{A/B-1} \frac{1}{ 1 -c^{A} d^{A/B }}.  \end{equation}
 In particular, by specializing \eqref{eq:eq2} and \eqref{eq:eq3} to $d=1$, we have
that  for any positive integer $A$ and $c \in [0,1)$: 
\begin{equation} \label{eq:eq1}
  \sum_{\ell = 1}^{\infty}  c^{\ell}  \left \lfloor \frac{\ell}{A} \right \rfloor =  \frac{c^A}{(1-c^A)(1-c)},
\end{equation}
and
\begin{equation} \label{eq:eq1b}
  \sum_{\ell = 1}^{\infty}  c^{\ell}  \left \lfloor \frac{\ell+1}{A} \right \rfloor  -   \sum_{\ell = 1}^{\infty}  c^{\ell}  \left \lfloor \frac{\ell}{A} \right \rfloor=  \frac{c^{A-1}}{(1-c^A)}.
\end{equation}
Also, for any $A>2$ and $B \ge 2$ so that $B$ divides $A$ and any $c\in[0,1)$ and $d\in[0,1]$,
 \begin{equation} \label{eq:eq2new}
  \sum_{\ell = 1}^{\infty}  c^{\ell}  \left \lfloor \frac{\ell}{A} \right \rfloor d^{ \left \lceil \frac{\ell}{B} \right \rceil } =\frac{c^A d^{A/B}}{ (1 -c^{A} d^{A/B})}\left (1 + \frac{cd}{(1-c^B d)}  \cdot  \frac{ 1- c^{B}}{1-c} \right)  \end{equation}
  and 
\begin{equation}  \label{eq:eq3new2}
  \sum_{\ell = 1}^{\infty}  c^{\ell}  \left \lfloor \frac{\ell+1}{A} \right \rfloor d^{ \left \lceil \frac{\ell}{B} \right \rceil } -\sum_{\ell = 1}^{\infty}  c^{\ell}  \left \lfloor \frac{\ell}{A} \right \rfloor d^{ \left \lceil \frac{\ell}{B} \right \rceil } = c^{A -1}d^{A/B} \frac{1}{ 1 -c^{A} d^{A/B }}.   \end{equation}
Similarly, for $B\ge 2$ and $c\in[0,1)$ and $d\in[0,1]$,
\begin{equation}  \label{eq:eq3new3}
  \sum_{\ell = 1}^{\infty}  c^{\ell}\ell d^{ \left \lceil \frac{\ell}{B} \right \rceil } = 
\end{equation}

To prove \eqref{eq:eq2}, represent the summation index  as
  \begin{equation}
 \ell=\left\lfloor\frac{\ell}{A} \right\rfloor  A +  \left\lfloor\frac{i}{B} \right\rfloor B  + k ,
 \end{equation}
 for $i= \ell \mod A$, which lies in $i\in \{0, 1,\ldots, A-1\}$, and $k=\ell \mod B$, which lies in $\{0,1,\ldots, B-1\}$. We can then 
 write
 \begin{IEEEeqnarray}{rCl}
    \sum_{\ell = 1}^{\infty}  c^{\ell}  \left \lfloor \frac{\ell}{A} \right \rfloor d^{ \left \lfloor \frac{\ell}{B} \right \rfloor }
  &= &   \sum_{j= 1}^{\infty}  \sum_{f = 0}^{ \frac{A}{B}-1}  \sum_{k = 0}^{B-1} c^{j A + fB+ k} \cdot   j \cdot d ^{ j A/B + f} \\
  &=&  \sum_{j= 1}^{\infty}  j \left(c^{A} d^{A/B} \right)^j \; \cdot \sum_{f = 0}^{ \frac{A}{B}-1 }  \left(c^{B}d\right)^f \; \cdot \sum_{k = 0}^{B-1} c^{ k}   \\
&= & \frac{c^A d^{A/B}}{ (1 -c^{A} d^{A/B})^2} \cdot  \frac{ 1- c^{A}d^{A/B}}{1-c^B d}  \cdot  \frac{ 1- c^{B}}{1-c} \\
&= & \frac{c^A d^{A/B}}{ (1 -c^{A} d^{A/B})(1-c^B d)}  \cdot  \frac{ 1- c^{B}}{1-c} ,
  \end{IEEEeqnarray}
  which proves \eqref{eq:eq2}.

  To prove  \eqref{eq:eq3}, we notice that the two sums in \eqref{eq:eq3} only differ in the terms $\ell={A}-1+i A$ for all indices $i=0,1,\ldots$. In fact, we have
    \begin{IEEEeqnarray}{rCl}
    \sum_{\ell = 1}^{\infty}  c^{\ell}  \left \lfloor \frac{\ell+1}{A} \right \rfloor d^{ \left \lfloor \frac{\ell}{B} \right \rfloor } -   \sum_{\ell = 1}^{\infty}  c^{\ell}  \left \lfloor \frac{\ell}{A} \right \rfloor d^{ \left \lfloor \frac{\ell}{B} \right \rfloor }  &= &  \sum_{ \substack{\ell = A-1 + i A\\ i \in \{0,1,\ldots \}}} c^\ell d^{ \left \lfloor \frac{\ell}{B} \right \rfloor } \\
  &=&  \sum_{i= 0}^{\infty}  c^{A-1} \left(c^A \right)^i d^{\frac{A}{B}-1} \left( d^{A/B}\right)^i \\
&= & c^{A-1}  d^{\frac{A}{B}-1} \sum_{i= 0}^{\infty}  \left(c^A  d^{A/B}\right)^i   \\
&= & \frac{ c^{A-1}  d^{\frac{A}{B}-1} }{1-c^A  d^{A/B}}.
  \end{IEEEeqnarray}

To prove \eqref{eq:eq2new}, we use the fact that $\lceil \frac{\ell}{B} \rceil = \lfloor \frac{\ell -1}{B} \rfloor +1$ which results in
\begin{IEEEeqnarray}{rCl}
\sum_{\ell = 1}^{\infty}  c^{\ell}  \left \lfloor \frac{\ell}{A} \right \rfloor d^{ \left \lceil \frac{\ell}{B} \right \rceil }
& = & d\sum_{\ell = 1}^{\infty}  c^{\ell}  \left \lfloor \frac{\ell}{A} \right \rfloor d^{ \left \lfloor \frac{\ell-1}{B} \right \rfloor } \\
& \stackrel{(i)}{ = }& d\sum_{\ell '= 0}^{\infty}  c^{\ell'+1}  \left \lfloor \frac{\ell' +1}{A} \right \rfloor d^{ \left \lfloor \frac{\ell'}{B} \right \rfloor } \\
&\stackrel{(ii)}{ = } & dc \sum_{\ell '= 1}^{\infty}  c^{\ell'}  \left \lfloor \frac{\ell' +1}{A} \right \rfloor d^{ \left \lfloor \frac{\ell'}{B} \right \rfloor } \\
& \stackrel{(iii)}{=}& \frac{c^A d^{A/B}}{ (1 -c^{A} d^{A/B})}\left (1 + \frac{cd}{(1-c^B d)}  \cdot  \frac{ 1- c^{B}}{1-c} \right)
\end{IEEEeqnarray}
where  $(i)$ is obtained by  setting  $\ell' = \ell -1$;  step $(ii)$ holds because   $ \left \lfloor \frac{\ell' +1}{A} \right \rfloor = 0$ when $A>2$; and step $(iii)$ holds by \eqref{eq:eq2} and \eqref{eq:eq3}.


\begin{IEEEeqnarray}{rCl}
\lefteqn{\sum_{\ell = 1}^{\infty}  c^{\ell}  \left \lfloor \frac{\ell+1}{A} \right \rfloor d^{ \left \lceil \frac{\ell}{B} \right \rceil } -\sum_{\ell = 1}^{\infty}  c^{\ell}  \left \lfloor \frac{\ell}{A} \right \rfloor d^{ \left \lceil \frac{\ell}{B} \right \rceil } } \notag \\
& = &  \sum_{\ell = 1}^{\infty}  c^{\ell}  \left \lfloor \frac{\ell+1}{A} \right \rfloor d^{ \left \lfloor \frac{\ell-1}{B} \right \rfloor +1} -\sum_{\ell = 1}^{\infty}  c^{\ell}  \left \lfloor \frac{\ell}{A} \right \rfloor d^{ \left \lfloor \frac{\ell -1}{B} \right \rfloor +1 } \\
& = & cd \left (\sum_{\ell' = 1}^{\infty}  c^{\ell'}  \left \lfloor \frac{\ell'+2}{A} \right \rfloor d^{ \left \lfloor \frac{\ell'}{B} \right \rfloor } -\sum_{\ell' = 1}^{\infty}  c^{\ell'}  \left \lfloor \frac{\ell'+1}{A} \right \rfloor d^{ \left \lfloor \frac{\ell'}{B} \right \rfloor  } \right). \\
\end{IEEEeqnarray}
Since the two sums in the above equation only differ in the terms $\ell' = A-2+ iA$ for all indices $i = 0,1, \ldots, $ similarly to the proof of \eqref{eq:eq3} we have
\begin{IEEEeqnarray}{rCl}
cd\left (\sum_{\ell' = 1}^{\infty}  c^{\ell'}  \left \lfloor \frac{\ell'+2}{A} \right \rfloor d^{ \left \lfloor \frac{\ell'}{B} \right \rfloor } -   \sum_{\ell' = 1}^{\infty}  c^{\ell'}  \left \lfloor \frac{\ell' + 1}{A} \right \rfloor d^{ \left \lfloor \frac{\ell'}{B} \right \rfloor } \right) &= &  cd \sum_{ \substack{\ell' = A-2 + i A\\ i \in \{0,1,\ldots \}}} c^{\ell'} d^{ \left \lfloor \frac{\ell'}{B} \right \rfloor } \\
  &=& cd \sum_{i= 0}^{\infty}  c^{A-2} \left(c^A \right)^i d^{\frac{A}{B}-1} \left( d^{A/B}\right)^i \\
&= & c^{A-1}  d^{\frac{A}{B}} \sum_{i= 0}^{\infty}  \left(c^A  d^{A/B}\right)^i   \\
&= & \frac{ c^{A-1}  d^{\frac{A}{B}} }{1-c^A  d^{A/B}},
\end{IEEEeqnarray}
which proves \eqref{eq:eq3new2}.
\medskip

  Notice next that for  any positive $B$ and any $c \in [0,1)$ and $d\in [0,1]$,
\begin{equation} \label{eq:eqsumnew1}
  \sum_{\ell = 1}^{\infty}  c^{\ell}  \ell d^{ \left \lfloor \frac{\ell}{B} \right \rfloor } = \frac{B c^B d}{(1-c^Bd)^2} \cdot \frac{1-c^B}{1-c} + \frac{(B-1)c^{B+1} - Bc^{B}+c}{(1-c^Bd)(c-1)^2}, 
\end{equation}
which is proved as follows. Represent the summation index as
\begin{equation}
\ell = \left \lfloor \frac{\ell}{B} \right \rfloor B + k
\end{equation}
for $k = \ell \mod B$, which lies in $\{0,1,\ldots, B-1\}$. We then write
\begin{IEEEeqnarray}{rCl}
 \sum_{\ell = 1}^{\infty}  c^{\ell} \ell d^{ \left \lfloor \frac{\ell}{B} \right \rfloor }
  &= &   \sum_{j= 1}^{\infty}  \sum_{k = 0}^{B-1} c^{jB+ k} \cdot  ( jB + k) \cdot d ^{j} \\
  &=&   B \sum_{j= 1}^{\infty}  \sum_{k = 0}^{B-1} c^{jB+ k} \cdot  j \cdot d ^{j} +  \sum_{j= 1}^{\infty}  \sum_{k = 0}^{B-1}k \cdot  c^{jB+ k} \cdot d ^{j}\\
 &=&   B \sum_{j= 1}^{\infty} j (c^Bd)^j  \sum_{k = 0}^{B-1} c^{k} +  \sum_{j= 1}^{\infty} (c^B d)^j  \sum_{k = 0}^{B-1}k \cdot  c^{k} \\
& = & \frac{B c^B d}{(1-c^Bd)^2} \cdot \frac{1-c^B}{1-c} + \frac{(B-1)c^{B+1} - Bc^{B}+c}{(1-c^Bd)(c-1)^2}
\end{IEEEeqnarray}
which proves \eqref{eq:eqsumnew1}. 


\medskip
We conclude this appendix with a few summation formulas that can be proved similarly to the above proofs. Notice that:
\begin{IEEEeqnarray}{rCl}
 \sum_{\ell = 1}^{\infty}  c^{\ell}  \left \lfloor \frac{\ell}{A} \right \rfloor d^{ \left \lfloor \frac{\ell}{B} \right \rfloor } \mathbbm{1}\{\ell \; \text{is even} \; \}&=&\frac{c^A d^{A/B}}{ (1 -c^{A} d^{A/B})(1-c^B d)}  \cdot  \frac{ 1- c^{B}}{1-c^2},  \label{eq:90}\\
 \sum_{\ell = 1}^{\infty}  c^{\ell}  \ell d^{ \left \lfloor \frac{\ell}{B} \right \rfloor } \mathbbm{1}\{\ell \; \text{is even} \; \}&=& \frac{B c^B d}{(1-c^Bd)^2} \cdot \frac{1-c^B}{1-c^2} + \frac{(B-2)c^{B+2} - Bc^{B}+2c^2}{2(1-c^Bd)(c^2-1)^2}. \label{eq:91}
\end{IEEEeqnarray}
For even values of $A$ and $B$ an integer divisor of $A$, we further have
 \begin{equation} \label{eq:eq3new}
  \sum_{\ell = 1}^{\infty}  c^{\ell}  \left \lfloor \frac{\ell+1}{A} \right \rfloor d^{ \left \lfloor \frac{\ell}{B} \right \rfloor } \mathbbm{1}\{\ell \; \text{is even} \; \} -\sum_{\ell = 1}^{\infty}  c^{\ell}  \left \lfloor \frac{\ell}{A} \right \rfloor d^{ \left \lfloor \frac{\ell}{B} \right \rfloor } \mathbbm{1}\{\ell \; \text{is even} \; \} = 0.  \end{equation}

Finally, notice that  $ \left \lceil \frac{\ell}{2} \right \rceil  +  \left \lfloor \frac{\ell}{2} \right \rfloor = \ell$ and $d^{ \left \lceil \frac{\ell}{2} \right \rceil }= d^\ell \cdot \frac{1}{d}^{ \left \lfloor \frac{\ell}{2} \right \rfloor}$. Replacing in  \eqref{eq:eq2}, \eqref{eq:eq3}, and \eqref{eq:eqsumnew1}, the parameter $c$ by $cd$, the parameter $d$ by  $d^{-1}$, and $B$ by $2$, we obtain  that for any even $A$
\begin{IEEEeqnarray}{rCl} \label{eq:125}
 \sum_{\ell = 1}^{\infty}  c^{\ell}  \left \lfloor \frac{\ell}{A} \right \rfloor d^{ \left \lceil \frac{\ell}{2} \right \rceil } &=&\frac{c^A d^{A/2} (1+cd)}{ (1 -c^{A} d^{A/2})(1-c^2 d)},  \\
 \sum_{\ell = 1}^{\infty}  c^{\ell}  \ell d^{ \left \lceil \frac{\ell}{2} \right \rceil } &=& \frac{cd(2c + c^2d+1)}{(1-c^2d)^2} , \label{eq:138}\\
\sum_{\ell = 1}^{\infty}  c^{\ell}  \left \lfloor \frac{\ell+1}{A} \right \rfloor d^{ \left \lceil \frac{\ell}{2} \right \rceil } -\sum_{\ell = 1}^{\infty}  c^{\ell}  \left \lfloor \frac{\ell}{A} \right \rfloor d^{ \left \lceil \frac{\ell}{2} \right \rceil }  &=& c^{A -1}d^{A/2} \frac{1}{ 1 -c^{A} d^{A/2 }}.
\end{IEEEeqnarray}

\section{Proofs of  Theorems~\ref{theorem1} and \ref{theorem2}} \label{App:A}
  
  \subsection{MG results with chosen message  assignments} \label{App:A-B}
  
In this subsection, we characterize total MG pairs that are achievable over Wyner's symmetric network for \emph{chosen} message assignments. These results will be used  in the next-following sections to  derive achievable  per-user  MG pairs under the \emph{random} message-arrival models 1 and 2. 
 For ease of readability of future sections, we  denote the number of users in a subnet by $\ell$.
  
  In all our schemes  URLLC messages are only sent by non-consecutive Txs. We thus will make this assumption without further mentioning it in the titles of the sections. 

  \subsubsection{Sending  eMBB messages at  Txs~$2, \D, \D+4, 2\D+2,  2\D+6,3\D+4,\ldots$}\label{sec:full_S}
  
We will start with an example of the message assignment. Choose  $\mathcal{T}_{\textnormal{URLLC}}$ as the set   of odd integers 
\begin{equation}\label{eq:TxURLLC}
\mathcal{T}_{\textnormal{URLLC}} \subseteq \{1,3, 5, \ldots\},
\end{equation}
and
 silence Tx/Rx pairs 
\begin{equation}\label{eq:Tsi}
\mathcal{T}_{\textnormal{silent}} \triangleq\big \{c(\D + 2)\big\}_{c=0} ^{\lfloor \frac{\ell}{\D+2}\rfloor}. 
\end{equation} 
Choose  the set of eMBB Txs as 
\begin{equation}\label{eq:Tsl}
\mathcal{T}_{\textnormal{eMBB}} = \mathcal{K} \backslash\left( \mathcal{T}_{\textnormal{silent}}\cup \mathcal{T}_{\textnormal{URLLC}}\right).
\end{equation}
It can be verified that  with this selection no adjacent Txs send URLLC messages and the network is split into subnets, so that in each subnet there is a master Rx  (namely Rx~$(c+\frac{1}{2})\D$ for some integer $c\in \{0, 1,\ldots,  \lceil \frac{\ell}{\D+2}\rceil-1\}$)  that can be reached by any other eMBB Rx in the same subnet in $\D/2-1$ cooperation hops. Our  assignment \eqref{eq:TxURLLC}--\eqref{eq:Tsl} thus satisfies  conditions C1 and C2 in Subsection~\ref{sec:basic-both}. 
By  \eqref{eq:rates_Model1},  the proposed message assignment thus  allows to achieve a  total sum MG of   
\begin{IEEEeqnarray}{rCl}   \label{eq:avg_exp_MG}
 \M_{\textnormal{sum,full}}(\ell) \triangleq \ell- \left \lfloor \frac{ \ell}{\D+2}\right \rfloor .
\end{IEEEeqnarray}
Since every second Tx sends a URLLC message, the total URLLC MG is
\begin{IEEEeqnarray}{rCl}   
\M_{\textnormal{URLLC}}(\ell)  = \left \lceil \frac{\ell}{2} \right \rceil.
\end{IEEEeqnarray}

\begin{remark}\label{rem:more_general}
The proposed scheme can  easily be adapted to any other  URLLC Tx-set $\mathcal{T}_{\textnormal{URLLC}}$ 
 that  contains no two successive indices and that for any $c=0, \ldots, \lfloor \frac{ \ell}{\D+2}\rfloor$ satisfies\footnote{If Condition~\eqref{eq:as} is violated, then the scheme in Subsection~\ref{sec:basic-both} does require an extra cooperation round, thus violating the condition on $\D$. For this reason, we propose a second scheme for a different message assignment 
  in the next Subsection~\ref{sec:red_S}. }
\begin{equation}\label{eq:as} 
c(\D+2)+2, c(\D+2)+\D \notin \mathcal{T}_{\textnormal{URLLC}} ,
\end{equation}
while the definitions in \eqref{eq:Tsi} and \eqref{eq:Tsl} are mantained.
The scheme in Subsection~\ref{sec:basic-both} only needs to be adapted in the sense that in any subnet $c$ the Txs $c(\D+2)+1, c(\D+2)+\D+1$ should act as URLLC Txs even in case they are transmitting eMBB messages. This is possible by our assumption \eqref{eq:as} that the neighbors of these Txs have eMBB messages to send. The same total sum MG is attained as before, i.e., \eqref{eq:avg_exp_MG}. The  total URLLC MG is simply 
\begin{equation}\label{eq:SURLLC}
\M_{\textnormal{URLLC,set}}(\ell)\triangleq | \mathcal{T}_{\textnormal{URLLC}}|.
\end{equation}
\end{remark}
\medskip

%

  \subsubsection{Sending  an eMBB message at    Tx~2 or at Tx~$\D-1$ on a subnet of size $\ell \leq \D$} \label{sec:red_S}
  In this case, no Tx has to be silenced, i.e., we choose
     \begin{equation}\label{eq:Tf1}
\mathcal{T}_{\textnormal{silent}} \triangleq \emptyset, 
\end{equation} 
   and we can achieve a total sum-MG of $\M_{\textnormal{sum,full}}(\ell)=\ell$.

Moreover, by assumption  there are no consecutive URLLC Txs. The key idea is to treat either Tx $1$ or Tx $\D$ as a URLLC Tx, irrespective of whether they have URLLC or eMBB messages to transmit. Specifically, if $2 \in \mathcal{T}_{\textnormal{eMBB}}$ we treat Tx 1 as if it was sending a URLLC message and if $\D-1 \in \mathcal{T}_{\textnormal{eMBB}}$ we treat Tx $\D$ as if it were sending a URLLC message. 
We can then run the scheme in Subsection~\ref{sec:basic-both} because conditions C1 and C2 are satisfied. 
%
%
%
In particular, if $2 \in \mathcal{T}_{\textnormal{eMBB}}$, then Rx~$\frac{\D}{2}+1$  can act as a  master Rx that jointly decodes all eMBB messages of Txs $2,\ldots, \ell$. (Recall that in this case Tx 1 acts as a URLLC Tx even it has an eMBB message to send).  If $\ell < \D $ or $\D-1\in \mathcal{T}_{\textnormal{eMBB}}$, then Rx~$\frac{\D}{2}$ can act as a master Rx that jointly decodes all eMBB messages of Txs $1,\ldots, \min\{\ell, \D-1\}$. (Recall that in this case if Tx $\D$ exists, it acts as a URLLC Tx, even if it has an eMBB message to send.)

  \subsection{Achievability Result  for $\S^{(\U)}=0$ under Model 1} \label{App: A-C}

In Wyner's symmetric network, the random user activity cuts the network into smaller non-interfering subnets. The  total eMBB MG of the network  is then simply the sum  of the  eMBB MGs achieved in the various subnets. Moreover, since in each subnet we send only eMBB messages, we are in the situation of Subsection~\ref{sec:full_S}, and the total eMBB  MG  is given by \eqref{eq:avg_exp_MG}. 
We  can then compute  the expected per-user eMBB MG over the entire network as
  \begin{IEEEeqnarray}{rCl} \label{eq:sonlys}
 \S^{(\e)}&=& \varlimsup_{K\to \infty} \frac{1}{K}  \sum_{k=1}^{K}  \sum_{\ell = 1}^{K-k+1} P_{\ell,k}   \cdot  \M_{\textnormal{sum,full}}(\ell)
 \\
&=&  \varlimsup_{K\to \infty} \frac{1}{K}  \sum_{\ell =1}^{K}  \sum_{k = 1}^{K-\ell+1} P_{\ell,k}   \cdot  \M_{\textnormal{sum,full}}(\ell) , \label{eq:limit1}
\end{IEEEeqnarray}
where $P_{\ell,k}$ denotes the probability that a new subnet starts at user $k$ and is of size $\ell$, which equals
 \begin{equation}
P_{\ell, k} = \begin{cases} \rho^{\ell} (1-\rho)^2 & \ell < K-k+1, \; k \geq 2  \\ \rho^{\ell} (1-\rho) &( \ell=K-k+1, \; k\geq  2 )\; \; \\
&   \textnormal{or}  \; \; ( \ell < K-k+1, \; k=1)  \\
\rho^K & k=1, \; \ell=K \end{cases} .\label{eq:Plk0}
\end{equation}

  Plugging  \eqref{eq:Plk0}  into \eqref{eq:limit1}, we obtain
  \begin{IEEEeqnarray}{rCl} \label{eq:sonlys4}
\S^{(\e)} 
&=& \varlimsup_{K\to \infty} \left[  \frac{1}{K} \sum_{\ell = 1}^{K-1}  \rho^{\ell} (1 -\rho)\left(  (K-\ell-1) (1-\rho) + 2 \right) \M_{\textnormal{sum,full}}(\ell)   + \frac{\rho^K}{K} \cdot \M_{\textnormal{sum,full}}(K) \right]. \label{eq:lasT} \IEEEeqnarraynumspace
 \end{IEEEeqnarray} 
 
 In the remainder of this subsection, we analyze the asymptotic behavior of the expression in \eqref{eq:lasT}, where recall that $\rho\in(0,1)$. 
  
We observe  the trivial bound{
\begin{equation}\label{eq:boundS_ell}
 \M_{\textnormal{sum,full}}(\ell ) \leq \ell,
 \end{equation}
and thus since $\rho^K\to 0$ as $K\to \infty$ when $\rho\in(0,1)$, the last term
in \eqref{eq:lasT} vanishes as $K\to \infty$, i.e., 
\begin{IEEEeqnarray}{rCl}\label{eq:first_limit}
\lim_{K\to \infty} \frac{\rho^K}{K} \cdot \left (K- \left \lfloor \frac{K}{\D+2}\right \rfloor \right)& =&0.
\end{IEEEeqnarray}
The trivial bound in \eqref{eq:boundS_ell} also implies  \begin{subequations}\label{eq:sumsdd}
\begin{IEEEeqnarray}{rCl}
0\leq  \sum_{\ell = 1}^{K-1}  \rho^{\ell} (1 -\rho)^2  (\ell +1)  \M_{\textnormal{sum,full}}(\ell) &\leq &  \sum_{\ell = 1}^{K-1}   \rho^{\ell} (1 -\rho)^2  (\ell ^2+\ell) \\
0\leq  \sum_{\ell = 1}^{K-1}2  \rho^{\ell} (1 -\rho)  \M_{\textnormal{sum,full}}(\ell)  &\leq &  \sum_{\ell = 1}^{K-1}2   \rho^{\ell} (1 -\rho) \ell,
\end{IEEEeqnarray} 
\end{subequations}
and since the sums on the right-hand sides of \eqref{eq:sumsdd} converge, see \eqref{eq:sum_divide} and \eqref{eq:sum_sq_Kdivide}, the following limit holds: 
\begin{IEEEeqnarray}{rCl}\label{eq:second_limit}
\lim_{K \to \infty} \frac{1}{K} \sum_{\ell = 1}^{K-1}  \rho^{\ell} (1 -\rho) \left( (-\ell -1) (1-\rho)+2\right)\cdot   \M_{\textnormal{sum,full}}(\ell) =0.
\end{IEEEeqnarray} 
Plugging Limits \eqref{eq:first_limit} and \eqref{eq:second_limit} into \eqref{eq:lasT},  we obtain an eMBB per-user MG  of
\begin{IEEEeqnarray}{rCl} 
\S^{(\e)} &=  & \varlimsup_{K\to \infty}  \frac{1}{K} \sum_{\ell = 1}^{K-1}  K  \rho^{\ell} (1 -\rho)^2  \M_{\textnormal{sum,full}}(\ell)  = \sum_{\ell = 1}^{\infty}   \rho^{\ell} (1 -\rho)^2  \left (\ell - \left \lfloor \frac{\ell}{\D+2}\right \rfloor \right)  \\
&= &\rho - \frac{(1-\rho)\rho^{\D+2}}{1- \rho^{\D+2}}, \label{eq:ssmax0}
\end{IEEEeqnarray}
where the last equation holds by \eqref{eq:sumx}  and \eqref{eq:eq1}.

 \subsection{Converse Bound under  Model 1}  \label{App: A-D}
 The steps described in the previous section can also be used to derive a converse bound on $\S^{(\U)}+\S^{(\e)}$ that holds for all  achievable MG pairs. In fact, $\M_{\textnormal{Sum,full}}$  also represents an upper bound on the  total MG of any subnet of length $\ell$. Following the same arguments as above but where we replace $\S^{(\e)}$ by $\S^{(\U)}+\S^{(\e)}$, we obtain the upper bound
 \begin{IEEEeqnarray}{rCl} \label{eq:ssmax3}
\S^{(\U)}+ \S^{(\e)} & \leq &   \rho - \frac{(1-\rho)\rho^{\D+2}}{1- \rho^{\D+2}}.
\end{IEEEeqnarray}
Combined with the trivial bound 
\begin{equation}
\S^{(\U)} \leq \frac{\rho\rho_f}{2},
\end{equation}
this proves  Theorem~\ref{theorem2} for $\rho\in(0,1)$.

For $\rho=1$ the sum-rate bound follows from the findings in \cite{Roy}, which assumes that all Txs are eMBB.
\medskip 

\subsection{Achievability Result for Large $\S^{(\U)}$ under Model 1} \label{App: A-E}
As before, the user activity pattern $\mathbf{A}$ splits the network into multiple disjoint  subnets. To ensure that each URLLC message can be transmitted at same rate, we split the blocklength into two equally-long phases and time-share two schemes over the two phases. In Phase 1 we only send URLLC messages on odd Txs and in Phase 2 only  from even Txs.

Setting \begin{IEEEeqnarray}{rCl}\label{eq:k1k2}
 	\mathcal K_1&\triangleq& \{1,3, \ldots, K-1\}, \\
 	\mathcal K_2  &\triangleq& \{2,4, \ldots, K\},
 \end{IEEEeqnarray}
in Phase~1 we choose
\begin{IEEEeqnarray}{rCl}\label{eq:TURLLC1}
\mathcal{T}_{\textnormal{URLLC},1}\triangleq  	\mathcal{K}_1 \cap \Tf
\end{IEEEeqnarray}
and in Phase 2  
\begin{IEEEeqnarray}{rCl}\label{eq:TURLLC2}
\mathcal{T}_{\textnormal{URLLC},2}\triangleq  \mathcal{K}_2 \cap \Tf.
\end{IEEEeqnarray}

The URLLC message assignments in both phases satisfy the condition that no two consecutive Txs send URLLC messages and are thus valid. In each of the phases, and for each of the subnets created by deactivated users, we can use one of the two schemes in Subsections   \ref{sec:full_S} or \ref{sec:red_S}, depending on the phase and the URLLC user activity pattern in the subnet. In particular, for a subnet that starts at user $k$  and is of length $\ell$, in Phase $i\in\{1,2\}$ we use:
\begin{itemize}
\item If $k \in \mathcal K_i$ or  if   $k+(c-1)(\D+2)+1, k+c(\D+2)-3 \in \mathcal{K}_{\textnormal{eMBB}}$  for all $c = 1,\ldots,\left \lceil \frac{\ell}{\D+2}\right\rceil$, then we use the scheme in Subsection~\ref{sec:full_S} and achieve a total sum-MG of $ \M_{\textnormal{sum,full}}(\ell)$.
\item  Else if $\ell \leq  \D$, we  use the scheme described in Subsection  \ref{sec:red_S} to transmit. Notice  that the way that we defined the URLLC Tx-sets  $\mathcal{T}_{\textnormal{URLLC},i}$ in \eqref{eq:TURLLC1} and \eqref{eq:TURLLC2} and since $k\notin \mathcal{K}_i$ and $\D$  even,  Tx $k+\D-2$ (if $\ell \geq \D-1$) sends an eMBB message and we satisfy the condition of Subsection~\ref{sec:red_S}. In this case, we again achieve a total sum-MG of $\M_{\textnormal{sum,full}}(\ell)$ over the entire network.
\item Else,   we  desactivate  user $k+\D$, use the scheme in Subsection  \ref{sec:red_S} to transmit on the subnet consisting of the first $\D$ Tx-Rx pairs $k, k+1, \ldots, k+\D-1$  and the scheme in Subsection  \ref{sec:full_S} to transmit on the subnet consisting of the  Tx-Rx pairs $k+\D+1, \ldots, k+\ell-1$ (where we notice that for this latter subnet Condition~\eqref{eq:as} is satisfied in Phase $i$ because $k\notin \mathcal{K}_i$ and therefore also $k+c(\D+2) , k+c(\D+2)+\D\notin \mathcal{T}_{\text{URLLC},i}$ for all positive integers $c$). In this case, since we silenced Tx $k+\D$ (instead of $k+\D+1$), our scheme achieves at total sum-MG of 
\begin{equation}
\M_{\textnormal{sum,red}}(\ell)\triangleq \ell- \left\lfloor \frac{ \ell+1}{\D+2}\right \rfloor.
\end{equation}

\end{itemize}

As previously, the overall scheme  achieves a per-user URLLC MG of 
  \begin{IEEEeqnarray}{rCl} \label{eq:SFF1}
 \S^{(\U)}
&=& \frac{\rho \rho_f }{2}.
\end{IEEEeqnarray}
Moreover, by above considerations, in any subnet that starts at Tx~$k$  and is of size $\ell> \D$, the following  expected sum MG is achievable over the two phases:
\begin{IEEEeqnarray}{rCl} \label{eq:G} 
\bar{\M}_{\textnormal{sum}}(\ell)   & \triangleq & \frac{1}{2}  \M_{\textnormal{sum,full}}(\ell)   + \frac{1}{2}  \M_{\textnormal{sum,full}}(\ell)  \cdot   \prod_{j\in \mathcal{L}} \Pr[ B_j  =1]  + \frac{1}{2}\M_{\textnormal{sum,red}}(\ell) \cdot \left( 1-     \prod_{j\in \mathcal{L}_k} \Pr[ B_j  =1] \right) \nonumber \\\\ 
&=& \left (\ell - \left \lfloor \frac{\ell}{\D+2} \right \rfloor  \right )  \left (\frac{1}{2}  + \frac{(1- \rho_f)^{L}}{2} \right) +  \left (\ell - \left \lfloor \frac{\ell +1}{\D+2} \right \rfloor  \right ) \left (\frac{1}{2}-  \frac{(1- \rho_f)^{L }}{2} \right).
\end{IEEEeqnarray}
where 
\begin{subequations}\label{eq:Lset}
\begin{equation}
\mathcal{L}_k\triangleq \left\{k+(c-1)(\D+2)+1, k+c(\D+2)-3\right\}_{c=1}^{\lceil \frac{\ell}{\D+2}\rceil} \cap \{k,k+1,\ldots, k+L-1\}
\end{equation}
and  
\begin{equation}
L\triangleq |\mathcal{L}_k|= \begin{cases}2\left \lceil \frac{\ell}{\D+2}\right \rceil & \textnormal{ if }\ell \mod \D+2 \in\{0, \D, \D+1\} \\ \\
2\left \lfloor \frac{\ell}{\D+2}\right \rfloor &   \textnormal{ if }\ell \mod \D+2 =1 \\ \\
2\left \lfloor \frac{\ell}{\D+2}\right \rfloor+1 & \textnormal{ otherwise}.
\end{cases}
\end{equation}
\end{subequations}
If the subnet is of size $\ell \leq \D$ then the total sum MG  $\bar{\M}_{\textnormal{sum}}(\ell) =\ell$ is achievable. So \eqref{eq:G} is achievable also for $\ell \leq \D$. 

Similarly to \eqref{eq:sonlys}--\eqref{eq:lasT} we  then obtain
 	\begin{IEEEeqnarray}{rCl}
 		\S^{(\e)} + \S^{(\U)}
 		&=& \varlimsup_{K\to \infty} \frac{1}{K} \sum_{k=1}^{K}  \sum_{\ell = 1}^{K-k+1}  \ P_{\ell,k}\cdot  \bar{\M}_{\textnormal{sum}}(\ell)  \\
		&=& \varlimsup_{K\to \infty} \frac{1}{K} \sum_{\ell=1}^{K}  \sum_{k = 1}^{K-\ell+1}   P_{\ell,k}\cdot  \bar{\M}_{\textnormal{sum}}(\ell)  \\
&=& \varlimsup_{K\to \infty} \left[  \frac{1}{K} \sum_{\ell = 1}^{K-1}  \rho^{\ell} (1 -\rho)\left(  (K-\ell-1) (1-\rho) + 2 \right)\bar{ \M}_{\textnormal{sum}}(\ell) 
 + \frac{\rho^K}{K} \cdot \bar{\M}_{\textnormal{sum}}(K) \right]. \nonumber \\
 \label{eq:last2}
 	\end{IEEEeqnarray}
We continue to analyze the asymptotic expression in \eqref{eq:last2}, where recall that we assume $\rho \in(0,1)$.
Since $\bar{\M}_{\textnormal{sum}}(\ell)\leq \ell$, following similar steps as in \eqref{eq:boundS_ell}--\eqref{eq:ssmax0}, we can conclude  that 
\begin{IEEEeqnarray}{rCl}
   \S^{(\e)} + \S^{(\U)} & =& \sum_{\ell = 1}^{\infty}    \rho^{\ell} (1 -\rho)^2 \bar{ \M}_{\textnormal{sum}}(\ell)  \\
   &\stackrel{(i)}{\geq}& (1-\rho)^2 \sum_{\ell = 1}^{\infty} \rho^{\ell} \ell  -  \frac{(1-\rho)^2}{2}  \sum_{\ell = 1}^{\infty}\rho^{\ell} \left \lfloor \frac{\ell}{\D+2} \right \rfloor \left (1 + (1-\rho_f)^{2 \lceil\frac{\ell}{\D+2} \rceil} \right ) \notag \\
&&- \frac{(1-\rho)^2}{2} \sum_{\ell = 1}^{\infty} \rho^{\ell} \left \lfloor \frac{\ell+1}{\D+2} \right \rfloor \left (1 - (1-\rho_f)^{2\lceil\frac{\ell}{\D+2} \rceil} \right )  \\
   &=& (1-\rho)^2 \sum_{\ell = 1}^{\infty} \rho^{\ell} \ell  -  \frac{(1-\rho)^2}{2}  \sum_{\ell = 1}^{\infty}\rho^{\ell} \left \lfloor \frac{\ell}{\D+2} \right \rfloor \notag \\
&&- \frac{(1-\rho)^2}{2} \sum_{\ell = 1}^{\infty} \rho^{\ell} \left( \left \lfloor \frac{\ell+1}{\D+2}  \right \rfloor  - \left \lfloor \frac{\ell}{\D+2}  \right \rfloor   \right)\left (1 - (1-\rho_f)^{2\lceil\frac{\ell}{\D+2} \rceil} \right )  \\
&\stackrel{(ii)}{=}& (1-\rho)^2  \sum_{\ell = 1}^{\infty} \rho^{\ell} \ell  
-  (1-\rho)^2  \sum_{\ell = 1}^{\infty}\rho^{\ell} \left \lfloor \frac{\ell}{\D+2} \right \rfloor 
 -  \frac{(1-\rho)^2}{2} \cdot \frac{  \rho^{ \D+1}   }{1-  \rho^{ \D+2} }
 \notag \\
 &&+\frac{(1-\rho)^2}{2} \frac{  \rho^{ \D+2-1} (1-\rho_f)^{2\frac{\D+2}{ \D+2}}  }{1-  \rho^{ \D+2}  (1-\rho_f)^{2\frac{\D+2}{ \D+2} } } \\ 
&\stackrel{(iii)}{=} & \rho - \frac{ \rho^{\D+2}(1-\rho)}{1- \rho^{\D+2}} -    \frac{(1-\rho)^2}{2} \cdot \frac{  \rho^{ \D+1}   }{1-  \rho^{ \D+2} }
+\frac{(1-\rho)^2}{2} \frac{  \rho^{ \D+1} (1-\rho_f)^{2 }  }{1-  \rho^{ \D+2}  (1-\rho_f)^{2 } }\\ 
&\stackrel{(iv)}{=}& \rho - \frac{ \rho^{\D+1}(1-\rho^2)}{2(1- \rho^{\D+2})} 
+\frac{(1-\rho)^2}{2} \frac{  \rho^{ \D+1} (1-\rho_f)^{2 }  }{1-  \rho^{ \D+2}  (1-\rho_f)^{2} },
  \end{IEEEeqnarray}
  where  in $(i)$ we use that $\bar{\M}_{\textnormal{sum,full}}(\ell)$ is decreasing in $L$ and we can upper bound $L$ by $2\left \lceil \frac{\ell}{\D+2}\right \rceil$;  in $(ii)$  the third term is the result of using the limiting expression \eqref{eq:eq1b} for $c=\rho$ and $A=\D+2$, and the fourth term is the result of using the limiting expression \eqref{eq:eq3new2} for $c=\rho$, $A=\D+2$ and  $B=\D+2$;  in step $(iii)$ we used the  limits \eqref{eq:sum2} and Limit \eqref{eq:eq1}; and in step $(iv)$ we combined the second and third term into a single expression.   Using \eqref{eq:SFF1}, we finally obtain
  \begin{IEEEeqnarray}{rCl}
   \S^{(\e)}  & =& \rho - \frac{\rho \rho_f}{2} -  \frac{(1-\rho^2) \rho^{\D+1}}{2(1-\rho^{\D+2})}  +  \frac{(1-\rho)^2  \rho^{\D+1}(1-\rho_f)^{2 } }{2 \left(1- \rho^{\D+2} (1-\rho_f)^{2}\right)}.\label{eq:sslast}
   \end{IEEEeqnarray}

\section{Proofs   of Theorems \ref{theorem3} and \ref{theorem4} } \label{App:B1}

\subsection{Achievability for   $\S^{(\U)} = 0$ under Model 2:} \label{App: A-F}
We  now consider Model 2, where each user either has an eMBB or a URLLC message to transmit but never both.
For $\S^{(\U)}=0$, this is  equivalent to considering URLLC Txs as inactive. In this case,  the probability that a subnet starts at Tx $k$ and is of size $\ell$ now equals
 \begin{equation}
P_{\ell, k}^{(2)} = \begin{cases} \rho^{\ell}(1-\rho_f)^{\ell}(1-\rho(1-\rho_f))^2 & \ell < K-k+1, \; k \geq 2  \\ 
 \rho^{\ell}(1-\rho_f)^{\ell}(1-\rho(1-\rho_f))&( \ell=K-k+1, \; k\geq  2 )\; \; \\
&   \textnormal{or}  \; \; ( \ell < K-k+1, \; k=1)  \\
\rho^K (1-\rho_f)^{K}& k=1, \; \ell=K \end{cases} .\label{eq:Plk}
\end{equation}

In each subnet we can employ the coding scheme in Subsection~\ref{sec:full_S} (but without URLLC messages) and thus achieve a total eMBB MG  of $\M_{\textnormal{sum,full}}(\ell)$, when $\ell$ denotes the length of the subnet. Therefore (similarly to \eqref{eq:sonlys}) we have
  \begin{IEEEeqnarray}{rCl}
 \S^{(\e)} &=& \varlimsup_{K\to \infty} \frac{1}{K}  \sum_{k=1}^{K}  \sum_{\ell = 1}^{K-k+1} P_{\ell,k}^{(2)}   \cdot  \M_{\textnormal{sum,full}}(\ell)
 \\
&=&  \varlimsup_{K\to \infty} \frac{1}{K}  \sum_{\ell =1}^{K}  \sum_{k = 1}^{K-\ell+1} P_{\ell,k}^{(2)}   \cdot  \M_{\textnormal{sum,full}}(\ell) \\
&=& \varlimsup_{K\to \infty} \Bigg[  \frac{1}{K} \sum_{\ell = 1}^{K-1}  \rho^{\ell}(1-\rho_f)^{\ell}(1-\rho(1-\rho_f)) \left(  (K-\ell-1) (1-\rho(1-\rho_f))  + 2 \right) \M_{\textnormal{sum,full }}(\ell)  \nonumber \\
&& \hspace{2cm}
 + \frac{\rho^K(1-\rho_f)^K}{K} \cdot \M_{\textnormal{sum,full}}(K) \Bigg ]. \label{eq:sonlys2}
  \end{IEEEeqnarray} 
 For $\rho_f=1$, obviously $\S^{(\e)}=0$, which can also be verified on \eqref{eq:sonlys2}.
 In the following we assume  that $\rho_f\in[0,1)$. Recall  also that  $\rho\in(0,1)$.
 
 To analyze the limiting expression \eqref{eq:sonlys2}  we start  by noticing that $\M_{\textnormal{sum,full}}(\ell) \leq \ell $ implies}
  \begin{IEEEeqnarray}{rCl}
  \varlimsup_{K\to \infty} \frac{\rho^K(1-\rho_f)^K}{K} \cdot \M_{\textnormal{sum,full}}(K) &= &0, \label{eq:first_limitt} \\
\lim_{K \to \infty} \frac{1}{K} \sum_{\ell = 1}^{K-1}  \rho^{\ell} (1-\rho_f)^{\ell}(1-\rho(1-\rho_f)) \left( (-\ell -1) (1-\rho(1-\rho_f)) +2\right)\cdot   \M_{\textnormal{sum,full}}(\ell) &=&0,\IEEEeqnarraynumspace \label{eq:second_limitt}
 \end{IEEEeqnarray}
 where the second limit \eqref{eq:second_limitt} holds by the same arguments that we used to prove \eqref{eq:second_limit}.
 Substituting \eqref{eq:first_limitt} and \eqref{eq:second_limitt} into \eqref{eq:sonlys2}, we then obtain
  \begin{IEEEeqnarray}{rCl}
   \S^{(\e)}& =&   \sum_{\ell = 1}^{\infty}  \rho^{\ell}(1-\rho_f)^{\ell}(1-\rho(1-\rho_f))^2  \M_{\textnormal{sum,full }}(\ell) \\
   &=& \rho(1-\rho_f) - \frac{(1-\rho(1-\rho_f))\rho^{\D+2}(1-\rho_f)^{\D+2}}{ 1- \rho^{\D+2} (1- \rho_f)^{\D+2} }, \label{eq:lastsss}
 \end{IEEEeqnarray}
   where the last equality holds by \eqref{eq:sum2} and \eqref{eq:eq1}.

\subsection{Converse Bound under   Model 2} \label{App: A-G}
 The steps described in the previous section can also be used to derive a converse bound on $\S^{(\e)}$ that holds for all  achievable eMBB MGs. In fact, $\M_{\textnormal{Sum,full}}(\ell)$  also represents an upper bound on the  total eMBB MG of any subnet of length $\ell$. Following the same arguments as above  we obtain the upper bound
 \begin{IEEEeqnarray}{rCl} \label{eq:ssmax2}
\S^{(\e)} & \leq &  \rho(1-\rho_f) - \frac{(1-\rho(1-\rho_f))\rho^{\D+2}(1-\rho_f)^{\D+2}}{ 1- \rho^{\D+2} (1- \rho_f)^{\D+2} }.
\end{IEEEeqnarray}
Combined with the trivial bound 
\begin{equation}
\S^{(\U)} \leq \frac{\rho\rho_f}{2},
\end{equation}
this proves  Theorem~\ref{theorem4}.
  
 \subsection{Achievability for Large  $\S^{(\U)}$ under Model 2:} \label{App: A-H}
 As in Subsection~\ref{App: A-E}, to ensure that each URLLC message can be transmitted at the same rate, we split the blocklength into two equally-long phases and time-share two schemes over the two phases. In Phase 1 we only send URLLC messages from odd Txs and in Phase 2 only  from even Txs. We thus reconsider the two URLLC Tx sets $\mathcal{T}_{\textnormal{URLLC},1}$ and $\mathcal{T}_{\textnormal{URLLC},2}$ defined in \eqref{eq:TURLLC1} and \eqref{eq:TURLLC2}.  Though we use the same URLLC Tx-sets as in Subsection~\ref{App: A-E}, now under Model 2 the network is split into smaller subnets, because not  only deactivated users decompose  the network but also URLLC users that are not contained in the URLLC Tx set $\mathcal{T}_{\textnormal{URLLC},i}$ of the respective Phase $i$. (Under Model 1 these Txs still participated in the communication because besides their URLLC message they also had an eMBB message to send.) We thus have different  subnets in the two phases, even though the user activity pattern is the same.
 
 In particular,  in Phase $i\in\{1,2\}$ and for $k\in \mathcal{K}_i$ a subnet starts at Tx $k$ and is of length $\ell$,
 \begin{enumerate}
 \item for $\ell$ even, if Tx $k-1$ is URLLC or deactivated;  Txs~$k, k+2, \ldots, k+\ell-2$ are active;  Txs $k+1,k+3,\ldots, k+\ell-1$ are eMBB and  active; and Tx $k+\ell$ is deactivated; 
 \item  for $\ell$ odd,  if  Tx $k-1$ is URLLC or deactivated;  Txs~$k, k+2, \ldots, k+\ell-1$ are  active;  Txs $k+1,k+3,\ldots, k+\ell-2$ are eMBB and  active; and Tx $k+\ell$ is URLLC or  deactivated. 
 \end{enumerate}
Similarly, for $k\notin \mathcal{K}_i$ a subnet starts at Tx $k$ and is of length $\ell$, 
 \begin{enumerate}
 \item[3)] for $\ell$ even, if Tx $k-1$ is deactivated; Txs~$k, k+2, \ldots, k+\ell-2$ are eMBB and active;  Txs $k+1,k+3,\ldots, k+\ell-1$ are active; and Tx $k+\ell$ is URLLC or  deactivated;
 \item[4)]  for $\ell$ odd, if Tx $k-1$ is deactivated;  Txs~$k, k+2, \ldots, k+\ell-1$ are eMBB and active;  Txs $k+1,k+3,\ldots, k+\ell-2$ are active; and Tx $k+\ell$ is  deactivated.
 \end{enumerate}
 The probability that in Phase $i$ a subnet starts at Tx $k$ and is of length $\ell$ is thus given by
    \begin{equation} \label{eq:plk}
 \P_{\ell, k,i}^{(2)}= \begin{cases} 
 \rho^{\ell}(1-\rho_f)^{ \left \lfloor \frac{\ell}{2} \right \rfloor} \cdot  a_{k,\ell} ,&  \textnormal{ if } k \in \mathcal{K}_{i}\\
\rho^{\ell}(1-\rho_f)^{ \left \lceil \frac{\ell}{2} \right \rceil}  \cdot b_{k,\ell},& \textnormal{ if } k \notin \mathcal{K}_{i}
 \end{cases}, \qquad \ell=1,2,\ldots,
  \end{equation} 
with
\begin{IEEEeqnarray}{rCl}
a_{k,\ell} \triangleq  \begin{cases}  (1-\rho(1-\rho_f)  )^2,& \ell < K-k+1, \; k \geq 2, \;  \ell \textnormal{ odd} \\
 (1-\rho(1-\rho_f) )(1-\rho) ,& \ell < K-k+1, \; k \geq 2, \;  \ell \textnormal{ even}  \\
  1-\rho(1-\rho_f)  ,&   ( \ell=K-k+1, \; k\geq  2 ) \\
 & \textnormal{or } 
(  \ell < K-k+1, \; k=1, \;  \ell \textnormal{ odd})\\
 1-\rho ,&  \ell <K, \; k=1, \; \ell \textnormal{ even}  \\
1 ,& k=1, \; \ell=K \end{cases}
\end{IEEEeqnarray}
and 
\begin{IEEEeqnarray}{rCl}
b_{k,\ell} \triangleq\begin{cases}  (1-\rho)^2,& \ell < K-k+1, \; k \geq 2, \;  \ell \textnormal{ odd} \\
 (1-\rho(1-\rho_f) )(1-\rho) ,& \ell < K-k+1, \; k \geq 2, \;  \ell \textnormal{ even}  \\
  1-\rho ,&   ( \ell=K-k+1, \; k\geq  2 ) \\
 & \textnormal{or } 
(  \ell < K-k+1, \; k=1, \;  \ell \textnormal{ odd})\\
 1-\rho(1-\rho_f) ,&  \ell <K, \; k=1, \; \ell \textnormal{ even}  \\
1 ,& k=1, \; \ell=K. \end{cases}
\end{IEEEeqnarray}

In subnets of the form 1)--2), the scheme in Subsection~\ref{sec:full_S} can be applied and  a  total sum-MG of $\M_{\textnormal{sum,full}}(\ell)$  achieved. The same holds in subnets of the form 3)--4) if  $\ell \leq\D$ or if  in a  subnet starting at Tx $k$ all Txs in the set $\mathcal L_k$ as defined in \eqref{eq:Lset} send eMBB messages. Otherwise, Tx $k+\D$ can be deactivated (notice that by our assumptions in 3) and 4), Tx $k+\D-2$ is not in $\mathcal{T}_{\textnormal{URLLC}}$) and the scheme in Subsection~\ref{sec:red_S} can be applied to the first $\D$ Tx/Rx pairs in the subnet and the scheme in Subsection~\ref{sec:full_S} to the remaining non-silenced Tx/Rx pairs in the subnet. In this case, a sum-MG of  $\M_{\textnormal{sum,red}}(\ell)$ is achieved. 
For the  expected  total per-user MG we thus obtain
 	\begin{IEEEeqnarray}{rCl}
\lefteqn{ 	\S^{(\e)} + \S^{(\U)}} \; \nonumber \\
 		&\mw{\geq}& \varlimsup_{K\to \infty} \frac{1}{2} \sum_{i =1}^2  \frac{1}{K}  \bigg[ \sum_{k \in \mathcal K_i}  \sum_{\ell = 1}^{K-k+1}    P_{\ell,k,i}^{(2)}\cdot  \M_{\textnormal{sum,full}}(\ell)   \nonumber \\
		&& \hspace{3cm}+ \sum_{k \notin \mathcal K_i}  \sum_{\ell = 1}^{K-k+1}    P_{\ell,k,i}^{(2)}\cdot  \left( \M_{\textnormal{sum,full}}(\ell) (1-\rho_f)^{2 \lceil\frac{\ell}{\D+2} \rceil} + \M_{\textnormal{sum,red}}(\ell) \cdot \left (1- (1-\rho_f)^{2 \lceil\frac{\ell}{\D+2} \rceil} \right)\right)  \bigg] \IEEEeqnarraynumspace \label{eq:first_ineq}\\
		&=& \varlimsup_{K\to \infty}  \frac{1}{K} \sum_{k=1}^K  \sum_{\ell = 1}^{K-k+1} \bar{\M}_{\textnormal{sum}}(\ell,k) \\
		& =&  \varlimsup_{K\to \infty}  \frac{1}{K} \sum_{\ell=1}^K  \sum_{k= 1}^{K-\ell+1}  \bar{\M}_{\textnormal{sum}}(\ell,k)\\
				& =&  \varlimsup_{K\to \infty}  \frac{1}{K} \left[ \sum_{\ell=1}^{K-2}  \sum_{k= 2}^{K-\ell} \bar{\M}_{\textnormal{sum}}(\ell,k) +  \sum_{\ell=1}^{K-1}   \left( \bar{\M}_{\textnormal{sum}}(\ell,1) + \bar{\M}_{\textnormal{sum}}(\ell,K-\ell+1) \right)+  \bar{\M}_{\textnormal{sum}}(1,K) \right]
		\end{IEEEeqnarray}
		where 
 	\begin{IEEEeqnarray}{rCl}
		 \bar{\M}_{\textnormal{sum}}(\ell,k)& \triangleq &\frac{1}{2} P_{\ell,k, 1+\mathbbm{1}\{k \textnormal{ even}\}}^{(2)}\cdot  \M_{\textnormal{sum,full}}(\ell)  \nonumber \\
		 && +\frac{1}{2} P_{\ell,k, 1+\mathbbm{1}\{k \textnormal{ odd}\} }^{(2)}  \left( \M_{\textnormal{sum,full}}(\ell)   (1-\rho_f)^{2 \lceil\frac{\ell}{\D+2} \rceil} + \M_{\textnormal{sum,red}}(\ell) \left (1- (1-\rho_f)^{2 \lceil\frac{\ell}{\D+2} \rceil} \right)\right) \IEEEeqnarraynumspace
		\end{IEEEeqnarray}
		In the following we analyze above asymptotic limit. 
		
		 If $\rho_f=1$, then $\P_{\ell, k,i}^{(2)}=0$ except for $\ell=1$ and $k\in\mathcal{K}_i$. In this case, from \eqref{eq:first_ineq} we have
 	\begin{IEEEeqnarray}{rCl}
 	\S^{(\e)} + \S^{(\U)}
 		&=& \varlimsup_{K\to \infty} \frac{1}{2} \sum_{i =1}^2  \frac{1}{K} \sum_{k \in \mathcal K_i}    P_{1,k,i}^{(2)}\cdot  \M_{\textnormal{sum,full}}(1)    \\
		 	&=& \varlimsup_{K\to \infty} \frac{1}{2}  \frac{1}{K} K \rho  \cdot 1 \\
			& =& \frac{\rho}{2}.
\end{IEEEeqnarray}	
Since in the described scheme, we generally have
\begin{equation}\label{eq:SFF}
\S^{(\U)}=\frac{\rho \rho_f}{2},
\end{equation}
trivially  we obtain that for $\rho_f=1$  $\S^{(\e)}=0$ and $\S^{(\U)}$ as in \eqref{eq:SFF} is achievable.

If  $\rho_f=0$,  we are back to the case with only eMBB messages studied in  Subsection~\ref{App: A-F}.

For the remainder of this section, we assume that both $\rho, \rho_f\in(0,1)$. We  notice that $P_{\ell,k,i}\leq \rho^\ell$ and $\M_{\textnormal{sum,full}}(\ell)\leq \ell$, which implies that  $\bar{\M}_{\textnormal{sum}}(\ell,k) \leq \rho^\ell \ell$. Therefore,   by \eqref{eq:sum_divide}
\begin{equation}
\varlimsup_{K\to \infty}  \frac{1}{K} \left[   \sum_{\ell=1}^{K-1}   \left( \bar{\M}_{\textnormal{sum}}(\ell,1) +  \bar{\M}_{\textnormal{sum}}(\ell,K-\ell+1) \right)+  \bar{\M}_{\textnormal{sum}}(1,K) \right] =0,
\end{equation}
and thus 
	\begin{IEEEeqnarray}{rCl}\label{eq:simpl10}
	\S^{(\e)} + \S^{(\U)}	&=&\varlimsup_{K\to \infty}  \frac{1}{K} \sum_{\ell=1}^{K-2}  \sum_{k= 2}^{K-\ell} \bar{\M}_{\textnormal{sum}}(\ell,k).
	\end{IEEEeqnarray}
	For $\ell \in \{1,\ldots, K-2\}$ and $k\in\{2,\ldots,K-\ell\}$ 
	\begin{IEEEeqnarray}{rCl}\label{eq:Mbar}
	\bar{\M}_{\textnormal{sum}}(\ell,k) &=&  \frac{1}{2}   \rho^{\ell}(1-\rho_f)^{ \left \lfloor \frac{\ell}{2} \right \rfloor}  (1-\rho(1-\rho_f) )(1-\rho+\mathbbm{1}\{ \ell \textnormal{ odd}\} \rho \rho_f )   \M_{\textnormal{sum,full}}(\ell)  \nonumber \\
	&& + \frac{1}{2}   \rho^{\ell}(1-\rho_f)^{ \left \lceil \frac{\ell}{2} \right \rceil}  (1-\rho +\mathbbm{1}\{ \ell \textnormal{ even}\} \rho \rho_f )(1-\rho)   \nonumber \\
	&& \qquad \cdot \left( \M_{\textnormal{sum,full}}(\ell) (1-\rho_f)^{2 \lceil\frac{\ell}{\D+2} \rceil}+ \M_{\textnormal{sum,red}}(\ell) \left (1- (1-\rho_f)^{2 \lceil\frac{\ell}{\D+2} \rceil} \right) \right)  \IEEEeqnarraynumspace
	\end{IEEEeqnarray}
	which  does not depend on the index $k$. We can thus simplify the expression in \eqref{eq:simpl10} to
	\begin{IEEEeqnarray}{rCl}
	\S^{(\e)} + \S^{(\U)}	&=&\varlimsup_{K\to \infty}  \sum_{\ell=1}^{K-2}  \frac{K-\ell-1}{K}\bar{\M}_{\textnormal{sum}}(\ell,2)\\
	&=& \varlimsup_{K\to \infty}  \sum_{\ell=1}^{K-2} \bar{\M}_{\textnormal{sum}}(\ell,2) \label{eq:lastSsum}
	\end{IEEEeqnarray}
	where the second equality holds because $ \bar{\M}_{\textnormal{sum}}(\ell,2)  \leq \rho^\ell \ell$ and thus the sum $\sum_{\ell=1}^{K-2}  \frac{\ell+1}{K}\bar{\M}_{\textnormal{sum}}(\ell,2)$ vanishes as $K\to \infty$, see  \eqref{eq:sum_divide}. 
	
	Substituting \eqref{eq:Mbar} into \eqref{eq:lastSsum}, we obtain
  \begin{IEEEeqnarray}{rCl}\label{eq:180n}
\S^{(\e)} +\S^{(\U)} & =& \sum_{\ell=1}^{\infty} \bar{\M}_{\textnormal{sum}}(\ell,2)   \\ 
&=& -  \frac{1}{2}  \sum_{\ell = 1}^{\infty}   \rho^{\ell} (1-\rho_f)^{\lfloor \frac{\ell}{2}\rfloor}  \left ( \ell - \left \lfloor\frac{\ell}{\D+2} \right \rfloor\right)(1-\rho(1-\rho_f)) \rho\rho_f \mathbbm{1}\{\ell \; \text{is even} \; \} \nonumber \\
&& +   \frac{1}{2}  \sum_{\ell = 1}^{\infty}   \rho^{\ell} (1- \rho_f)^{\lfloor\frac{\ell}{2} \rfloor}   \left ( \ell - \left \lfloor\frac{\ell}{\D+2} \right \rfloor\right)(1-\rho(1-\rho_f))^2\nonumber \\
&& +  \frac{1}{2}  \sum_{\ell = 1}^{\infty}   \rho^{\ell} (1-\rho_f)^{\lceil \frac{\ell}{2}\rceil}  (1-\rho) \rho \rho_f \left ( 1-(1-\rho_f)^{2 \lceil\frac{\ell}{\D+2} \rceil}\right) \left (\ell - \left \lfloor\frac{\ell +1}{\D+2} \right \rfloor \right) \mathbbm{1}\{\ell \; \text{is even} \; \}\nonumber \\
&& + \frac{1}{2} \sum_{\ell = 1}^{\infty}   \rho^{\ell} (1-\rho_f)^{\lceil \frac{\ell}{2}\rceil}  (1-\rho)^2\left ( 1- (1-\rho_f)^{2 \lceil\frac{\ell}{\D+2} \rceil}\right) \left (\ell - \left \lfloor\frac{\ell +1}{\D+2} \right \rfloor \right)  \nonumber \\
&& + \frac{1}{2}  \sum_{\ell = 1}^{\infty}   \rho^{\ell} (1-\rho_f)^{\lceil \frac{\ell}{2}\rceil}  (1-\rho)\rho\rho_f  (1-\rho_f)^{2 \lceil\frac{\ell}{\D+2} \rceil}\left (\ell - \left \lfloor\frac{\ell}{\D+2} \right \rfloor \right) \mathbbm{1}\{\ell \; \text{is even} \; \}\nonumber \\
&& +  \frac{1}{2} \sum_{\ell = 1}^{\infty}   \rho^{\ell} (1-\rho_f)^{\lceil \frac{\ell}{2}\rceil}  (1-\rho)^2 (1-\rho_f)^{2 \lceil\frac{\ell}{\D+2} \rceil} \left (\ell - \left \lfloor\frac{\ell}{\D+2} \right \rfloor \right) \label{eq:178}  \\ 
& \stackrel{(i)}{=} & - \frac{\rho^2 \rho_f^2}{2(1- \rho^2(1-\rho_f))}\left ( \frac{2\rho^2(1-\rho_f)}{1- \rho^2(1-\rho_f)}- \frac{(\rho^2(1-\rho_f))^{\frac{\D+2}{2}}}{1 - (\rho^2(1-\rho_f))^{\frac{\D+2}{2}}}\right) \notag \\
&& + \frac{(1- \rho(1-\rho_f))^2}{2(1- \rho^2(1-\rho_f))}\left ( \rho + \frac{2\rho (1-\rho_f)(1+\rho)}{1- \rho^2(1-\rho_f)} - \frac{(1+\rho)(\rho^2(1-\rho_f))^{\frac{\D+2}{2}}}{1 - (\rho^2(1-\rho_f))^{\frac{\D+2}{2}}}\right) \notag \\
&& + \frac{(1-\rho)^2}{2(1-\rho^2(1-\rho_f))} \left ( \rho (1- \rho_f) (2 \rho + 2\rho^2(1-\rho_f) +1) - \frac{\rho^{\D+1}(1- \rho_f)^{\frac{\D+2}{2}}(1+ \rho)}{1- (\rho^2(1-\rho_f))^{\frac{\D+2}{2}}}\right) \notag \\
&& + \frac{(1-\rho)^2 \rho^{\D+1} (1-\rho_f)^{\frac{\D+6}{2}}}{2(1- \rho^{\D+2} (1- \rho_f)^{\frac{\D+6}{2}})} \\
& = & \frac{\rho(1-\rho_f)}{(1-\rho^2(1-\rho_f))^2} \left ( (1+ \rho)(1- \rho(1-\rho_f))^2 - \rho^3 \rho_f^2 \right) \notag \\
&& + \frac{\rho \left ((1- \rho(1-\rho_f))^2 + (1- \rho)^2 (1- \rho_f) (2 \rho + 2\rho^2 (1- \rho_f) + 1) \right)}{2( 1- \rho^2(1- \rho_f))} \notag \\
&& - \frac{\rho^{\D+2} (1- \rho_f)^{\frac{\D+2}{2}}}{2 (1- \rho^2 (1-\rho_f)) (1- \rho^{\D+2} (1- \rho_f)^{\frac{\D+2}{2}})} \left ( (1+\rho) \left ( (1- \rho(1- \rho_f))^2 + \frac{(1-\rho)^2}{\rho}\right)\right) \notag \\
&& + \frac{(1-\rho)^2 \rho^{\D+1} (1-\rho_f)^{\frac{\D+6}{2}}}{2(1- \rho^{\D+2} (1- \rho_f)^{\frac{\D+6}{2}})},
\end{IEEEeqnarray}
where in step $(i)$ the first summation term of \eqref{eq:178} is calculated by \eqref{eq:90} for $c = \rho, d = 1- \rho_f, B = 2$ and $A = \D +2$ and by \eqref{eq:91} for $c = \rho, d =1- \rho_f $ and $B = 2$; the second summation term of \eqref{eq:178} is calculated by \eqref{eq:eq2}  for $c = \rho, d = 1- \rho_f, B = 2$ and $A = \D +2$ and by  \eqref{eq:eqsumnew1} for $c = \rho, d =1- \rho_f $ and $B = 2$; the third and the fifth summation terms of \eqref{eq:178} are first combined together using the fact that for even values of $\ell$ the terms $\lfloor \frac{\ell + 1}{\D+2}\rfloor - \lfloor \frac{\ell }{\D+2}\rfloor =0 $  and $\lceil \frac{\ell}{2}\rceil = \frac{\ell}{2}$  and the resulting term is calculated by \eqref{eq:sum2} for $c = \rho \sqrt{1- \rho_f}$ and by \eqref{eq:eq1} for $c =  \rho \sqrt{1- \rho_f}$ and $A = \D+2$ under the assumption that  $\ell$ is even; and the fourth and the sixth summation terms of \eqref{eq:178} are first combined together and the resulting term  is calculated by \eqref{eq:125} and \eqref{eq:138} for $c = \rho$, $d = 1- \rho_f$, $A = \D+2$ and $B = 2$ and  following the proof  of \eqref{eq:eq3new2}. Using \eqref{eq:SFF1}, we finally obtain
\begin{IEEEeqnarray}{rCl} \label{eq:198}
\S^{(\e)} & = & \frac{\rho(1-\rho_f)}{(1-\rho^2(1-\rho_f))^2} \left ( (1+ \rho)(1- \rho(1-\rho_f))^2 - \rho^3 \rho_f^2 \right) \notag \\
&& + \frac{\rho \left ((1- \rho(1-\rho_f))^2 + (1- \rho)^2 (1- \rho_f) (2 \rho + 2\rho^2 (1- \rho_f) + 1) \right)}{2( 1- \rho^2(1- \rho_f))} \notag \\
&& - \frac{\rho^{\D+2} (1- \rho_f)^{\frac{\D+2}{2}}}{2 (1- \rho^2 (1-\rho_f)) (1- \rho^{\D+2} (1- \rho_f)^{\frac{\D+2}{2}})} \left ( (1+\rho) \left ( (1- \rho(1- \rho_f))^2 + \frac{(1-\rho)^2}{\rho}\right)\right) \notag \\
&& + \frac{(1-\rho)^2 \rho^{\D+1} (1-\rho_f)^{\frac{\D+6}{2}}}{2(1- \rho^{\D+2} (1- \rho_f)^{\frac{\D+6}{2}})} - \frac{\rho\rho_f}{2}.
\end{IEEEeqnarray}

\section{Proofs of  Theorems~\ref{theorem5} and \ref{theorem6} } \label{App:B}

  
\subsection{Achievability Results for $\S^{(\U)} = 0$ Under Model $1$}\label{App: B-B}

When   $\S^{(\U)}=0$ and only eMBB messages are transmitted, only Rx-Cooperation is as powerful as Tx-Cooperation. In particular, in any subnet of size $\ell$ one can  achieve a total eMBB MG of $M_{\textnormal{sum},\text{full}}(\ell)$. Following the same steps as in Appendix~\ref{App: A-C}, one can thus conclude achievability of the per-user MG pair $\S^{(\U)}=0$ and $\S^{(\e)}$ as in \eqref{eq:ssmax0}:
\begin{equation}
\S^{(\e)}  = \rho - \frac{(1-\rho)\rho^{\D+2}}{1- \rho^{\D+2}}.
\end{equation}

\subsection{Achievability Results for Large $\S^{(\U)}$ under Model~$1$} \label{App: B-C}
Notice that, when only Rxs can cooperate, it is not possible to precancel the interference of eMBB transmissions on URLLC Txs and thus eMBB Tx interfering URLLC transmissions should be deactivated. To achieve $\S^{(\U)}=\frac{\rho\rho_f}{2}$,  we schedule each URLLC Tx in $\mathcal{K}_{\text{URLLC}}$ to send its URLLC message over half of the blocklength, during which it is not interfered by any other communication. We thus split the blocklength in two equally-long phases, where in Phase 1 we only send URLLC messages from odd Txs and in Phase 2 only URLLC messages from even Txs. 
We thus reconsider the two URLLC Tx sets $\mathcal{T}_{\textnormal{URLLC},1}$ and $\mathcal{T}_{\textnormal{URLLC},2}$  in \eqref{eq:TURLLC1} and \eqref{eq:TURLLC2} and in the scheme of Phase $i$ we deactivate the users adjacent to $\mathcal{T}_{\textnormal{URLLC},i}$. This way we split the network into small \emph{eMBB-subnets} over which we  only send eMBB messages, and thus can achieve an eMBB total MG of $\M_{\textnormal{sum,full}}$ as described in the previous Subsection~\ref{App: B-B}.  
 
 So in Phase $i\in\{1,2\}$ and for $k\in \mathcal{K}_i$, an eMBB-subnet of length $\ell$ starts at Tx $k$:
 \begin{enumerate}
 \item for $\ell$ even, if Tx $k-1$ is inactive or its left neighbor Tx~$k-2$ is URLLC ;  Txs~$k, k+2, \ldots, k+\ell-2$ are eMBB and  active; Txs~$k+1, k+3, \ldots, k+\ell-1$ are  active; and Tx $k+\ell$ is inactive; 
 \item  for $\ell$ odd,  if  Tx $k-1$ is  inactive or  Tx~$k-2$ is URLLC;  Txs~$k, k+2, \ldots, k+\ell-2$ are eMBB and  active; Txs~$k+1, k+3, \ldots, k+\ell-1$ are  active; and Tx $k+\ell$ is inactive or its right neighbor Tx~$k+\ell+1$  is URLLC . 
 \end{enumerate}
Similarly, for $k\notin \mathcal{K}_i$ an eMBB-subnet of length $\ell$  starts at Tx $k$:
 \begin{enumerate}
 \item[3)] for $\ell$ even, if Tx $k-1$ is inactive;  Txs~$k, k+2, \ldots, k+\ell-2$ are   active; Txs~$k+1, k+3, \ldots, k+\ell-1$ are eMBB and  active; and Tx $k+\ell$ is inactive or its right neighbor Tx~$k+\ell+1$ is URLLC;
 \item[4)]  for $\ell$ odd, if Tx $k-1$ is inactive;  Txs~$k, k+2, \ldots, k+\ell-2$ are   active; Txs~$k+1, k+3, \ldots, k+\ell-1$ are eMBB and  active; and Tx $k+\ell$ is  inactive.
 \end{enumerate}
 The probability that in Phase $i$ an eMBB-subnet starts at Tx $k$ and is of length $\ell \ge 2$ is thus given by 
 \begin{equation} \label{eq:plk2}
\tilde \P_{\ell, k,i}= \begin{cases} 
 \rho^{\ell}(1-\rho_f)^{ \lceil \frac{\ell}{2} \rceil} \cdot  \tilde a_{k,\ell} ,&  \textnormal{ if } k \in \mathcal{K}_{i}\\
\rho^{\ell}(1-\rho_f)^{ \lfloor \frac{\ell}{2} \rfloor}  \cdot \tilde b_{k,\ell},& \textnormal{ if } k \notin \mathcal{K}_{i}
 \end{cases}, \qquad \ell=1,2,\ldots,
  \end{equation} 
with
\begin{IEEEeqnarray}{rCl} \label{eq:akl}
\tilde a_{k,\ell} \triangleq  \begin{cases}  (1-\rho(1-\rho_f)  )^2,& \ell < K-k+1, \; k \geq 2, \;  \ell \textnormal{ odd} \\
 (1-\rho(1-\rho_f) )(1-\rho) ,& \ell < K-k+1, \; k \geq 2, \;  \ell \textnormal{ even}  \\
  1-\rho(1-\rho_f)  ,&   ( \ell=K-k+1, \; k\geq  2 ) \\
 & \textnormal{or } 
(  \ell < K-k+1, \; k=1, \;  \ell \textnormal{ odd})\\
 1-\rho ,&  \ell <K, \; k=1, \; \ell \textnormal{ even}  \\
1 ,& k=1, \; \ell=K \end{cases}
\end{IEEEeqnarray}
and 
\begin{IEEEeqnarray}{rCl}\label{eq:bkl}
\tilde b_{k,\ell} \triangleq\begin{cases}  (1-\rho)^2,& \ell < K-k+1, \; k \geq 2, \;  \ell \textnormal{ odd} \\
 (1-\rho(1-\rho_f) )(1-\rho) ,& \ell < K-k+1, \; k \geq 2, \;  \ell \textnormal{ even}  \\
  1-\rho ,&   ( \ell=K-k+1, \; k\geq  2 ) \\
 & \textnormal{or } 
(  \ell < K-k+1, \; k=1, \;  \ell \textnormal{ odd})\\
 1-\rho(1-\rho_f) ,&  \ell <K, \; k=1, \; \ell \textnormal{ even}  \\
1 ,& k=1, \; \ell=K. \end{cases}
\end{IEEEeqnarray}

Under this scheme, $\S^{(\U)} = \frac{\rho \rho_f}{2}$ and 
 	\begin{IEEEeqnarray}{rCl} \label{eq:first_ineq2}
 		\S^{(\e)}&=& \varlimsup_{K\to \infty} \frac{1}{K} \sum_{k=1}^{K}  \sum_{\ell = 1}^{K-k+1}  \ \tilde \P_{\ell,k,i}\cdot  \M_{\textnormal{sum,full}}(\ell)   \\
&=& \varlimsup_{K\to \infty} \frac{1}{K} \sum_{\ell=1}^{K}  \sum_{k = 1}^{K-\ell+1}  \ \tilde \P_{\ell,k,i}\cdot  \M_{\textnormal{sum,full}}(\ell)   \\
		 & = &  \varlimsup_{K\to \infty}  \frac{1}{K} \cdot \frac{1}{2} \sum_{i = 1}^2 \Bigg[   \sum_{\ell=1}^{K-2}  \sum_{k= 2}^{K-\ell}  \tilde \P_{\ell,k, i}  \M_{\textnormal{sum,full}}(\ell)    \notag \\
&&\hspace{2.5cm} +  \sum_{\ell=2}^{K-1}  \M_{\textnormal{sum,full}}(\ell) \left(  \tilde \P_{\ell,1, i} + \tilde \P_{\ell,K-\ell+1, i}  \right)+  \tilde \P_{1,K, i} \cdot \M_{\textnormal{sum,full}}(K)\Bigg].\label{eq:first_ineq2d}
 	\end{IEEEeqnarray}
	In the following we analyze above asymptotic limit.
	
	If  $\rho_f=0$, trivially $\S^{(\U)}=0$ and we are back to Subsection~\ref{App: B-D}.
	 If $\rho_f=1$, then $\tilde \P_{\ell, k,i}=0$ except for $\ell=1$ when $k \notin \mathcal K_i$. In this case, from \eqref{eq:first_ineq2d} we have
 	\begin{IEEEeqnarray}{rCl}
 	\S^{(\e)} 
 		&=& \varlimsup_{K\to \infty}  \frac{1}{K}     \frac{1}{2} \sum_{i = 1}^2  \sum_{k \notin \mathcal K_i   }\tilde \P_{1,k,i}\cdot  \M_{\textnormal{sum,full}}(1)    \\
		 	&=& \varlimsup_{K\to \infty} \frac{1}{K}\frac{K}{2} \rho(1-\rho)^2 \cdot 1 \\
			& =& \frac{\rho(1-\rho)^2}{2},
\end{IEEEeqnarray}	
and
\begin{equation}\label{eq:SFF2}
\S^{(\U)}=\frac{\rho }{2}.
\end{equation}

For the remainder of this section, we assume that $\rho\in(0,1)$ and  $(1-\rho_f)\in(0,1)$. We  notice that $\tilde P_{\ell,k,i}\leq \rho^\ell$ and $\M_{\textnormal{sum,full}}(\ell)\leq \ell$, which implies that  $\bar{\M}_{\textnormal{sum}}(\ell,k) \leq \rho^\ell \ell$. Therefore,   by \eqref{eq:sum_divide}
\begin{equation}
\varlimsup_{K\to \infty}  \frac{1}{K}\cdot \frac{1}{2} \sum_{i = 1}^2 \left[    \sum_{\ell=2}^{K-1}  \M_{\textnormal{sum,full}}(\ell) \left(  \tilde \P_{\ell,1, i} + \tilde \P_{\ell,K-\ell+1, i}  \right)+  \tilde \P_{1,K, i} \cdot \M_{\textnormal{sum,full}}(K)\right] =0,
\end{equation}
and thus 
	\begin{IEEEeqnarray}{rCl}\label{eq:simpl1}
	\S^{(\e)}
	&=&\varlimsup_{K\to \infty}  \frac{1}{K} \cdot \frac{1}{2} \sum_{i = 1}^2 \left [   \sum_{\ell=1}^{K-2}  \sum_{k= 2}^{K-\ell}  \P_{\ell,k, i}  \M_{\textnormal{sum,full}}(\ell) \right].
	\end{IEEEeqnarray}
Following the same argument as in \eqref{eq:Mbar}--\eqref{eq:lastSsum}, we obtain
  \begin{IEEEeqnarray}{rCl}\label{eq:180n2new}
\S^{(\e)}  
&=& -  \frac{1}{2}  \sum_{\ell = 1}^{\infty}   \rho^{\ell} (1-\rho_f)^{ \lceil \frac{\ell}{2} \rceil} \left ( \ell - \left \lfloor\frac{\ell}{\D+2} \right \rfloor\right)(1-\rho(1-\rho_f)) \rho\rho_f \mathbbm{1}\{\ell \; \text{is even} \; \} \nonumber \\
&& +   \frac{1}{2}  \sum_{\ell = 1}^{\infty}   \rho^{\ell} (1-\rho_f)^{ \lceil \frac{\ell}{2} \rceil}  \left ( \ell - \left \lfloor\frac{\ell}{\D+2} \right \rfloor\right)(1-\rho(1-\rho_f))^2\nonumber \\
&& + \frac{1}{2}  \sum_{\ell = 1}^{\infty}   \rho^{\ell} (1-\rho_f)^{ \lfloor \frac{\ell}{2} \rfloor}   (1-\rho)\rho \rho_f\left (\ell - \left \lfloor\frac{\ell}{\D+2} \right \rfloor \right) \mathbbm{1}\{\ell \; \text{is even} \; \}\nonumber \\
&& +  \frac{1}{2} \sum_{\ell = 1}^{\infty}   \rho^{\ell} (1-\rho_f)^{ \lfloor \frac{\ell}{2} \rfloor}   (1-\rho)^2 \left (\ell - \left \lfloor\frac{\ell}{\D+2} \right \rfloor \right) \label{eq:210}  \\ 
&=& -  \frac{\rho^2\rho_f^2}{2}  \sum_{\ell = 1}^{\infty}   \rho^{\ell} (1-\rho_f)^{ \lfloor \frac{\ell}{2} \rfloor} \left ( \ell - \left \lfloor\frac{\ell}{\D+2} \right \rfloor\right) \mathbbm{1}\{\ell \; \text{is even} \; \} \nonumber \\
&& +   \frac{1}{2}  \sum_{\ell = 1}^{\infty}   \rho^{\ell} (1-\rho_f)^{ \lceil \frac{\ell}{2} \rceil}  \left ( \ell - \left \lfloor\frac{\ell}{\D+2} \right \rfloor\right)(1-\rho(1-\rho_f))^2\nonumber \\
&& +  \frac{1}{2} \sum_{\ell = 1}^{\infty}   \rho^{\ell} (1-\rho_f)^{ \lfloor \frac{\ell}{2} \rfloor}   (1-\rho)^2 \left (\ell - \left \lfloor\frac{\ell}{\D+2} \right \rfloor \right) \label{eq:211}  \\ 
&\stackrel{(i)}{=}&
 -  \frac{\rho^4 \rho_f^2(1-\rho_f)}{2(1-\rho^2(1-\rho_f)))} \left ( \frac{2}{(1-\rho^2(1-\rho_f))} - \frac{\rho^{\D}(1-\rho_f)^{\frac{\D}{2}}}{1-\rho^{\D+2}(1-\rho_f)^{\frac{\D+2}{2}}} \right)\nonumber \\
&& +   \frac{\rho (1-\rho_f)(1- \rho(1-\rho_f))^2}{2(1- \rho^2(1-\rho_f))} \left ( 2 \rho + 2 \rho^2(1- \rho_f) + 1 - \frac{\rho^{\D+1}(1-\rho_f)^{\frac{\D}{2}}(1+\rho(1-\rho_f))}{(1- \rho^{\D+2}(1-\rho_f)^{\frac{\D+2}{2}})}\right) \notag \\
&& + \frac{\rho (1-\rho)^2}{2(1-\rho^2(1-\rho_f))} \left ( \frac{2\rho(1-\rho_f)}{(1-\rho^2(1-\rho_f))} + 1- \frac{\rho^{\D+1}(1-\rho_f)^{\frac{\D+2}{2}}(1+\rho)}{(1- \rho^{\D+2}(1-\rho_f)^{\frac{\D+2}{2}})}\right).
\end{IEEEeqnarray}

\subsection{Proof of the Converse Result under Model~$1$} \label{App: B-conv}
 Fix $K$ and realizations of the sets $\Ta$, $\mathcal{K}_{\textnormal{eMBB}}$, and $\mathcal{K}_{\textnormal{URLLC}}$.  Following the steps in \cite[Section V]{HomaISIT2018}, we can prove that for each $k \in \Ta$,
 \begin{IEEEeqnarray}{rCl} \label{eq:ap1}
\lefteqn{R_k^{(\U)} + R_k^{(\e)} + R_{k+1}^{(\U)} }\notag \\
&&\hspace{1cm}\le \frac{1}{2}\log (1+ (|h_{k,k}|^2+|h_{k,k+1}|^2) \P) + \frac{1}{2} \log (|h_{k,k}|^2 + |h_{k,k+1}|^2)  \notag \\
 &&\hspace{1.5cm}+ \max \{- \log|h_{k,k+1}|, 0\} + \frac{\epsilon_n}{n},
 \end{IEEEeqnarray} 
 where $R_k^{(\U)} $ and $R_{k+1}^{(\U)}$ denote  the rates of the URLLC message at Rxs~$k$ and $k+1$, respectively. Recall that in our setup a URLLC rate $R_k^{(\U)}$ is   equal to  $0$ if  $k \in\mathcal{K}_{\textnormal{URLLC}}$   (i.e., with probability $1-\rho\rho_f$) and it is equal to the global URLLC rate $R^{(\U)}$ if $k+1\notin\mathcal{K}_{\textnormal{URLLC}}$ (i.e., with probability $\rho\rho_f$). Similarly for $R_{k+1}^{(\U)}$. 
  
Abbreviating  the right hand-side  of \eqref{eq:ap1} by $\Delta$, and  summing up this bound for all values of $k \in \mathcal{K}_{\textnormal{active}}$, we obtain
 \begin{IEEEeqnarray}{rCl}
 \sum_{k \in \Ta} \left (R_k^{(\U)}  + R_k^{(\e)} + R_{k+1}^{(\U)} \right ) \le |\Ta| \cdot \Delta.
 \end{IEEEeqnarray}
 Taking expectation over the random activity sets  $\Ta$, $\mathcal{K}_{\textnormal{eMBB}}$, and $\mathcal{K}_{\textnormal{URLLC}}$ of  \eqref{eq:ap1} and dividing by $K$, we further have
 \begin{equation} \label{eq:ap3}
 \mathbb E [\bar R^{(\e)} ] + R^{(\U)}\left(\rho \rho_f + \rho^2 \rho_f  \right) \le \rho \cdot \Delta,
 \end{equation}
 because the expected number of indices $k \in \mathcal{K}_{\textnormal{active}}$ for which $R_k^{(\U)} = R^{(\U)}$ equals $\rho\cdot \rho_f$ and the expected numbers of indices $k \in \mathcal{K}_{\textnormal{active}}$ for which $R_{k+1}^{(\U)} = R^{(\U)}$ equals $\rho^2\cdot \rho_f$. (For this latter formula, observe that  $R_{k+1}^{(\U)}=R^{(\U)}$ with probability $\rho \rho_f$, however it is counted only if $k\in  \mathcal{K}_{\textnormal{active}}$, which happens with probability $\rho$.) 
 
 Dividing by $\frac{1}{2} \log \P$  and letting $\P \to \infty$ proves Theorem~\ref{theorem6-Rx}. 

\subsection{Achievability Results for $\S^{(\U)} = 0$ under Model $2$} \label{App: B-D}
As mentioned, when we only wish to send eMBB messages, only Rx-cooperation suffices and we can achieve the same per-user MG pairs as with Tx and Rx cooperation.  Thus $\S^{(\e)} =   \rho(1-\rho_f) - \frac{(1-\rho(1-\rho_f))\rho^{\D+2}(1-\rho_f)^{\D+2}}{ 1- \rho^{\D+2} (1- \rho_f)^{\D+2} }$ as in \eqref{eq:ssmax2} is achievable.

\subsection{Achievability Results for Large $\S^{(\U)}$ Under Model~$2$} \label{App: B-E}
As before in Section~\ref{App: B-C},  we split the blocklength into two equally-long phases, where in Phase 1 we only send URLLC messages from odd Txs in $\mathcal{T}_{\textnormal{URLLC},1}$  (see \eqref{eq:TURLLC1}) and in Phase 2 only  from even Txs in  $\mathcal{T}_{\textnormal{URLLC},2}$ (\eqref{eq:TURLLC2}). In Phase $i$ we further  deactivate the users adjacent to $\mathcal{T}_{\textnormal{URLLC},i}$ and send only eMBB messages over the remaining   eMBB-subnets.

 So in Phase $i\in\{1,2\}$ and for $k\in \mathcal{K}_i$ an eMBB-subnet starts at Tx $k$ and is of length $\ell$,
 \begin{enumerate}
 \item for $\ell$ even, if Tx $k-1$ is inactive or its left neighbor Tx~$k-2$ is URLLC ;  Txs~$k, k+1, \ldots, k+\ell-1$ are eMBB and  active;
 \item  for $\ell$ odd,  if  Tx $k-1$ is  inactive or  Tx~$k-2$ is URLLC;  Txs~$k, k+1, \ldots, k+\ell-1$ are eMBB and  active;  and Tx $k+\ell$ is inactive or its right neighbor Tx~$k+\ell+1$  is URLLC . 
 \end{enumerate}
Similarly, for $k\notin \mathcal{K}_i$ an eMBB-subnet starts at Tx $k$ and is of length $\ell$, 
 \begin{enumerate}
 \item[3)] for $\ell$ even, if Tx $k-1$ is inactive;  Txs~$k, k+1, \ldots, k+\ell-1$ are  eMBB and  active; and Tx $k+\ell$ is inactive or its right neighbor Tx~$k+\ell+1$ is URLLC;
 \item[4)]  for $\ell$ odd, if Tx $k-1$ is inactive;  Txs~$k, k+1, \ldots, k+\ell-1$ are  eMBB and  active; and Tx $k+\ell$ is  inactive.
 \end{enumerate}
 The probability that in Phase $i$ an eMBB-subnet starts at Tx $k$ and is of length $\ell \ge 2$ is thus given by 
 \begin{equation} \label{eq:plk2}
\tilde \P_{\ell, k,i}^{(2)}= \begin{cases} 
 \rho^{\ell}(1-\rho_f)^{ \ell} \cdot  \tilde a_{k,\ell} ,&  \textnormal{ if } k \in \mathcal{K}_{i}\\
\rho^{\ell}(1-\rho_f)^{ \ell}  \cdot \tilde b_{k,\ell},& \textnormal{ if } k \notin \mathcal{K}_{i}
 \end{cases}, \qquad \ell=1,2,\ldots,
  \end{equation} 
with $\tilde a_{k,\ell}$ and $\tilde b_{k,\ell}$ are defined \eqref{eq:akl} and \eqref{eq:bkl}. 

Under this scheme, $\S^{(\U)} = \frac{\rho \rho_f}{2}$ and 
 	\begin{IEEEeqnarray}{rCl} \label{eq:first_ineq2}
 		\S^{(\e)}&=& \varlimsup_{K\to \infty} \frac{1}{K} \sum_{k=1}^{K}  \sum_{\ell = 1}^{K-k+1}  \ \tilde \P_{\ell,k,i}^{(2)}\cdot  \M_{\textnormal{sum,full}}(\ell)   \\
&=& \varlimsup_{K\to \infty} \frac{1}{K} \sum_{\ell=1}^{K}  \sum_{k = 1}^{K-\ell+1}  \ \tilde \P_{\ell,k,i}\cdot  \M_{\textnormal{sum,full}}(\ell)   \\
		 & = &  \varlimsup_{K\to \infty}  \frac{1}{K} \cdot \frac{1}{2} \sum_{i = 1}^2 \Bigg[   \sum_{\ell=1}^{K-2}  \sum_{k= 2}^{K-\ell} \tilde \P_{\ell,k, i}^{(2)}  \M_{\textnormal{sum,full}}(\ell)  \nonumber \\
		 && \hspace{3.3cm}+  \sum_{\ell=2}^{K-1}  \M_{\textnormal{sum,full}}(\ell) \left(  \tilde \P_{\ell,1, i}^{(2)}  + \tilde \P_{\ell,K-\ell+1, i}^{(2)}   \right)+  \tilde \P_{1,K, i}^{(2)}   \M_{\textnormal{sum,full}}(K)\Bigg].\label{eq:first_ineq2} \IEEEeqnarraynumspace
 	\end{IEEEeqnarray}
	In the following we analyze the above asymptotic limit.

%

If  $\rho_f=0$, trivially $\S^{(\U)}=0$ and we are back to Subsection~\ref{App: B-D}. If $\rho_f=1$ then $\S^{(\e)}=0$.

For the remainder of this section, we assume that $\rho,\rho_f\in(0,1)$. We  notice that $\tilde \P_{\ell,k,i}^{(2)}\leq \rho^\ell$ and $\M_{\textnormal{sum,full}}(\ell)\leq \ell$, which implies that  $\bar{\M}_{\textnormal{sum}}(\ell,k) \leq \rho^\ell \ell$. Therefore,   by \eqref{eq:sum_divide}
\begin{equation}
\varlimsup_{K\to \infty}  \frac{1}{K}\cdot \frac{1}{2} \sum_{i = 1}^2 \left[    \sum_{\ell=1}^{K-1}  \M_{\textnormal{sum,full}}(\ell) \left(  \tilde \P_{\ell,1, i}^{(2)} + \tilde \P_{\ell,K-\ell+1, i}^{(2)}  \right)+  \tilde \P_{1,K, i}^{(2)}  \cdot \M_{\textnormal{sum,full}}(K)\right] =0,
\end{equation}
and thus 
	\begin{IEEEeqnarray}{rCl}\label{eq:simpl1}
	\S^{(\e)}
	&=&\varlimsup_{K\to \infty}  \frac{1}{K} \cdot \frac{1}{2} \sum_{i = 1}^2   \sum_{\ell=1}^{K-2}  \sum_{k= 2}^{K-\ell}  \tilde \P_{\ell,k, i}^{(2)}   \cdot \M_{\textnormal{sum,full}}(\ell) .
	\end{IEEEeqnarray}
Following the same arguments as in \eqref{eq:Mbar}--\eqref{eq:lastSsum}, we obtain
  \begin{IEEEeqnarray}{rCl}\label{eq:180n2new}
\S^{(\e)}  
&=& -  \frac{1}{2}  \sum_{\ell = 1}^{\infty}   \rho^{\ell} (1-\rho_f)^{\ell} \left ( \ell - \left \lfloor\frac{\ell}{\D+2} \right \rfloor\right)(1-\rho(1-\rho_f)) \rho\rho_f \mathbbm{1}\{\ell \; \text{is even} \; \} \nonumber \\
&& +   \frac{1}{2}  \sum_{\ell = 1}^{\infty}   \rho^{\ell} (1-\rho_f)^{ \ell}  \left ( \ell - \left \lfloor\frac{\ell}{\D+2} \right \rfloor\right)(1-\rho(1-\rho_f))^2\nonumber \\
&& + \frac{1}{2}  \sum_{\ell = 1}^{\infty}   \rho^{\ell} (1-\rho_f)^{\ell}   (1-\rho)\rho \rho_f\left (\ell - \left \lfloor\frac{\ell}{\D+2} \right \rfloor \right) \mathbbm{1}\{\ell \; \text{is even} \; \}\nonumber \\
&& +  \frac{1}{2} \sum_{\ell = 1}^{\infty}   \rho^{\ell} (1-\rho_f)^{ \ell}   (1-\rho)^2 \left (\ell - \left \lfloor\frac{\ell}{\D+2} \right \rfloor \right) \label{eq:210}  \\ 
&=& -  \frac{\rho^2\rho_f^2}{2}  \sum_{\ell = 1}^{\infty}   \rho^{\ell} (1-\rho_f)^{ \ell} \left ( \ell - \left \lfloor\frac{\ell}{\D+2} \right \rfloor\right) \mathbbm{1}\{\ell \; \text{is even} \; \} \nonumber \\
&& +   \frac{(1-\rho(1-\rho_f))^2 + (1-\rho)^2}{2}  \sum_{\ell = 1}^{\infty}   \rho^{\ell} (1-\rho_f)^{ \ell}  \left ( \ell - \left \lfloor\frac{\ell}{\D+2} \right \rfloor\right) \\
&=&
 - \frac{\rho^4\rho_f^2(1-\rho_f)^2}{2(1-\rho^2(1-\rho_f)^2)} \left ( \frac{2}{1-\rho^2(1-\rho_f)^2} - \frac{\rho^{\D}(1-\rho_f)^{\D}}{1-\rho^{\D+2}(1-\rho_f)^{\D+2}}\right)\nonumber \\
&& +  \frac{\rho(1-\rho_f) \left ( (1-\rho(1-\rho_f))^2 + (1-\rho)^2 \right)}{2(1- \rho(1-\rho_f))}\left (\frac{1}{1-\rho(1-\rho_f)} - \frac{\rho^{\D+1}(1-\rho_f)^{\D+1}}{1-\rho^{\D+2}(1-\rho_f)^{\D+2}} \right).\IEEEeqnarraynumspace
\end{IEEEeqnarray}


   \subsection{Proof of the Converse Results under Model~$2$} \label{App: B-conv2}
 Fix $K$ and realizations of the sets $\mathcal{K}_{\textnormal{eMBB}}$ and $\mathcal{K}_{\textnormal{URLLC}}$. Following the steps in \cite[Section V]{HomaISIT2018}, we can prove that for each $k \in \mathcal{K}_{\textnormal{eMBB}}$,
 \begin{IEEEeqnarray}{rCl} \label{eq:ap22}
\lefteqn{R_k^{(\e)} + R_{k+1}^{(\U)} \le \frac{1}{2}\log (1+ (|h_{k,k}|^2+|h_{k,k+1}|^2) \P) + \frac{\epsilon_n}{n}  }\notag \\
 &&\hspace{2.2cm}+ \frac{1}{2} \log (|h_{k,k}|^2 + |h_{k,k+1}|^2)+ \max \{- \log|h_{k,k+1}|, 0\}, \IEEEeqnarraynumspace
 \end{IEEEeqnarray}
 where $R_{k+1}^{(\U)}$ is the rate of the URLLC messages at Rx~$k+1$, which  is either $0$  (when $k \notin \mathcal{K}_{\textnormal{URLLC}}$, i.e., with probability $1-\rho \rho_f$) or equal to the global URLLC rate $R^{(\U)}$ (when $k \in \mathcal{K}_{\textnormal{URLLC}}$, i.e., with probability $\rho \rho_f$).

Abbreviating  the right-hand side of \eqref{eq:ap22} by $\tilde \Delta$, and  summing up this bound for all values of $k \in \mathcal{K}_{\textnormal{eMBB}}$, we obtain
 \begin{IEEEeqnarray}{rCl}
 \sum_{k \in \mathcal{K}_{\textnormal{eMBB}}} \left (R_k^{(\e)} + R_{k+1}^{(\U)} \right ) \le |\mathcal{K}_{\textnormal{eMBB}}| \cdot \tilde \Delta,
 \end{IEEEeqnarray}
 or equivalently
  \begin{IEEEeqnarray}{rCl}
 K \bar R^{(\e)} + \sum_{\substack{k \in \mathcal{K}_{\textnormal{eMBB}} \\  k+1 \in \Tf}} R^{(\U)}   \le |\mathcal{K}_{\textnormal{eMBB}}| \cdot  \tilde \Delta. \label{eq:ap2-2}
 \end{IEEEeqnarray}
 Dividing both  sides of  \eqref{eq:ap2-2} by $K$ and taking expectation  over the random user activity sets $\mathcal{K}_{\textnormal{eMBB}}$ and $\mathcal{K}_{\textnormal{URLLC}}$, we obtain:
 \begin{equation} \label{eq:ap3}
 \mathbb E [\bar R^{(\e)} ] + R^{(\U)}\left (\rho^2 \rho_f (1-\rho_f) \right )\le \rho(1-\rho_f) \cdot  \tilde \Delta.
 \end{equation}
 Dividing by $\frac{1}{2} \log \P$  and letting $\P \to \infty$ and $K\to \infty$, then proves Theorem~\ref{theorem8-Rx}.

\section{Hexagonal Network with Rx-Cooperation Only: Proof of Theorems \ref{theorem7} and \ref{theorem8}} \label{App:C}

\subsection{Achievability Result for $\S^{(\U)} = 0$ under Model $1$}\label{App: C-A}
Substituting  $\mathcal{T}_{\textnormal{eMBB}}'=\mathcal{K}$ and $\alpha=0$  into \eqref{eq:Mod1-onlyRx} establishes achievability of the eMBB per-user MG  $\S^{(\e)}=\rho$. 

\subsection{Achievability Result for Large $\S^{(\U)}$ under Model $1$}\label{App: C-B}
Reconsider the partition  $\mathcal K_1, \mathcal K_2, \mathcal K_{3} \subseteq \mathcal{K}$  given in Fig.~\ref{fig6.b}, for which each of the three sets  $\mathcal{K}_i$ only contains non-interfering cells: 
\begin{equation}
k' \notin \mathcal I_{\Rx, k''} \quad \text{and}\quad  k'' \notin \mathcal I_{\Rx, k'}, \quad  \forall k', k'' \in \ai.
\end{equation}
Notice that the three sets $\mathcal K_1, \mathcal K_2, \mathcal K_{3}$ are of equal size. 

We time-share three schemes, where in each scheme $i\in\{1,2,3\}$ only Txs in subset $\mathcal{T}_i$ are scheduled to send URLLC messages if they have any. 
So in scheme $i$, we set \begin{subequations}\label{eq:sethexa2}
\begin{IEEEeqnarray}{rCl}
 \mathcal{T}_{\textnormal{URLLC},i}& := &\mathcal K_i \cap \mathcal{K}_{\textnormal{URLLC}}.
\end{IEEEeqnarray}
Then we silence all Tx/Rx pairs that interfere at Rxs in $ \mathcal{T}_{\textnormal{URLLC},i}$: 
\begin{equation}
\mathcal{T}_{\textnormal{silent},i} := \{\tilde k \in \mathcal K \colon   \exists k' \in \mathcal{T}_{\textnormal{URLLC},i} \textnormal{ so that } k\in\mathcal{I}_{k'}\},
\end{equation} 
and schedule the remaining Txs to send eMBB messages if they are active 
\begin{equation}
\mathcal{T}_{\textnormal{eMBB},i} := \mathcal{K}_{\textnormal{active}}\backslash ( \mathcal{T}_{\textnormal{URLLC},i} \cup \mathcal{T}_{\textnormal{silent},i}).
\end{equation}
\end{subequations}
 
 Since we can use an unlimited number of cooperation rounds, each scheduled eMBB Tx  can send at full MG. The probability of a user $k$ to be scheduled as an eMBB user in scheme $i$ is: 
 \begin{itemize}
 \item If $k\in \mathcal{K}_i$ it is $\rho(1-\rho_f)$. (This is the probability that Tx $k$ has an eMBB but no URLLC message to send.)
 \item If $k\notin \mathcal{K}_i$ it is $\rho(1-\rho\rho_f)^3$.
 (This is the probability that none of the  3 adjacent Txs in $\mathcal{K}_i$ has a URLLC message to send and Tx $k$ is active.) 
 \end{itemize}

By above considerations, the scheme thus achieves the per-user MG pair 
  \begin{IEEEeqnarray}{rCl}
 \S^{(\U)}   &=&  \frac{\rho \rho_f}{3}\\
 \S^{(\e)}  &= &  \frac{2 \rho (1- \rho\rho_f)^3}{3} + \frac{\rho (1- \rho_f)}{3}.
\end{IEEEeqnarray}

Time-sharing the described scheme with the scheme in Appendix~\ref{App: C-A}  proves Theorem~\ref{theorem7}. 

\subsection{Achievability Results for $\S^{(\U)} = 0$ Under Model $2$} \label{App: C-D}
Substituting $\mathcal{T}_{\textnormal{eMBB}}'=\mathcal{K}$ and $\alpha=0$  into \eqref{eq:SS_Model2-OnlyRx} establishes achievability of the eMBB per-user MG  $\S^{(\e)}=\rho$.

\subsection{Achievability Results for Large $\S^{(\U)}$  Under Model $2$} \label{App: C-E}

We again time-share three schemes, where 
 in scheme $i$, we again choose 
  \begin{subequations}\label{eq:sethexa2}
\begin{IEEEeqnarray}{rCl}
 \mathcal{T}_{\textnormal{URLLC},i}& := &\mathcal K_i \cap \mathcal{K}_{\textnormal{URLLC}}
\end{IEEEeqnarray}
and \begin{equation}
\mathcal{T}_{\textnormal{silent},i} := \{\tilde k \in \mathcal K \colon   \exists k' \in \mathcal{T}_{\textnormal{URLLC},i} \textnormal{ so that } k\in\mathcal{I}_{k'}\}.
\end{equation} 
Under Model 2 however we can only schedule users that have eMBB messages to send:
\begin{equation}
\mathcal{T}_{\textnormal{eMBB},i} := \mathcal{K}_{\textnormal{eMBB}} \backslash ( \mathcal{T}_{\textnormal{URLLC},i} \cup \mathcal{T}_{\textnormal{silent},i}).
\end{equation}
\end{subequations}
 
 Since we can use an unlimited number of cooperation rounds,  each scheduled eMBB Tx  can send at full MG. The probability that a user $k$ will be scheduled as an eMBB user in scheme $i$ is: 
 \begin{itemize}
 \item If $k\in \mathcal{K}_i$ it is $\rho(1-\rho_f)$. (This is the probability that Tx $k$ has an eMBB but no URLLC message to send.)
 \item If $k\notin \mathcal{K}_i$ it is $\rho(1-\rho_f)(1-\rho\rho_f)^3$.
 (This is the probability that none of the  3 adjacent Txs in $\mathcal{K}_i$ has a URLLC message to send and Tx $k$ has an eMBB message to send.) 
 \end{itemize}

By above considerations, the scheme thus achieves the per-user MG pair 
  \begin{IEEEeqnarray}{rCl}
 \S^{(\U)}   &=&  \frac{\rho \rho_f}{3}\\
 \S^{(\e)}  &= & \rho(1-\rho_f ) \frac{2 (1- \rho\rho_f)^3}{3} + \frac{\rho (1- \rho_f)^2}{3}.
\end{IEEEeqnarray}
Time-sharing this scheme with the scheme in Appendix~\ref{App: C-A}  proves Theorem~\ref{theorem7}.

 \end{document}